\documentclass[a4paper,11pt]{article}
\usepackage[DIV10]{typearea}
\usepackage[T1]{fontenc}
\usepackage[bookmarksopen,colorlinks]{hyperref}
\usepackage{graphicx}
\usepackage[english]{babel}
\usepackage{amssymb}
\usepackage{geometry}
\usepackage{color}
\usepackage{xspace}
\usepackage{textcomp}
\usepackage{epic}
\usepackage{slashed}
\usepackage{eepic}
\usepackage{amsmath,bm}
\usepackage{epstopdf}
\usepackage{ifthen}
\usepackage{wasysym} 
\usepackage[footnotesize]{caption2}

\hypersetup{
   pdfstartview={FitH},
   pdfpagemode={UseOutlines}
}

\DeclareMathSymbol{\Gamma}{\mathalpha}{letters}{"00}
\DeclareMathSymbol{\Delta}{\mathalpha}{letters}{"01}
\DeclareMathSymbol{\Theta}{\mathalpha}{letters}{"02}
\DeclareMathSymbol{\Lambda}{\mathalpha}{letters}{"03}
\DeclareMathSymbol{\Xi}{\mathalpha}{letters}{"04}
\DeclareMathSymbol{\Pi}{\mathalpha}{letters}{"05}
\DeclareMathSymbol{\Sigma}{\mathalpha}{letters}{"06}
\DeclareMathSymbol{\Upsilon}{\mathalpha}{letters}{"07}
\DeclareMathSymbol{\Phi}{\mathalpha}{letters}{"08}
\DeclareMathSymbol{\Psi}{\mathalpha}{letters}{"09}
\DeclareMathSymbol{\Omega}{\mathalpha}{letters}{"0A}
\DeclareMathSymbol{\varGamma}{\mathalpha}{operators}{"00}
\DeclareMathSymbol{\varDelta}{\mathalpha}{operators}{"01}
\DeclareMathSymbol{\varTheta}{\mathalpha}{operators}{"02}
\DeclareMathSymbol{\varLambda}{\mathalpha}{operators}{"03}
\DeclareMathSymbol{\varXi}{\mathalpha}{operators}{"04}
\DeclareMathSymbol{\varPi}{\mathalpha}{operators}{"05}
\DeclareMathSymbol{\varSigma}{\mathalpha}{operators}{"06}
\DeclareMathSymbol{\varUpsilon}{\mathalpha}{operators}{"07}
\DeclareMathSymbol{\varPhi}{\mathalpha}{operators}{"08}
\DeclareMathSymbol{\varPsi}{\mathalpha}{operators}{"09}
\DeclareMathSymbol{\varOmega}{\mathalpha}{operators}{"0A}
\DeclareMathOperator{\sgn}{sign}
\renewcommand{\vec}[1]{\boldsymbol{#1}}
\newcommand{\D}{\mathrm{d}}

\def\beq{\begin{equation}}
\def\eeq{\end{equation}}
\def\bea{\begin{eqnarray}}
\def\eea{\end{eqnarray}}
\def\bi{\begin{itemize}}
\def\ei{\end{itemize}}

\newcommand{\stau}{{\widetilde{\tau}}}		
\newcommand{\stauR}{{\stau_\text{R}}}
\newcommand{\stauL}{{\stau_\text{L}}}
\newcommand{\stopo}{{\widetilde{t}}}
\newcommand{\sboto}{{\widetilde{b}}}		
\newcommand{\snutau}{{\widetilde{\nu}_\tau}}
\newcommand{\sq}{{\widetilde{q}}}
\newcommand{\go}{{\widetilde{g}}}

\newcommand{\mstau}{m_{\stau_1}} 
\newcommand{\mstop}{m_{\stopo_1}}
\newcommand{\msbot}{m_{\sboto_1}}
\newcommand{\msnutau}{m_{\snutau}}
\newcommand{\mstautwo}{m_{\stau_2}} 
   
\newcommand{\mne}{{m_{\s{\chi}^0_1}}}
  
\newcommand{\msq}{{m_{\widetilde{q}}}}
\newcommand{\mgo}{{m_{\widetilde{g}}}}

\newcommand{\thest}{\theta_{\stau}}

\newcommand{\MEV}{\ensuremath{\,\textnormal{MeV}}}
\newcommand{\GEV}{\ensuremath{\,\textnormal{GeV}}}
\newcommand{\TEV}{\ensuremath{\,\textnormal{TeV}}}
\newcommand{\SEC}{\ensuremath{\,\textnormal{s}}}

\newcommand{\fb}{\ensuremath{\,\textnormal{fb}}}

\newcommand{\ifb}{\ensuremath{\,\textnormal{fb}^{-1}}}

\newcommand{\s}[1]{\widetilde{#1}}

\newcommand{\Eqref}[1]{eq.~\eqref{#1}}
\newcommand{\Figref}[1]{figure~\ref{#1}}

\newcommand{\MO}{{\textsc{micrOMEGAs}}}
\newcommand{\FH}{{\textsc{FeynHiggs}}}
\newcommand{\HB}{{\textsc{HiggsBounds}}}

\newcommand{\sveff}{\langle \sigma_\text{eff}\,v_{\text{M\o l}} \rangle}
\newcommand{\sveffx}{\langle \sigma_\text{eff}\,v \rangle_x}
\newcommand{\Tf}{T_{\text{f}}}

\newcommand{\Mp}{M_{\text{Pl}}}

\newcommand{\xf}{x_{\text{f}}}

\newcommand{\GeV}{\ensuremath{\textnormal{GeV}}}

\newcommand{\cm}{\ensuremath{\textnormal{cm}}}

\newcommand{\be}{\beta}

\definecolor{evgray}{gray}{.48}
\definecolor{orange}{RGB}{255,127,0}
\definecolor{purple}{RGB}{147,112,219}
\definecolor{dgreen}{RGB}{0,180,10}

\begin{document}

\date{\mbox{ }}

\title{ 
{\normalsize  
October $10^{\text{th}}$, 2013 \hfill\mbox{}\\}
\vspace{1cm}
\bf 
A survey for low stau yields in the MSSM
\\[6mm]}

\author{Jan Heisig$^1$, J\"{o}rn Kersten$^2$, Boris Panes$^3$, Tania Robens$^4$  \\[2mm] 
{\small
${}^1${\it Institute for Theoretical Particle Physics and Cosmology,
RWTH Aachen,
Germany}}
\\
{\small
${}^2${\it II.~Institute for Theoretical Physics, University of Hamburg,
Germany}}
\\
{\small
${}^3${\it Instituto de F\'{i}sica, Universidade de S\~{a}o Paulo, Brazil}}
\\
{\small
${}^4${\it IKTP, TU Dresden,
Germany}}
\\
{\small\tt heisig@physik.rwth-aachen.de, joern.kersten@desy.de, }
\\
{\small\tt bapanes@if.usp.br, tania.robens@tu-dresden.de}
}

\maketitle

\thispagestyle{empty}

\vspace{1cm}

\begin{abstract}
\noindent
We study the implications of LHC results for the abundance of long-lived
staus after freeze-out from thermal equilibrium in a super-WIMP dark
matter scenario.  We classify regions in the MSSM parameter space
according to the stau yield, considering all possible co-annihilation 
effects as well as the effects of resonances and large Higgs-sfermion 
couplings.  Afterwards, we examine the viability of these regions after 
imposing experimental and theoretical constraints, in particular a Higgs 
mass around $125\GEV$ and null-searches for heavy stable charged 
particles (HSCP) at the LHC\@. We work in a pMSSM framework and 
perform a Monte Carlo scan over the parameter space.  To interpret 
the HSCP searches in our scenario, we consider all potentially important 
superparticle production processes, developing a fast estimator for NLO 
cross sections for electroweak and strong production at the LHC\@.  
After applying all constraints, we find that stau yields below $10^{-14}$ 
occur only for resonant annihilation via a heavy Higgs in combination 
with either co-annihilation or large left-right stau mixing. We encounter 
allowed points with yields as low as $2\times10^{-16}$, thus satisfying 
limits from big bang nucleosynthesis even for large stau lifetimes.
\end{abstract}

\clearpage

\tableofcontents

\section{Introduction}

One way to tackle the cosmological gravitino problem
\cite{Falomkin:1984eu,Ellis:1984eq} caused by late gravitino decays in
$R$-parity-conserving supersymmetry (SUSY) is to make the gravitino the
lightest supersymmetric particle (LSP) and thus stable
\cite{Ellis:1984er,Bolz:1998ek}.  This leads to an attractive scenario
where the gravitino accounts for the dark matter
\cite{Fayet:1981sq,Pagels:1981ke} whose density can match the observed
one for a relatively high reheating temperature after inflation
for gravitino masses in the GeV range \cite{Ellis:1984er,Bolz:1998ek}.
However, the next-to-LSP (NLSP) tends to be long-lived due to the very 
weak coupling of the gravitino. In this case, late decays of the NLSP
\cite{Moroi:1993mb} and catalysis effects \cite{Pospelov:2006sc}
can endanger the success of big bang nucleosynthesis (BBN); this is
sometimes called the NLSP decay problem.  The standard
models of particle physics and cosmology successfully describe BBN\@.
In the scenario with gravitino dark matter and a stau NLSP\@, $\stau_1$,
the preservation of this success translates into bounds on the stau lifetime
$\tau_{\stau_1}$ and the stau yield $Y$ after freeze-out from thermal equilibrium.  
For example, if $\tau_{\stau_1} \gtrsim 10^5\SEC$, which corresponds to gravitino 
masses $m_{\widetilde G} \gtrsim 300\GEV$ for a $1\TEV$ stau 
(or $m_{\widetilde G} \gtrsim 20\GEV$ for $\mstau=300\GEV$),
the yield is required to be
smaller than roughly $10^{-15}$ \cite{Jedamzik:2007qk,Kawasaki:2008qe}.
While the stau is in thermal equilibrium in the hot early universe, its
abundance decreases rapidly with time once the temperature falls below
its mass.  It freezes out from thermal equilibrium at a time determined
by the cross section for the annihilation of staus and possibly other
superparticles into Standard Model (SM) particles.  As a consequence,
smaller stau yields correspond to larger annihilation cross sections.

In this paper, we provide a classification of parameter space regions 
according to the stau yield, with emphasis on a thorough survey for
regions where the yield is exceptionally low, i.e., much smaller than
$10^{-13}$, which is the order of magnitude generically expected for a
stau with mass around $100\GEV$ \cite{Asaka:2000zh}.
We work in the framework of the phenomenological Minimal
Supersymmetric SM (pMSSM),
whose parameters are defined at low energies and thus immediately
applicable to the calculation of the annihilation cross section,
without first calculating the running from some high scale.
Systematically varying these parameters, we also allow all other
superparticles to become nearly mass-degenerate with the stau.  Then
co-annihilation effects can lead to large annihilation cross sections
and correspondingly small stau yields.

We first consider cases without significant left-right sfermion
mixing.  Then the order of magnitude of the annihilation cross section
is set by gauge couplings.
The results can therefore equally be adopted for the case of a smuon or
selectron NLSP\@.
In a second part we allow significant left-right mixing for the
third-generation squarks and finally for the staus themselves.  In such
cases, sfermion-Higgs couplings can become much larger than the gauge
couplings, leading to a strong enhancement of the annihilation cross
section.  In our survey we will encounter the known exceptional regions 
with resonance effects~\cite{Pradler:2008qc}
and enhanced Higgs-stau couplings \cite{Ratz:2008qh,Pradler:2008qc}, as
well as regions with interesting co-annihilation effects that had not
been studied in detail yet.  We also investigate which combinations
of effects produce interesting results.

Afterwards, we determine which parts of the parameter space are
currently allowed.  To this end, we perform a Monte Carlo scan and apply
the relevant experimental and theoretical constraints.  We require a
$CP$-even Higgs with a mass around $125\GEV$.  We also consider the
results of searches for the heavy Higgs particles of the MSSM\@.  In
order to take into account the LHC searches for heavy stable charged
particles (HSCP) and $R$-hadrons, we compute the
next-to-leading-order-corrected and next-to-leading-log-resummed cross
sections for the electroweak and strong production of superparticles
and reinterpret the cross section upper limits for the 7 and $8\TEV$ runs
reported by CMS \cite{CMS1305.0491}
within our scenario.  We consider the flavor and electroweak precision
observables that are most relevant for the scenario, namely the $W$ mass
as well as the branching ratios of $b \to s\gamma$ and
$B^0_s \to \mu^+\mu^-$.  As to theoretical constraints, we
impose the absence of charge- and color-breaking minima in
the scalar potential.
Combining all results will finally allow us to determine the range of
stau yields that is allowed by present constraints.

Our discussion reveals the most important parameters the stau yield
depends upon and is thus a first step towards answering the question
whether one could infer the stau yield from measurements at colliders.
Together with a measurement of the stau lifetime, a determination of
the yield could indicate whether the standard picture of the early
universe is consistent or whether it must be extended.
A number of such extensions have been proposed, for example, dilution of
the NLSP density by late-time entropy production \cite{Buchmuller:2006tt}.

Although the gravitino-LSP scenario serves as our primary motivation to
consider long-lived stau NLSPs, we would like to stress that this is not
the only possibility.  A long stau lifetime can occur in a number of
scenarios with different super-weakly interacting LSPs.  As our study is
independent of the exact properties of the LSP, it can be applied to all
these scenarios.  However, the precise constraints from BBN do vary for
different LSPs, which is why we will not consider them here.

The paper is organized as follows.  Based on considerations about the
freeze-out of MSSM sparticles, we will work out several general insights
into the dependence of the stau yield on the model parameters in
section~\ref{sec:surveygeneral}.  This discussion will set the stage for
the deductive classification of the pMSSM parameter space in
section~\ref{sec:systscan}.  Afterwards, we will describe a scan in the
17-dimensional pMSSM parameter space with a stau NLSP, applying current
theoretical and experimental constraints.  Their effects on the yield
will then be presented in section~\ref{sec:yieldMCscan}. We will conclude in
section~\ref{sec:Conclusion}.

\section{The freeze-out abundance of staus} \label{sec:surveygeneral}

In this section we will briefly review the physics of sparticle freeze-out and
mention the specific assumptions for our setup.
Suitably rewriting the solution of the Boltzmann equation
we deduct guidelines for the systematic survey of the parameter space
which we perform in section \ref{sec:systscan}.

The abundance of staus around the time of BBN is considered to arise from the freeze-out 
\cite{Lee:1977ua,Binetruy:1983jf,Bernstein:1985th,Srednicki:1988ce,Griest:1990kh,
Gondolo:1990dk,Kolb:1990vq,Edsjo:1997bg} of the stau NLSP, the lightest sparticle of the 
MSSM\@. We assume that all MSSM particles have been in thermal equilibrium
at some point during a hot phase in the early universe and that $R$-parity is exactly 
conserved. When the temperature of the universe decreases below the mass of the stau, 
the stau number density decreases exponentially. This exponential decrease is 
maintained as long as pair annihilation of staus are efficient enough to keep their number 
density close to the equilibrium number density. 
At the freeze-out temperature, which is typically of the order
$\Tf \sim \mstau/25$ \cite{Griest:1990kh}, the stau decouples from the thermal bath and 
its number density freezes out. Consequently, the number density
changes only because of the expansion of the universe, so the stau yield
\begin{equation}
Y \equiv \frac{n_{\stau_1}}{s}
\end{equation}
remains (approximately) constant, where $s$ is the entropy density.
Afterwards, for the considered case of a metastable stau NLSP, it decays into the 
LSP once the Hubble parameter becomes comparable to the decay rate.

This simple picture changes slightly when other MSSM sparticles are close in mass
with the stau {giving rise} to co-annihilation effects \cite{Binetruy:1983jf,Griest:1990kh}. 
In this case the annihilation of these sparticles competes with
their decay into the stau, and a simultaneous freeze-out of
several sparticles can occur. We assume that all heavier MSSM sparticles eventually decay
into the stau NLSP and not directly into the LSP\@. Moreover, we will require that all 
other MSSM sparticles have a lifetime smaller than $\sim10^{-2}\SEC$ (correspondingly,
$\Gamma\gtrsim10^{-22}\GEV$) ensuring that none of these 
decays take place during or after BBN\@. Consequently, all changes on
BBN solely depend on the stau yield. 
With these assumptions the desired number density of staus is simply the sum of the 
number densities of all relic sparticles that survive freeze-out. 
Although our requirement on the lifetime of the other sparticles 
will set a lower limit on the mass degeneracy of co-annihilating sparticles with the stau, 
in this section we will blithely consider exact mass degeneracy as a limiting case, 
keeping in mind that a certain separation is in fact required. 
The effects of taking all current limits into account are discussed in 
section~\ref{sec:yieldMCscan}.

An approximate solution of the Boltzmann equations can be written as \cite{Gondolo:1990dk,
Edsjo:1997bg}
\beq
 \frac{1}{Y(x_{0})} -  \frac{1}{Y(x_{\text{f}})} =
 \int_{x_{\text{f}}}^{x_{0}}\D x\,\pi\Mp\sqrt{\frac{8 \bar g}{45} }\,\frac{\mstau}{x^{2}}\sveff\,,
 \label{eq:boltzsolY}
\eeq
where $x=\mstau/T$ and $x_{\text{f}}$, $x_{0}$ are the corresponding quantities at the point 
of freeze-out and at the desired point of observation, respectively. Furthermore, $\sveff$ is 
the thermally averaged cross section times M\o ller velocity and $\bar g$ is a degrees of 
freedom parameter as defined in appendix \ref{sec:cosmo}. 
Besides, $\Mp=(8\pi G_{\text{N}})^{-1/2}$ is the reduced Planck mass.
The stau yield at freeze-out, $Y(x_{\text{f}})$, can only be computed by solving the Boltzmann 
equation numerically. For this, we will make use of  \MO~2.4.5~\cite{Belanger:2008sj}. 
Note that this program automatically sets $x_0 = \infty$, which is a good approximation as 
long as the stau is not extremely short-lived, $\tau_{\stau_1} \gtrsim 10^{-4} \SEC$.
An approximate solution can be found by neglecting the term $1/Y(x_{\text{f}})$ 
\cite{Gondolo:1990dk}. For the following discussion it is instructive to rewrite
\Eqref{eq:boltzsolY}, where we closely follow \cite{Berger:2008ti}. 
We can express the yield as 
\beq
  Y \propto \frac{\mstau}{\int_{x_{\text{f}}}^{x_{0}} \D x \, \sveffx} \,,
\label{eq:Ycoviprop}
\eeq  
where we have introduced the dimensionless thermally averaged cross section \cite{Edsjo:1997bg}
\beq
\sveffx =
\frac{
\sum_{ij}\int_{x_{i}+x_{j}}^{\infty}\!\D z \,z^{2} g_ig_j\tilde\sigma_{ij}
K_{1}(z) 
}{
4\left(\sum_i g_{i} x_i^2K_{2}(x_{i})\right)^2
}\,.
\label{eq:veffcovi}  
\eeq
Here, $i,j$ run over all supersymmetric particles (including the stau and the antistau)
involved in the (co-)annihilation with mass $m_i=x_iT$ and internal degrees of 
freedom~$g_i$. Besides, $K_{n}$ are the modified Bessel functions of the second kind 
of order $n$ and $z=\sqrt{s}/T$. Note that $\sveffx = x^{-2}\mstau^2\sveff$. 

The rescaled cross section $ \tilde{\sigma}$ is connected to the (usual) annihilation cross 
section by
\beq
  \tilde{\sigma}_{ij}= 
  \frac{
  \left(s - (m_i+m_j)^2\right)\left(s - (m_i-m_j)^2\right)
  }{s}
\,\sigma_{ij}\,;
\label{eq:sigma_dimensionless}
\eeq 
it is a function of dimensionless quantities only,
\beq
\tilde{\sigma}_{ij} =  
\tilde{\sigma}_{ij}\left(\frac{x_{i}}{z},\frac{x_{j}}{z},\{a_{\text{SUSY}}\},\{a_{\text{SM}}\}\right)\,,
\eeq
where $\{a_{\text{SUSY}}\}$ denotes a set of SUSY parameters each normalized by $\mstau$,
\beq a_{\text{SUSY}} = m_{\text{SUSY}}/\mstau\,,\eeq
and $\{a_{\text{SM}}\}$ is a set of SM parameters, normalized in the same way.\\

Considering eqs.~\eqref{eq:Ycoviprop} and \eqref{eq:veffcovi} there are three interesting 
observations that can be made and that are important for the subsequent discussion.
\begin{enumerate}
\item
\label{item:scalingbeh}
For fixed $a_{\text{SUSY}}$, and if $\tilde{\sigma}_{ij}$ is asymptotically independent of 
$a_{\text{SM}}$ for $a_{\text{SM}}\to0$ (which corresponds to the limit of large SUSY 
masses)\footnote{%
In this sense, $a_{\text{SM}}\,\rightarrow\,0$ should be read as $\mstau\,\rightarrow\,\infty$; 
in particular, electroweak gauge and Yukawa couplings are considered constant in this 
limit. Note that $a_\text{SM}\to0$ implies vanishing left-right sfermion mixing. However, 
we will see in section~\ref{sec:scalingdev} that the dependence of $Y$ on $\mstau$ can be 
close to \Eqref{eq:Ypropsimpl} even for significant left-right mixing.
}, 
$ \tilde{\sigma}_{ij}$ only depends on the ratio $x/z$. From this, it 
follows that the yield is simply proportional to the stau mass,
\beq
\label{eq:Ypropsimpl}
Y\propto\mstau\,,
\eeq
up to effects induced by the dependence of the choice of $x_{\text{f}}$ and the correction 
$1/Y(x_{\text{f}})$ on the stau mass, as well as effects of order $a_{\text{SM}}$.
\item
For a fixed initial state $i,j$, opening up additional final state channels can only enhance 
the cross section $\tilde{\sigma}_{ij}$ and thus lower the yield. Here, $m_A$ plays an
important role, as it determines the masses of the heavy Higgses, that could be either below 
or above their pair-production threshold. All other $R$-parity even particles are considerably 
lighter than the stau, if we take current direct search limits into account which require 
$\mstau\,\gtrsim\,340\GEV$ \cite{CMS1305.0491}.
\item
In contrast to opening up new final state channels, introducing
additional initial state channels in the presence of co-annihilation effects can 
either raise or lower the yield depending on the involved cross sections and the
additionally introduced degrees of freedom. For simplicity, we consider the 
limiting case of an exact mass degeneracy where the co-annihilation effects are
maximal. In this case
\beq
\sveffx
\simeq
\frac{\sum_{ij}\,\langle \sigma_{ij}
v \rangle_x g_i g_j}{\left(\sum_i g_i\right)^2}\,,
\label{eq:sigeffdeg}
\eeq
where we have introduced
\beq
  \langle \sigma_{ij} v \rangle_{x} =
\frac{
\int_{2x}^{\infty}\!\D z \,z^{2} \tilde\sigma_{ij} K_{1}(z) 
}{
4 x^4K_2^2(x)
}\,.
\label{eq:sigijdeg}  
\eeq
Additional initial states can only lower the yield if they introduce
large cross sections 
$\langle \sigma_{ij} v \rangle_{x}$ in the numerator that are 
capable of overcompensating the introduction of additional terms
in the sum over the degrees of freedom in the denominator.
For instance, the mere introduction of more sparticles with similar
interactions as the stau cannot decrease the yield further\footnote{%
At least not significantly. The introduction of additional Majorana fermions
as co-annihilating sparticles
could in principle reduce the yield by up to a factor of 2 if
all combinations of initial sparticles which include at least 
one Majorana fermion provide the same cross section.}---in contrast, 
if there are combinations $i,j$ that lead to smaller cross sections
$\langle \sigma_{ij} v \rangle_{x}$ than 
$\langle \sigma_{\stau_1\stau_1^*} v \rangle_{x}$, the numerator
in \Eqref{eq:sigeffdeg} increases less than the denominator and
we obtain a net increase of the yield. This is the case in the
slepton co-annihilation region, as we will discuss below.
Considering co-annihilating sparticles $i,j$ that introduce
cross sections much larger than the stau-stau annihilation
cross section,
$\langle \sigma_{ij} v \rangle_{x}\gg\langle \sigma_{\stau_1\stau_1^*} v \rangle_{x}$,
we can approximate \Eqref{eq:sigeffdeg} by 
\beq
\sveffx
\simeq
\frac{\sum_{ij\neq\stau_1,\stau_1^*}\,\langle \sigma_{ij}
v \rangle_x g_i g_j}{\left(\sum_{i\neq\stau_1,\stau_1^*} g_i+2\right)^2}\,.
\label{eq:sigeffdegstrong}
\eeq
The introduction of more and more sparticles $i,j$ of the same kind
could only lead to an asymptotical increase of $\sveffx$ towards the
value we would obtain by neglecting the stau degrees of freedom in the
denominator altogether. However, this saturation would only be achieved
if all cross sections $\langle \sigma_{ij} v \rangle_{x}$ were equally large.
This is usually not the case. For instance, introducing an additional squark 
generation in a squark co-annihilation scenario effectively reduces $\sveffx$ 
due to the smaller inter-generational interactions, i.e., the smaller cross 
sections $\langle \sigma_{ij} v \rangle_{x}$ for $i$, $j$ belonging to different
generations.
This is an important fact which severely restricts the possibilities 
for exceptionally small stau yields as the result of co-annihilation 
effects and enables an economical discussion of the various 
combinations of co-annihilating sparticles that could occur in the 
general pMSSM\@. In particular, it allows us to study the different 
cases in an isolated way. 
\end{enumerate}

After we have discussed the general scaling behavior of the yield with the initial state
sparticle masses as well as the behavior under the introduction of additional initial 
and final states, in the following we will comment on the parameters that govern the 
size of the cross section $\tilde{\sigma}_{ij}$ for fixed initial and final states.

There are basically two ways how the free SUSY parameters $a_{\text{SUSY}}$ could
affect the cross section $\tilde{\sigma}_{ij}$.
One is the strength of the involved couplings. Besides the known SM gauge couplings
the MSSM contains the couplings of the sfermions $\s f$ to the Higgses which involve the 
trilinear soft terms $A_f$, $\tan\beta$ and the higgsino mass parameter $\mu$ as
a priori free parameters of the theory. Although subject to constraints (see section
\ref{sec:pMSSMsummary}) these couplings can be
very large \cite{Ratz:2008qh,Pradler:2008qc} 
and the resulting cross sections can even be larger than the ones for
processes dominated by the strong interaction.
All other couplings in the theory are
given by SM gauge couplings (multiplied by possible suppression factors $\le1$
due to mixings) or are proportional to the mass of the involved particle and thus 
do not introduce additional free parameters.
The second is the appearance of non-SM particles in the intermediate states of the
annihilation processes. On the one hand SUSY particles can appear in the
$t$-channel of the annihilation processes. On the other hand the heavy Higgses 
whose masses are determined by $m_A$ can appear in the $s$-channel.
Especially the latter effect can lead to a drastic enhancement of the cross section 
close to the resonant pole $m_A \simeq 2\mstau$ \cite{Pradler:2008qc}
(or $m_A \simeq2m_{\text{co-ann}}$ for the case of a co-annihilating particle).

With these general remarks in mind, we will now systematically explore the
different distinct regions in the pMSSM parameter space.

\section{A systematic survey in the pMSSM} \label{sec:systscan}

In this section we will give an overview of different regions in the pMSSM parameter space 
characterized by the physical processes governing the stau yield and the resulting ranges 
of values.

\begin{itemize}
\item
In section \ref{sec:staupair} we will consider the case of no sfermion mixings 
and no co-annihilation effects. In {the} case {of no sfermion mixings}, all 
couplings are determined by the known gauge couplings and masses of the
involved particles, as well as the field decomposition in the EW gaugino sector. 
Then the only parameters governing the yield are the stau mass and the 
masses and mixings of EWinos appearing in the $t$-channel diagrams. We 
consider a purely right-handed stau as well as a purely left-handed stau and 
investigate the dependence of the yield on the variation of both stau masses 
and the EWino mass parameters $M_1$, $M_2$ and $\mu$.

\item
In section \ref{sec:coannnomix} we allow for co-annihilation effects by 
approaching the region $a_i\lesssim1.1$,  where we successively
consider co-annihilation with sleptons, gauginos, and squarks. Naturally, 
the yield in this case additionally depends on the respective value(s) of 
$a_i$ and thereby the mass of the additional co-annihilating sparticle(s), 
as well as the handedness of the NLSP. Furthermore, for the case of EWino 
co-annihilation we vary $m_A$ and thus examine the effect of the opening 
up of additional Higgs final state channels and Higgs resonances. 

\item
In section \ref{sec:largemix} we finally allow for significant left-right mixing of 
the sfermions, thereby enabling large sfermion-Higgs interactions either in the 
third-generation squark sector (in a co-annihilating setup) or in the stau sector. 
In both cases, by varying $m_A$, effects of additional Higgs final state channels 
and Higgs resonances are studied. Besides the respective mixing angles and 
masses, in addition there is  a strong dependence for diagrams involving light 
or heavy $CP$-even Higgses in the final or intermediate state  on the 
$A$-parameters of the third-generation squarks as well as $\mu$ and  $\tan \beta$. 
\end{itemize}

If not stated otherwise, we set all rescaled mass parameters $a_i \equiv m_i/\mstau$ 
that are not under consideration to $a_i=4$, which we consider sufficiently large to 
ensure that processes containing these particles are significantly suppressed and 
generically do not contribute to the stau yield. All spectra in this section are 
calculated using \textsc{SuSpect}~2.41~\cite{Djouadi:2002ze} at leading order.\footnote{%
For the study of idealized cases in this section we switched off the higher-order
corrections in the spectrum generation by setting \texttt{ICHOICE(7)=0}. 
The computation of the yield in the Monte Carlo scan in section~\ref{sec:yieldMCscan} 
contains the full radiative corrections provided by \textsc{SuSpect}.
}

\subsection{Stau pair annihilation in the absence of left-right mixing} \label{sec:staupair}

In this subsection we consider stau annihilation in the absence of large 
left-right mixing and co-annihilation effects with other sparticles.\footnote{%
For a left-handed lighter stau we have to take sneutrino co-annihilation 
effects into account.
}
If not stated otherwise, in order to achieve purely right- or left-handed mass 
eigenstates we choose a low value for $\tan\beta$ ($\tan\beta=2$) and enforce the 
cancelation $A_{\tau} = \mu\tan\beta$, so that $X_{\tau}\equiv\,A_\tau-\mu\,\tan\be$ 
is zero and thus the stau mass matrix is diagonal, cf.\ appendix~\ref{app:con}.

\begin{figure}[thp]
\centering
\setlength{\unitlength}{1\textwidth}
\begin{picture}(0.8,0.35)
 \put(-0.024,0){ 
  \put(-0.03,0.025){\includegraphics[scale=1.1]{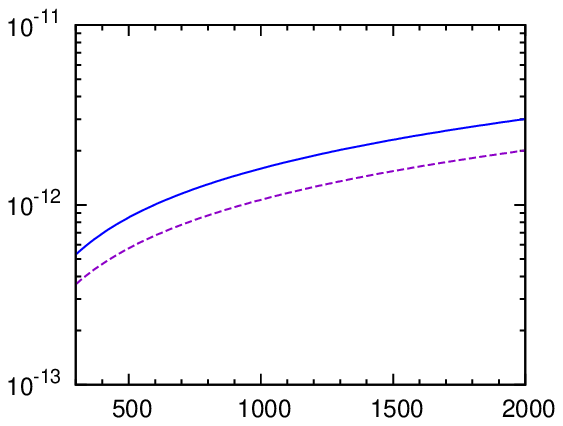}} 
   \put(0.11,0.23){\footnotesize $\stau_1 = \stau_{\text{R}}$}
   \put(0.3,0.196){\footnotesize $\stau_1 = \stau_{\text{L}}$}
  \put(0.19,0.0){\footnotesize $\mstau\,[\GEV\,]$}
  \put(-0.03,0.187){\rotatebox{90}{\footnotesize $Y$}}
  }
 \put(0.45,0.0){ 
  \put(-0.03,0.025){\includegraphics[scale=1.1]{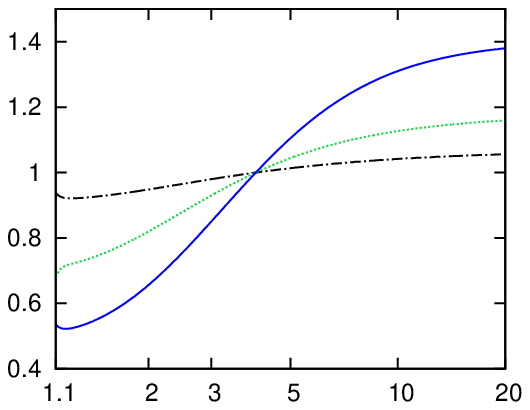}} 
   \put(0.11,0.098){\tiny $\stau_1 = \stau_{\text{R}}, M_1$}
   \put(0.28,0.2){\tiny $\stau_1 = \stau_{\text{L}}, M_1$}
   \put(0.294,0.255){\tiny $\stau_1 = \stau_{\text{L}}, M_2$}
  \put(0.18,0.0){\footnotesize $M_i/\mstau$}
  \put(-0.015,0.08){\rotatebox{90}{\footnotesize $Y(M_i)/Y (M_i\simeq4\mstau)$}}
  }
\end{picture}
\caption{%
Left panel:~Stau yield $Y$ as a function of the stau mass $\mstau$ for a right-handed 
lighter stau as well as for a left-handed lighter stau mass-degenerate with the tau 
sneutrino (to be considered as a limiting case for realistic spectra). All other SUSY 
mass parameters are set to $4\mstau$. Right panel:~Effects of EWinos in the 
$t$-channel. The curves show the yield as a function of $M_i/\mstau$ for the bino mass 
parameter, $i=1$, with $\stau_1=\stauR$ (blue solid curve) and $\stau_1=\stauL$ 
(black dot-dashed curve) as well as for the wino mass parameter, $i=2$, with 
$\stau_1=\stauL$ (green dotted curve). All curves are normalized to their respective 
value at $M_i/\mstau=4$.
}
\label{fig:mstau-yield}
\end{figure}

In this case, the stau yield depends only on the stau mass and on the masses and 
mixings of EWinos appearing in the $t$-channel of the annihilation processes. 
Figure~\ref{fig:mstau-yield} shows the stau yield as a function of the stau
mass for a purely right-handed and purely left-handed lighter stau $\stau_1$. 
For this plot, the tau sneutrino mass is set to $\mstau$ by hand for 
$\stau_1 = \stau_{\text{L}}$.\footnote{%
Strictly speaking, this choice should be considered as a limiting case which is not 
a valid point in the MSSM, since $\msnutau < \mstau$ for a purely left-handed lighter 
stau. However, it approximates nearby valid points with an almost left-handed lighter 
stau and a slightly heavier sneutrino.
} 

As we expect from the discussion above, the stau yield has an almost linear 
dependence on the stau mass. In fact, the expressions
\beq
       Y_{\stau_1 = \stau_{\text{R}}}= 1.59\times 10^{-12} 
        \left(\frac{\mstau}{1\TEV}\right)^{0.9}
\label{eq:YstauR}
\eeq
for a right-handed lighter stau and 
\beq
        Y_{\stau_1 \simeq \stau_{\text{L}} } = 1.07\times 10^{-12} 
        \left(\frac{\mstau}{1\TEV}\right)^{0.9}
\label{eq:YstauL}
\eeq
for a left-handed lighter stau describe the results in the given range at a percent level fit accuracy. 

In \Eqref{eq:Ypropsimpl}, we have argued that we generically expect a constant scaling of the 
yield with respect to the stau mass if all other SUSY parameters are fixed and the effective cross
section is independent of SM-like scales, which are parametrized by $a_{\text{SM}}$.
Therefore, the deviation from this expected scaling behavior in the above expressions calls for a 
more thorough investigation. Indeed, if the cross section defined by \Eqref{eq:sigma_dimensionless} 
exists and is finite in the limit $a_\text{SM}\,\rightarrow\,0$, \Eqref{eq:Ycoviprop} is expected to hold 
\emph{independent} of the $x$-dependence of $\langle \sigma_\text{eff} v \rangle_x$. Consequently, 
these deviations must come from the approximations employed in deriving \Eqref{eq:Ycoviprop}, i.e., 
from the freeze-out approximation with constant $\xf$ and from neglecting $1/Y(\xf)$. Indeed, the 
value of $\xf$ chosen by \MO\ varies with $\mstau$.

Considering the case of the right-handed stau, for our choice, $m_i/\mstau \simeq 4$, the dominant 
annihilation processes are $\stau_1\stau_1\to\gamma\gamma$ ($38\%$), 
$\stau_1\stau_1\to\tau\tau$ via (the bino content of) neutralinos in the $t$-channel ($30\%$) and 
$\stau_1\stau_1\to \gamma Z$ ($23\%$) followed by $\stau_1\stau_1\to Z Z$ ($4\%$) and 
$\stau_1\stau_1\to WW$ ($1.5\%$).\footnote{%
Whenever we give percentages of contributions, we refer to the
importance of the respective process for the final yield as reported by \MO.
}
For the left-handed lighter stau the unavoidable co-annihilation processes involving the tau 
sneutrino are important. As it is typical for co-annihilation scenarios, many processes contribute 
similarly strong to the annihilation. For $m_i/\mstau \simeq 4$ the most important channels
(contribution more than 10\%) are $\stau_1\stau_1\to WW$ ($15\%$), 
$\s\nu_{\tau}\s\nu_{\tau}\to WW$ ($14\%$), $\stau_1\s\nu_{\tau}\to \gamma W$ ($13\%$) 
and $\s\nu_{\tau}\s\nu_{\tau}\to ZZ$ ($12\%$). 

Effects of varying the mass of the bino and wino appearing in the $t$-channel diagrams are
shown in the right panel of figure \ref{fig:mstau-yield}. The additional $t$-channel
diagrams have the effect of lowering the stau yield.
For the right-handed stau, when lowering $M_1$ the channel $\stau_1\stau_1\to\tau\tau$ 
becomes more important reaching a maximum of around 65\% for an approximate degeneracy 
of the bino-like neutralino and the stau, i.e., $ M_1/\mstau \simeq 1.1$ 
(where co-annihilation effects are still small). Since the other channels (listed above) are
not affected by the variation of $M_1$, their absolute contribution remains unchanged and
the yield drops by a factor of roughly 0.5, accordingly.
For $M_1/\mstau=1.1$, the pre-factor in \Eqref{eq:YstauR} would become $8.55\times 10^{-13}$,
in rough agreement with the results given in \cite{Steffen:2006hw,Asaka:2000zh}.
For a completely decoupled bino the yield pre-factor in \Eqref{eq:YstauR} would
change to $2.24\times 10^{-12}$ according to the missing channel $\stau_1\stau_1\to\tau\tau$.
For the left-handed stau the $t$-channel diagrams are less important in comparison. 
For the case of small $M_2$ (but again above the region where co-annihilation
is efficient) the $t$-channel processes $\stau_1\s\nu_{\tau}\to \tau \nu_{\tau}$, 
$\s\nu_{\tau}\s\nu_{\tau}\to \nu_{\tau} \nu_{\tau}$ and $\stau_1\stau_1\to \tau \tau$  
(each of which contributes around $13\%$) become the most important processes followed 
by $\stau_1\stau_1\to WW$ ($10\%$). In contrast, lowering the bino mass does not have a 
large effect on the yield of the left-handed stau; in particular, the respective $t$-channel 
processes do not become the leading contributions. The (absolute) yield for right-handed 
staus even becomes smaller than the yield for left-handed staus for $M_1/\mstau\simeq1.1$ 
\cite{Pradler:2008qc}. In all cases the exponent of the $\mstau$-dependence of the yield 
stays approximately constant when varying $M_1/\mstau$ or $M_2/\mstau$.

\subsection{Co-annihilation regions} \label{sec:coannnomix}

Co-annihilation effects can be important whenever the mass splitting $\Delta m$
between the stau NLSP and the next-heavier sparticle(s) is of the order of the
freeze-out temperature, $\Delta m/\mstau \simeq x_\text{f}^{-1}$.
Given that the typical freeze-out temperature corresponds to $x_{\text{f}}\simeq25$, 
co-annihilation effects are expected to be significant for relative mass degeneracies 
of around $5$--$10\%$ \cite{Griest:1990kh}.

We will now systematically investigate how the stau yield changes if additional 
co-annihilating sparticles are introduced. Further, exemplarily we show how simple 
estimates including the exact consideration of all degrees of freedom can successfully 
predict the relative change in the yields. In the following we consider
$Y_{\stau_1 = \stau_\text{R}}$ and $Y_{\stau_1 \simeq \stau_\text{L}}$
from eqs.~\eqref{eq:YstauR} and \eqref{eq:YstauL} as reference yields that we 
normalize our results to.  We choose $\mstau=1000\GEV$ and we systematically vary 
the mass ratios $m_i/\mstau$ for the different sparticle species $i$ in order to study 
co-annihilation effects of the sparticles in the MSSM in an isolated way. If not stated 
otherwise we vary the corresponding soft masses and plot the physical sparticle mass.
If several sparticle masses are governed by one parameter that is subject to variation, 
we plot the smallest among these sparticle masses, if not stated otherwise. For example, 
if we vary the soft mass of the left-handed sleptons, we plot the sneutrino mass.

\begin{figure}[tbhp]
\centering
\setlength{\unitlength}{1\textwidth}
\begin{picture}(0.85,0.76)
 \put(0.45,0){  
   \put(-0.03,0.025){\includegraphics[scale=1.1]{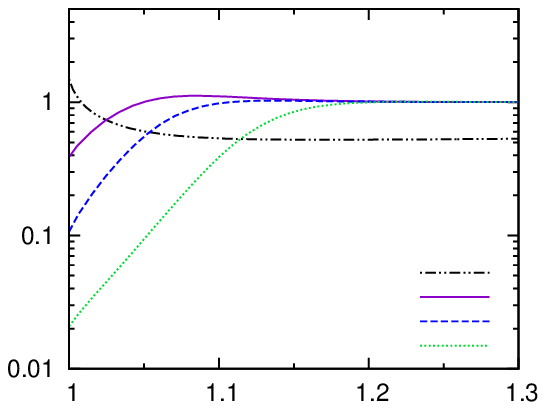}}
   \put(0.306,0.13){\tiny $\s B$}
   \put(0.306,0.112){\tiny $\s H$}
   \put(0.305,0.0933){\tiny $\s W$}
   \put(0.309,0.078){\tiny $\go$}
   \put(0.31,0.16){\footnotesize $\stau_1 = \stauR$}
  \put(0.2,0.005){\footnotesize $m_i/\mstau$}
  \put(-0.015,0.087){\rotatebox{90}{\footnotesize $Y(m_i)/Y (m_i\simeq4\mstau)$}}

  }
 \put(-0.024,0){ 
  \put(-0.03,0.025){\includegraphics[scale=1.1]{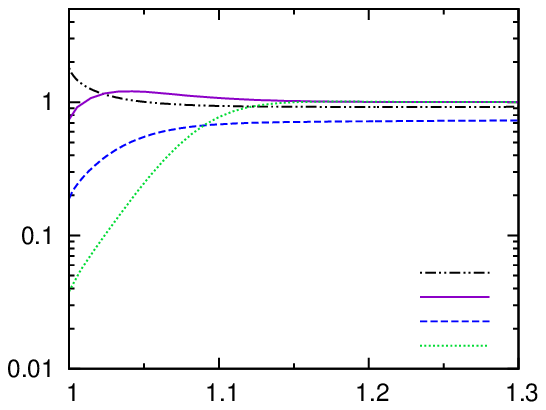}} 
   \put(0.306,0.13){\tiny $\s B$}
   \put(0.306,0.112){\tiny $\s H$}
   \put(0.305,0.0933){\tiny $\s W$}
   \put(0.309,0.078){\tiny $\go$}
   \put(0.31,0.16){\footnotesize $\stau_1 = \stauL$}
  \put(0.2,0.005){\footnotesize $m_i/\mstau$}
  \put(-0.015,0.087){\rotatebox{90}{\footnotesize $Y(m_i)/Y (m_i\simeq4\mstau)$}}
  }
 \put(0.45,0.38){ 
  \put(-0.03,0.025){\includegraphics[scale=1.1]{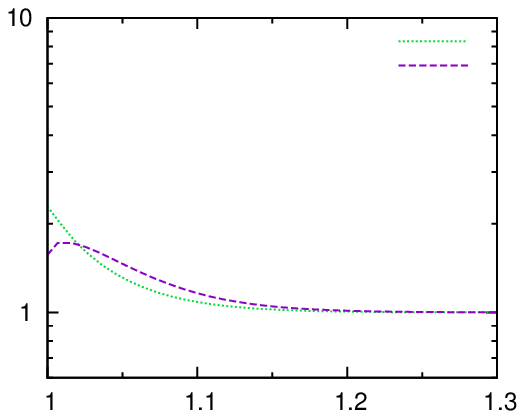}}
   \put(0.255,0.315){\tiny $\s e_{\text{R}},\s \mu_{\text{R}}$}
   \put(0.259,0.296){\tiny $\s e_{\text{L}},\s \mu_{\text{L}}$}
   \put(0.297,0.26){\footnotesize $\stau_1 = \stauR$}
  \put(0.2,0.0){\footnotesize $m_i/\mstau$}
  \put(-0.015,0.087){\rotatebox{90}{\footnotesize $Y(m_i)/Y (m_i\simeq4\mstau)$}}
  }
  \put(-0.024,0.38){ 
  \put(-0.03,0.025){\includegraphics[scale=1.1]{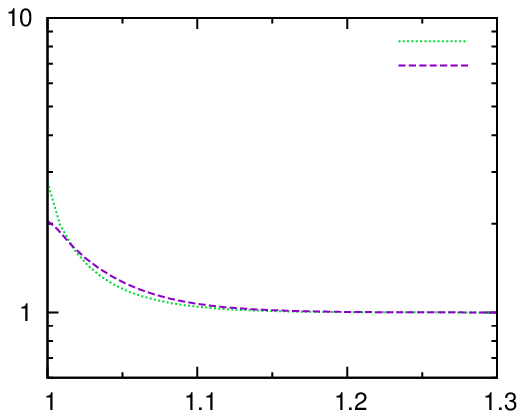} }
   \put(0.255,0.315){\tiny $\s e_{\text{R}},\s \mu_{\text{R}}$}
   \put(0.259,0.296){\tiny $\s e_{\text{L}},\s \mu_{\text{L}}$}
   \put(0.297,0.26){\footnotesize $\stau_1 = \stauL$}
  \put(0.2,0.005){\footnotesize $m_i/\mstau$}
  \put(-0.015,0.087){\rotatebox{90}{\footnotesize $Y(m_i)/Y (m_i\simeq4\mstau)$}}
  }
\end{picture}
\caption{%
Stau yield $Y$ for different co-annihilating sparticles $i$ normalized
to the respective stau yield at $m_i=4\mstau$. The plots are shown for 
a left-handed ligher stau $\stau_1\simeq\stauL$ (left panels) and for a 
right-handed lighter stau $\stau_1\simeq\stauR$ (right panels). Top:
degenerate sleptons. Bottom: degenerate gauginos.
}
\label{fig:slep-gaug-coann}
\end{figure}

\subsubsection{Co-annihilation with sleptons of the first and second generation}

The upper panels of figure \ref{fig:slep-gaug-coann} show the co-annihilation of staus 
with right- and left-handed sleptons of the first and second generation. The stau yield 
increases with an increasing importance of co-annihilation effects. This can be 
understood as follows. As an example, let us consider the case of a right-handed lighter 
stau which is mass-degenerate with the right-handed selectron and smuon. In the limit 
of complete degeneracy the denominator in \Eqref{eq:sigeffdeg} is enhanced by a factor 
of $9$. On the other hand, not all cross sections $\langle \sigma_{ij} v\rangle_{x}$ 
($i,j=1,2,3$ denote the generations) are equally large, as for $i\neq j$ only $t$-channel 
diagrams contribute. Accordingly, in the limit of a completely decoupled bino, the 
numerator in \Eqref{eq:sigeffdeg} increases only by a factor of $3$ and hence, since 
$Y\propto \sveffx^{-1}$,  the yield would increase by a factor of around $3$ with respect 
to the non-degenerate case. However, for the considered case of $M_1\simeq4\mstau$, 
the $t$-channel neutralino contribution is important. The six channels 
$\s\ell_i\s\ell_j\to\ell_i\ell_j$ contribute around $7.7\%$ each. For the choice 
$m_{\s B} \simeq1.1\mstau$,\footnote{$m_{\s B}$ denotes the mass of a bino-like lightest 
neutralino.} each channel contributes around $13\%$. Thus, the net increase of the stau 
yield is milder in the presence of the $t$-channel bino contributions and turns out to be 
2.3 and 1.8 for the former and latter choice for $m_{\s B}$, respectively.  This agrees with
the findings in \cite{Asaka:2000zh,Gherghetta:1998tq}.

\subsubsection{Co-annihilation with gauginos}

In this paragraph, we discuss the effects of co-annihilation of staus and gauginos from 
the electroweak as well as strong gauge groups. While the yield is generically lowered 
for sizeable co-annihilation cross sections, a special case is given for the 
co-annihilation with bino eigenstates, where we obtain an increase in the overall yield. 

The lower panels of figure \ref{fig:slep-gaug-coann} show the co-annihilation effects 
for gauginos. We vary $M_1,M_2,\mu$ and $M_3$ and plot the mass of the lightest 
EWino (which is the lightest neutralino) and the gluino, respectively. Hence, we 
consider (almost) pure gauge eigenstates in the EWino sector. The effects of large 
mixing in the EWino sector will be discussed in section \ref{sec:mAEwinos}. Due to the 
smaller annihilation cross section of the bino the net effect of bino co-annihilation 
increases the yield. This is, again, due to the increase in the degrees of freedom by a 
factor of 2 (for the right-handed lighter stau) or $3/2$ (for the left-handed lighter stau, 
accompanied by the tau sneutrino). At the same time the pair-annihilation of the binos 
is negligible and the associated co-annihilation of neutralinos with staus (or with tau 
sneutrinos) is sub-leading---in the limit of complete degeneracy the corresponding 
contributions add up to less than $25\%$ in the case of a right-handed light stau and 
$20\%$ in the case of a left-handed light stau. Consequently, for the right-handed stau 
we expect from eqs.~\eqref{eq:Ycoviprop} and \eqref{eq:sigeffdeg}
\begin{equation}
\frac{2g_\stauR^2 \langle \sigma_{\stauR\stauR^*} v \rangle_x}{(2g_\stauR+g_{\s{B}})^2}
\propto \frac{0.75}{Y(m_{\s{B}}=\mstau)} \,.
\end{equation}
(In this and the following estimates we make use of the fact that for annihilation 
processes without resonant or threshold effects the thermally averaged cross section 
can be expanded in $1/x$ where the leading contribution is independent of $x$ 
\cite{Gondolo:1990dk}.) At $m_{\s B} \simeq 1.1 \mstau$, where co-annihilation is 
already inefficient but the $t$-channel neutralino contributions are (still) maximal,
\begin{equation}
\frac{2g_\stauR^2 \langle \sigma_{\stauR\stauR^*}v \rangle_x}{(2g_\stauR)^2}
\propto \frac{1}{Y(m_{\s{B}}=1.1\mstau)} \,.
\end{equation}
Using $g_\stauR=1$ and $g_{\s B}=2$, we obtain
\begin{equation}
\frac{Y(m_{\s B}=\mstau)}{Y(m_{\s B}=1.1\mstau)} \simeq 3 \,,
\end{equation}
in good agreement with \Figref{fig:slep-gaug-coann} (lower right panel, dot-dashed 
curve).  For $\stau_1={\stauL}$, a similar estimate yields a net increase in the yield by 
a factor of around $2$ with respect to the yield at $m_{\s B} \simeq 1.1 \mstau$.

In contrast, for wino co-annihilation the annihilation of the wino-like neutralino and 
chargino among themselves is the dominant contribution.
For a right-handed lighter stau these contribute almost 100\% to the annihilation 
cross section while for a left-handed lighter stau they contribute more than 50\% followed 
by associated  co-annihilation processes of neutralino and chargino with the lighter stau 
and the tau sneutrino amounting to a contribution around 30\%. Hence, due to the larger 
annihilation cross sections of wino-like EWinos the stau yield is significantly reduced 
despite the 6 additional degrees of freedom introduced by the mass-degeneracy of one 
neutralino and chargino.

For higgsino co-annihilation the relative importance of annihilation and co-annihilation 
processes is not vastly different. However, due to the 8 additional degrees of freedom
the net reduction of the stau yield is less. In the case of a left-handed lighter
stau the yield even increases slightly at around $m_{\s H}/\mstau\simeq1.05$.

For the case of gluino co-annihilation the situation is even more pronounced than for the wino
case, since $\langle \sigma_{\go\go} v\rangle_x \gg \langle \sigma_{\stau_1\stau_1} v\rangle_x$
and $\sigma_{\go\stau}=0$.
In the case of a mass-degenerate gluino, eqs.~\eqref{eq:Ycoviprop} and \eqref{eq:sigeffdeg}
yield $Y\propto (g_\go+2g_{\stau})^2/g_\go^2$. Accordingly, the yields for a left-handed and 
a right-handed lighter stau differ only due to the extra degrees of freedom of the tau sneutrino,
$Y_{\stau_1 = \stauL}/Y_{\stau_1 = \stauR}\simeq 
(g_\go+2g_{\stau_1}+2g_{\s\nu_\tau})^2/(g_\go+2g_{\stau_1})^2\simeq1.2$.
(The relative yields in figure \ref{fig:slep-gaug-coann}, which are normalized to the respective 
yield without co-annihilation, show a larger difference, due to the difference in the reference 
yields.) Gluino pair annihilation processes are dominant up to a relative mass difference to 
the stau of 7\% and 6\% for a right-handed and left-handed lighter stau, respectively.

\subsubsection{Co-annihilation with squarks}

\begin{figure}[h!]
\centering
\setlength{\unitlength}{1\textwidth}
\begin{picture}(0.85,0.75)
 \put(-0.024,0){ 

  \put(-0.03,0.025){\includegraphics[scale=1.1]{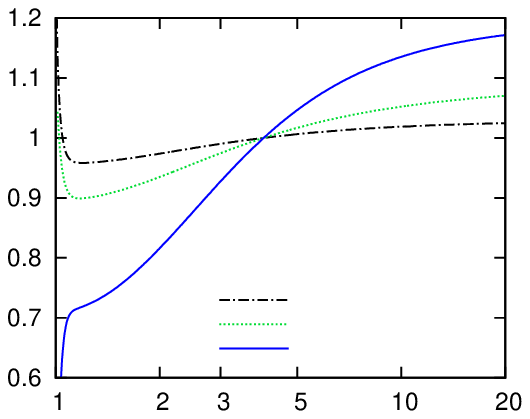}} 
   \put(0.29,0.157){\footnotesize $\stau_1 = \stauR$}
   \put(0.255,0.124){\tiny $\s t_{\text{R}}$ deg., $M_1$}
   \put(0.255,0.102){\tiny $\s b_{\text{L}},\s t_{\text{L}}$ deg., $M_2$}
   \put(0.255,0.08){\tiny $\s b_{\text{L}},\s t_{\text{L}}$ deg., $M_3$}
  \put(0.2,0.005){\footnotesize $m_i/\mstau$}
  \put(-0.015,0.087){\rotatebox{90}{\footnotesize $Y(m_i)/Y (m_i\simeq4\mstau)$}}
  }
 \put(0.45,0){  
  \put(-0.03,0.025){\includegraphics[scale=1.1]{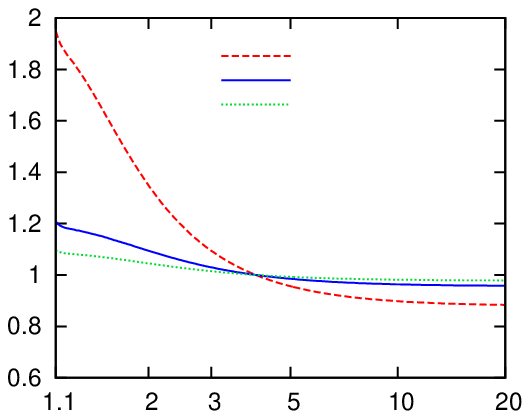}} 
   \put(0.183,0.228){\footnotesize $\stau_1 = \stauR$, $m_{\go}=\mstau$}
   \put(0.25,0.304){\tiny all squarks}
   \put(0.25,0.288){\tiny $m_{\s Q_{1,2}}$}
   \put(0.25,0.268){\tiny $m_{\s u_{1,2}}$}
  \put(0.2,0.005){\footnotesize $m_i/\mstau$}
  \put(-0.015,0.087){\rotatebox{90}{\footnotesize $Y(m_i)/Y (m_i\simeq4\mstau)$}}
  }
 \put(0.45,0.38){ 
    \put(-0.03,0.025){\includegraphics[scale=1.1]{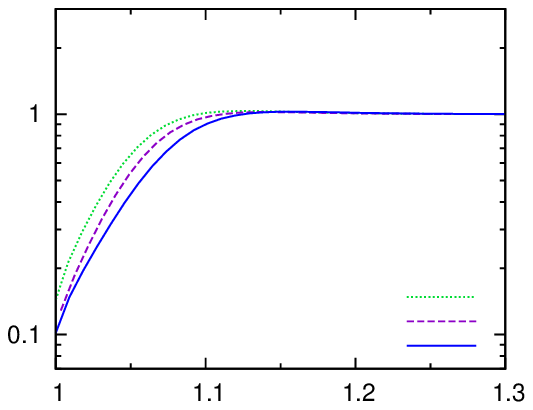}}
   \put(0.303,0.145){\footnotesize $\stau_1 = \stauR$}
   \put(0.29,0.114){\tiny $\s b_{\text{R}}$}
   \put(0.291,0.095){\tiny $\s t_{\text{R}}$}
   \put(0.264,0.075){\tiny $\s b_{\text{L}},\s t_{\text{L}}$}
  \put(0.2,0.005){\footnotesize $m_i/\mstau$}
  \put(-0.015,0.087){\rotatebox{90}{\footnotesize $Y(m_i)/Y (m_i\simeq4\mstau)$}}
  }
  \put(-0.024,0.38){ 
    \put(-0.03,0.025){\includegraphics[scale=1.1]{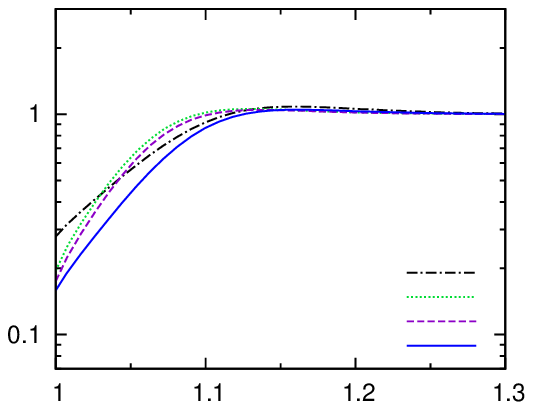}}
   \put(0.303,0.16){\footnotesize $\stau_1 = \stauR$}
   \put(0.144,0.132){\tiny all squarks of 1,2 gen.}
   \put(0.26,0.114){\tiny $\s d_{\text{R}},\s s_{\text{R}}$}
   \put(0.26,0.095){\tiny $\s u_{\text{R}},\s c_{\text{R}}$}
   \put(0.206,0.075){\tiny $\s d_{\text{L}},\s u_{\text{L}},\s s_{\text{L}},\s c_{\text{L}}$}
  \put(0.2,0.005){\footnotesize $m_i/\mstau$}
  \put(-0.015,0.087){\rotatebox{90}{\footnotesize $Y(m_i)/Y (m_i\simeq4\mstau)$}}
  }
\end{picture}
\caption{%
Upper panel:~Stau yield $Y$ for different co-annihilating squarks $i$ normalized
to the respective stau yield at $m_i = 4 \mstau$.  The plots show the case of a
right-handed lighter stau $\stau_1 = \stauR$. Lower left panel:~Effects of the variation
of the bino (black dot-dashed line), wino (green dotted line) and gluino (blue solid line)
mass parameter in a co-annihilation scenario with third-generation squarks. We adjusted 
$m_{\s u_3}$ and $m_{\s Q_3}$ such that the corresponding lighter sparticle (the stop
and sbottom, respectively) is exactly mass-degenerate with the stau.
Lower right panel:~Effects of the presence of squarks in the $t$-channel of gluino
co-annihilation diagrams. We adjust the gluino to be exactly mass-degenerate with the
stau and varied $m_{\s u_{1,2}}$ (green dotted curve), $m_{\s Q_{1,2}}$ (blue solid
curve) and all soft parameters of the three squark generations, namely $m_{\s u_{1,2,3}}$,
$m_{\s d_{1,2,3}}$ and $m_{\s Q_{1,2,3}}$ simultaneously (red dashed
curve).
}
\label{fig:sq-coann}
\end{figure}

In the upper left panel of \Figref{fig:sq-coann} we show the co-annihilation effects of the first 
two generation squarks for a right-handed stau. We vary the soft masses $m_{\s Q_{1,2}}$, 
$m_{\s u_{1,2}}$ and $m_{\s d_{1,2}}$ and plot the mass of the lightest among the squarks 
whose mass is dictated by the respective parameter. Although the involved strong 
interactions lead to relatively large cross sections (and in particular
$\sigma(\tilde{q}\tilde{q}\to X) \gg \sigma(\stau_{1}\stau_{1}\to X^{\prime})$)
the decrease in the yield is significantly less pronounced than in the case of gluino 
co-annihilation. We can understand this as follows, considering the case of a full 
degeneracy of the stau with the right-handed up-type squarks of the first two generations.
The dominant annihilation channels in this case are
$ \s u_\text{R} \s u_\text{R}, \s c_\text{R} \s c_\text{R} \rightarrow g g$
and contribute 72\% to the annihilation cross section. We compare this contribution with 
the case when there is no co-annihilation with squarks. In the latter case 
$\stau_{1} \stau_{1} \rightarrow \gamma \gamma$ contributes $38\%$ to the total 
annihilation cross section. From these numbers we can estimate the expected 
reduction of the yield. From eqs.~\eqref{eq:Ycoviprop} and \eqref{eq:sigeffdeg},
\begin{equation}
 \frac{4g_{\s u_\text{R}}^{2} \langle \sigma(\s u_\text{R} \s u_\text{R}^* \rightarrow g g) \,v \rangle_x
 }{
 (2g_{\stau_{1}}+ 4g_{\s u_\text{R}})^{2}
 }
\propto 
 \frac{0.72}{Y(m_{\s u_\text{R}}=\mstau)} 
\end{equation}
for the case of co-annihilation and
\begin{equation}
\frac{2g_{\stau_{1}}^{2}\langle \sigma(\stau_1 \stau_1^* \rightarrow \gamma \gamma) \,v \rangle_x}{(2g_{\stau_{1}})^{2}}
\propto 
\frac{0.38}{Y(m_{\s u_\text{R}}\simeq4\mstau)} 
\end{equation}
for the case without co-annihilation, and thus
\begin{equation}
 \frac{Y(m_{\s u_\text{R}}=\mstau)}{Y(m_{\s u_\text{R}}\simeq4\mstau)} \simeq
 5.2\,
\frac{\langle \sigma(\stau_1 \stau_1^* \rightarrow \gamma \gamma) \,v \rangle_x}{\langle \sigma(\s u_\text{R} \s u_\text{R}^* \rightarrow g g) \,v\rangle_x}
 \simeq 0.17 \,.
\end{equation}
We approximated the ratio between the two cross sections by the unaveraged cross 
sections in the non-relativistic regime as computed by \textsc{CalcHep} 
\cite{Belyaev:2012qa}. This estimate comes very close to the value that is displayed 
in the upper left plot of figure~\ref{fig:sq-coann} (purple dashed line).

Since for a close mass degeneracy the pair annihilation processes of squarks 
dominate over stau pair annihilation and associated squark-stau annihilation, the 
absolute stau yield for a left- and right-handed stau are virtually identical. This is why we
refrain from showing the corresponding plot for the left-handed stau in \Figref{fig:sq-coann}.
The main difference in such a plot would arise from the mere difference in the 
reference yields for left- and right-handed staus. 
   
The difference between the reductions of the yield for up- and down-type right-handed 
squarks arises solely from the different cross sections, since the number of degrees 
of freedom is exactly the same. The difference is induced from subdominant channels 
containing $\gamma\,g$ and $Z\,g$ final states. These contributions are 
sensitive to the charge of the corresponding squarks, leading to a smaller yield for 
the up-type squarks. 

In the case of degenerate left-handed squarks additional annihilation 
channels open up, namely the annihilation of up-type-down-type
squark pairs arising from diagrams with $t$-channel squarks or charginos 
as well as the four vertex contact interactions $\s u_{\text{L}} \s d_{\text{L}} 
\to Wg$. We found that the cross section for these processes containing electroweak
interactions are almost as large as those induced by strong interactions.
Furthermore, we observed a constructive interference between gluino and wino 
exchanging $t$-channel diagrams. In fact, this leads to a significant increase
of the effective thermally averaged cross section with respect to the case of right
handed squarks which overcompensate the doubling in the degrees of freedom. 
Finally, for the case $m_{\s Q_{1,2}}= m_{\s u_{1,2}}= m_{\s d_{1,2}}$ we observe 
a clear increase in  the yield as a mere result of an increase in the degrees of 
freedom relative to the cases considered before.

The upper right panel of \Figref{fig:sq-coann} shows the co-annihilation effects 
of the third-generation squarks. The relative behavior of the yield for the case
of $\s b_{\text{R}}$, $\s t_{\text{R}}$ and $(\s b, \s t)_{\text{L}}$ is comparable
to the yield in the corresponding cases for degenerate first two generations.
The overall decrease in the yield due to the co-annihilation effects is smaller
as the inter-generation initial states, for which the cross section
is considerably smaller, are absent.
It is interesting to note that channels with Higgs particles in the final states, which 
contain contact term interactions arising from the $F$- and $D$-terms in the scalar 
potential, are not suppressed in the absence of sfermion mixing. However, they do 
not play an important role even for the stop although the diagram arising from the
$F$-term is proportional to the Yukawa coupling squared.

Effects of gauginos appearing in the $t$-channel of the squark annihilation
processes are shown in the  lower left panel of \Figref{fig:sq-coann}. For
the blue solid and green dotted curve we fixed $m_{\s Q_3}$ such that
$m_{\s b_1}=\mstau$ ($m_{\s t_1}$ is slightly larger) and
varied $M_3/\mstau$ and $M_2/\mstau$, respectively. For the
black dot-dashed curve we fixed $m_{\s u_3}$ such that
$m_{\s t_1}=\mstau$ and varied $M_1/\mstau$.\footnote{%
Since the result is very sensitive to the mass of the co-annihilating particle, we enforced
the precise degeneracy of $\mstau$ and $m_{\s t_1}$ or $m_{\s b_1}$ by an 
iterative computation of the spectrum.
We will perform such an iterative procedure in
all cases where the result depends on the precise values of parameters 
that are required to be constant.
\label{fn:iterative}
}
All curves are normalized to the respective values at $a_i=4$.
As in the case of slepton annihilation, the $t$-channel contributions 
increase the effective annihilation cross section for small gaugino masses. 
However, the relative effect is smaller, cf.~right panel of \Figref{fig:mstau-yield}.
For $M_i/\mstau\lesssim1.1$ co-annihilation effects of gauginos become important.
For bino and winos these effects increase the yield due their smaller annihilation
cross sections relative to those of the squarks on the one hand and the additional 
degrees of freedom on the other hand.
For the gluino, co-annihilation effects lead to a further reduction of the yield despite the
additional degrees of freedom. In order to further understand the interplay between
squarks and gluinos, in the lower right panel of \Figref{fig:sq-coann} we fixed
$\mgo=\mstau$ and varied certain squark masses (according to our convention,
all others are kept at $4\mstau$ ). The contributions from $t$-channel squarks in the 
gluino-annihilation processes cause a destructive interference. For the green dotted 
and blue solid curve we vary the soft parameters $m_{\s u_{1,2}}$ ($m_{\s d_{1,2}}$ 
obviously give the same result) and $m_{\s Q_{1,2}}$, respectively. For the red dashed 
curve we varied all (bilinear)  soft mass parameters of the first- to third-generation 
squarks simultaneously. Interestingly, among scenarios with squark and gluino 
co-annihilation a scenario with a mass-degenerate gluino and decoupled squarks 
would have the smallest stau yield.

\subsubsection{Varying $m_A$ in the case of EWino co-annihilation} \label{sec:mAEwinos}

\begin{figure}[h!]
\centering
\setlength{\unitlength}{1\textwidth}
\begin{picture}(0.85,0.38)
\put(-0.024,-0.0){ 
  \put(-0.02,0.015){\includegraphics[scale=1.1]{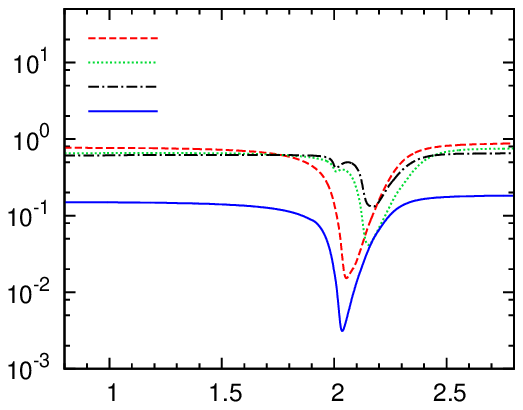} }
   \put(0.143,0.301){\tiny $M_1\!=\!\mu,$ deg.}
   \put(0.143,0.282){\tiny $M_2\!=\!\mu,$ 5\% off}
   \put(0.143,0.263){\tiny $M_1\!=\!\mu,$ 5\% off}
   \put(0.143,0.244){\tiny $M_2\!=\!\mu,$ deg.}
   \put(0.31,0.292){\footnotesize $\stau_1 \simeq \stauR$}
  \put(0.19,0.005){\footnotesize $m_A/\mstau$}
  \put(-0.015,0.087){\rotatebox{90}{\footnotesize $Y(m_i)/Y (m_i\simeq4\mstau)$}}
  }
\put(0.45,0.0){ 
  \put(-0.02,0.015){\includegraphics[scale=1.1]{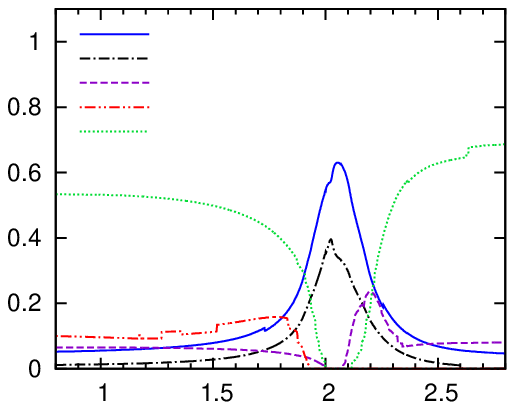} }
   \put(0.146,0.305){\tiny $\s\chi^{\pm}_1\s\chi^{\pm}_1\!\to\!\{t,b\}$}
   \put(0.146,0.286){\tiny $\s\chi^{0}_1\s\chi^{0}_1\!\to\!\{t,b\}$}
   \put(0.146,0.267){\tiny other  $\!\s\chi\s\chi\!\to\!\{t,b\}$}
   \put(0.146,0.248){\tiny $\s\chi\s\chi\!\to\!HX$}
   \put(0.146,0.229){\tiny others}
   \put(0.31,0.292){\footnotesize $\stau_1 \simeq \stauR$}
  \put(0.19,0.005){\footnotesize $m_A/\mstau$}
  \put(-0.02,0.1){\rotatebox{90}{\footnotesize rel.~contribution}}
  }
\end{picture}
\caption{%
Stau yield $Y$ in the presence of EWinos close in mass to the stau. We chose
the lightest EWinos to have maximal bino-higgsino and wino-higgsino mixing 
by taking $M_1=\mu$ and $M_2=\mu$, respectively. The bino and wino mass
parameters were adjusted to achieve either exact mass-degeneracy between 
the $\stau_1$ and the lightest neutralino (denoted by `deg.') or $\mne=1.05\mstau$ 
(denoted by `5\% off'). The right panel shows the relative annihilation contributions 
for a few classes of channels in the case $M_2=\mu$ and $\mne=\mstau$.
The abbreviation $\{t,b\}$ denotes all channels with only tops and/or bottoms in the 
final states. The abbreviation $HX$ denotes all channels with exactly one heavy 
Higgs field $H^0$, $H^{\pm}$ or $A^0$ in the final state.
}
\label{fig:EWino-mix-coan}
\end{figure}

We now vary the parameter $m_A$ in order to investigate the potential for 
changes in the cross sections of the EWino co-annihilation scenario due
to additional intermediate and final states, especially around the resonant
pole of an $s$-channel heavy Higgs and below the threshold for heavy
Higgs final states.
The EWino couplings to the Higgs are always of the type $\s H \s W \Phi$ or $\s H \s B \Phi$,
where $\s W$ and $\s B$ denote the wino and bino gauge eigenstates, $\s H$ the higgsino
gauge eigenstate and $ \Phi$ a Higgs. Hence, pair-annihilation of EWinos into Higgses requires
either a substantial higgsino admixture for the lightest EWinos or the presence of a 
higgsino-like EWino sufficiently light to significantly contribute in the $t$-channel. 
However, as we do not observe any change in the yield and the relative contributions when 
passing the heavy Higgs production threshold EWino pair annihilation into heavy Higgs final 
states is not an important channel (see \Figref{fig:EWino-mix-coan} at $m_A/\mstau\lesssim 1$).

Resonance effects occurring for $m_{\chi}\simeq2m_A$ are only important in the case of a 
large higgsino admixture in the lightest EWinos participating in the pair co-annihilation 
processes. Figure \ref{fig:EWino-mix-coan} shows the yield in a co-annihilation scenario 
where the lightest EWino is a bino-higgsino mixture ($M_1=\mu$, $M_2=4\mstau$) or a
wino-higgsino mixture ($M_2=\mu$, $M_1=4\mstau$). We vary $m_A$ and show the yield 
as a function of $m_A/\mstau$, for a complete degeneracy $\mne=\mstau$ and for a relative 
deviation of 5\%, $\mne=1.05\mstau$.
The resonant EWino co-annihilation can lower the yield by more than two orders of magnitude.
This is analogous to what happens in the $H/A$-funnel region of a neutralino LSP scenario
\cite{Drees:1992am,Arnowitt:1993mg,Baer:1997ai,Baer:2000jj,Ellis:2001msa,Lahanas:2001yr}.
These results were obtained for $\tan\beta=2$. However, we found very similar results with 
$\tan\beta=40$, although with a slightly shallower dip in the resonance. For $M_2=\mu$ as 
well as $\tan\beta=2$ and $\tan\beta=40$ the dominant annihilation channel is 
$\s\chi_1^\pm\to t\bar t$ and $\s\chi_1^\pm\to b\bar b$ in the resonance, respectively. 
The  small dip in the curves for $M_1=\mu$ slightly below the main resonance is caused 
by resonant annihilation of staus via the $CP$-even heavy Higgs.\footnote{%
In contrast to the scans we showed before, here we do not enforce exactly vanishing
stau mixing, since this would require very large values for $A_{\tau}$ for $\tan\beta=40$.  
Instead, we set $m_{\s e_3}\simeq1\TEV$, $m_{\s L_3}\simeq4\TEV$ and $A_{\tau}=0$,
accepting a very small left-right mixing.
}
For the case $M_1\simeq M_2\simeq\mu$, not shown here, the yield tends to be larger again
due to the additional degrees of freedom. 

The right panel of figure \ref{fig:EWino-mix-coan} shows the annihilation contributions for 
$M_2=\mu$ and $\mne=\mstau$.  The contribution $\s\chi\s\chi\!\to\!HX$ denotes
all channels with EWinos in the initial state and exactly one of the Higgs fields $H^0,H^{\pm},
A^0$ in the final state. (Channels with two Higgs fields in the final state contribute negligibly.) 
Independent of $m_A$, the contribution of channels with one light Higgs $h$ in the final states
is roughly a fifth of the remaining contributions denoted by `others'  in
the plot.

\subsection{Large sfermion mixings} \label{sec:largemix}

\subsubsection{Co-annihilation with mixed stops}

Still restricting ourselves to the case of small left-right mixing of the staus, we 
will now discuss the case of co-annihilation with squarks that acquire substantial
left-right mixing. The potentially large couplings of sfermions to the Higgses are 
proportional to the left-right mixing and proportional to the parameters appearing 
in the off-diagonal terms in the mass matrix.  We assume no particularly large $X_f$
in the first two generations and restrict the discussion to the third-generation sfermions,
i.e., to the case of a co-annihilating sbottom or stop. The couplings of  the sbottom and 
stop to the neutral, $CP$-even Higgses $h,H$ are summarized in 
appendix~\ref{sec:sfersferhiggs}\@. In the decoupling limit, $M_Z\ll m_A$, and for 
enhanced Higgs-sfermion couplings, these couplings can be approximated by
\begin{align}
C[h,\sboto_{1},\sboto_{1}] &\simeq
 \frac{g\, m_b}{2M_W}\left(A_b-\mu\tan\beta\right)\sin2\theta_{\s b}\equiv \hat C_{h,\sboto_{1}}\,, \\ 
C[H,\sboto_{1},\sboto_{1}] &\simeq
 \frac{g\, m_b}{2M_W}\left(A_b\tan\beta+\mu\right)\sin2\theta_{\s b}\equiv \hat C_{H,\sboto_{1}}\,, \\ 
C[h,\stopo_{1},\stopo_{1}]&\simeq
 \frac{g\, m_t}{2M_W}\left(A_t-\mu\cot\beta\right)\sin2\theta_{\s t} \equiv \hat C_{h,\s t_{1}}\,, \label{eq:clhstopdec}\\ 
C[H,\stopo_{1},\stopo_{1}]&\simeq
 \frac{g\, m_t}{2M_W}\left(A_t\cot\beta+\mu\right)\sin2\theta_{\s t}\equiv \hat C_{H,\s t_{1}}\,.
\end{align} 
We will here exemplarily
focus on the stop. We do not encounter potentially
larger enhancements of the couplings for sbottom co-annihilation. The smaller
Yukawa coupling, in contrast, tends to require larger SUSY parameters in order
to obtain the same coupling strength. Moreover, the couplings
of the sbottom and a stau are similar in the sense that they can become very large for large 
$\tan\beta$. In this concern it is more interesting to study the case of the stop being important
in a complementary corner of the parameter space, namely for smaller $\tan\beta$.
Furthermore, a significant left-right mixing of the stops is preferred from the requirement of 
large radiative corrections to the Higgs mass when interpreting the Higgs discovered at the 
LHC as the lighter neutral, $CP$-even Higgs $h$. 

\begin{figure}[h!]
\centering
\setlength{\unitlength}{1\textwidth}
\begin{picture}(0.8,0.76)
 \put(0.45,0.38){ 
  \put(-0.03,0.025){\includegraphics[scale=1.1]{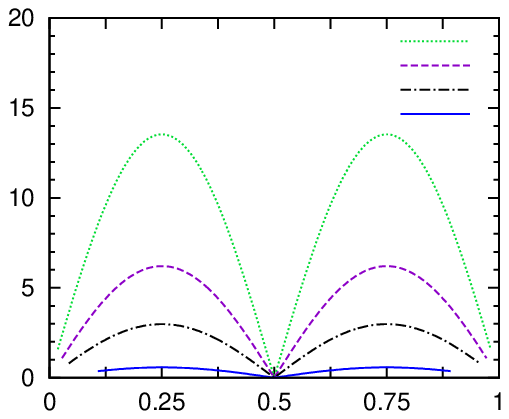}}
   \put(0.08,0.285){\footnotesize $\stau_1 = \stauR$}
   \put(0.194,0.313){\tiny $m_{\s t_2}/m_{\s t_1}\!\!=\!3.0$}
   \put(0.285,0.295){\tiny $2.0$}
   \put(0.285,0.277){\tiny $1.5$}
   \put(0.285,0.258){\tiny $1.1$}
  \put(0.2,0.005){\footnotesize $\theta_{\s t}/\pi$}
  \put(-0.02,0.15){\rotatebox{90}{\footnotesize $|X_t|/M_{\text{s}}$}}
  }
  \put(-0.024,0.38){ 
  \put(-0.03,0.025){\includegraphics[scale=1.1]{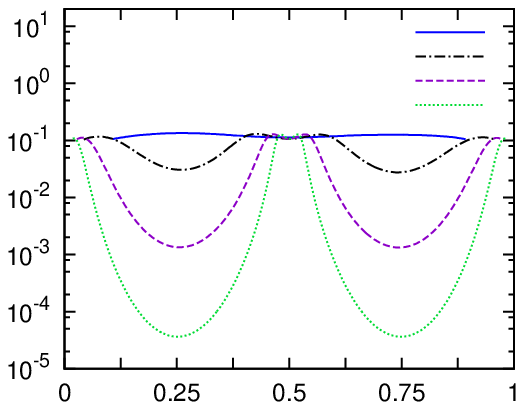} }
   \put(0.08,0.285){\footnotesize $\stau_1 = \stauR$}
   \put(0.194,0.313){\tiny $m_{\s t_2}/m_{\s t_1}\!\!=\!1.1$}
   \put(0.285,0.295){\tiny $1.5$}
   \put(0.285,0.277){\tiny $2.0$}
   \put(0.285,0.258){\tiny $3.0$}
  \put(0.2,0.005){\footnotesize $\theta_{\s t}/\pi$}
  \put(-0.035,0.087){\rotatebox{90}{\footnotesize $Y_{m_{\s t_1}\!=\,\mstau}/Y_{m_{\s t_1}\!\simeq\,4\mstau}$}}
  }
  \put(-0.024,0.0){ 
  \put(-0.03,0.025){\includegraphics[scale=1.1]{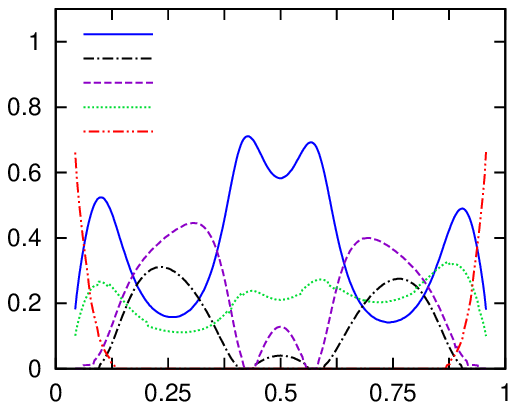}}
  \put(0.235,0.305){\footnotesize $m_{\s t_2}/m_{\s t_1}\!=1.5$}
    \put(0.144,0.313){\tiny $\s t\s t\!\to\! gg$}
   \put(0.144,0.294){\tiny $\s t\s t\!\to\! hh$}
   \put(0.144,0.275){\tiny $\s t\s t\!\to\! VV$}
   \put(0.144,0.256){\tiny $\s t\s t\!\to\,$others}
   \put(0.144,0.239){\tiny others}
  \put(0.2,0.005){\footnotesize $\theta_{\s t}/\pi$}
  \put(-0.02,0.11){\rotatebox{90}{\footnotesize rel.~contribution}}
  }
 \put(0.45,0.0){ 
  \put(-0.03,0.025){\includegraphics[scale=1.1]{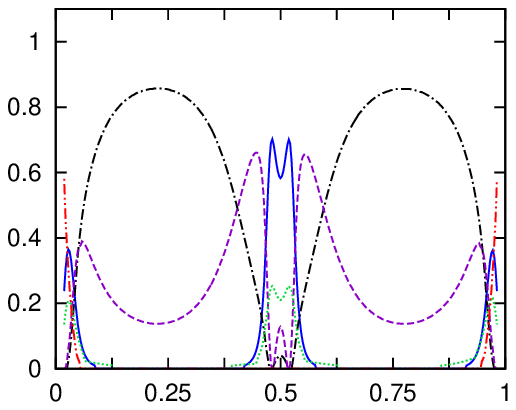} }
    \put(0.08,0.31){\tiny see left panel}
   \put(0.08,0.296){\tiny for legend}
  \put(0.25,0.305){\footnotesize $m_{\s t_2}/m_{\s t_1}\!=3$}
  \put(0.2,0.005){\footnotesize $\theta_{\s t}/\pi$}
  \put(-0.02,0.11){\rotatebox{90}{\footnotesize rel.~contribution}}
  }

\end{picture}
\caption{%
Upper panels:~Stau yield $Y$ (left panel) and stop mixing parameter $|X_t|/M_{\text{s}}$ 
(right panel) in the presence of a mass-degenerate stop as a function of $\theta_{\s t}$ for 
various choices of $m_{\s t_2}$. We set $\stau_1=\stauR$ and $\mstau = 1\TEV$. 
(The reference yield is $Y=1.59\times10^{-12}$.) Lower panels:~Relative contributions to 
the annihilation for $m_{\s t_2}/m_{\s t_1}=1.5$ (left panel) and $m_{\s t_2}/m_{\s t_1}=3$ 
(right panel). We do not display the curves for angles close to $0$ and $\pi$ where the mass 
of the lighter sbottom would run below the stop (and, consequently, the stau) mass. 
}
\label{fig:stop-mix-coan}
\end{figure}

Figure~\ref{fig:stop-mix-coan} shows the stau yield for a co-annihilating stop which is completely 
mass-degenerate with the stau NLSP\@. At tree-level, the leading contribution of the coupling of 
the stop to the light Higgs, \Eqref{eq:clhstopdec}, can be expressed solely by the spectrum 
parameters by using the analogon of \Eqref{eq:deltaM_NLOSP_LOSP} for the stop sector.
We chose $m_{\s t_1}=\mstau$ and varied $\theta_{\s t}$ for different choices of the mass of the
second stop, $m_{\s t_2}$. We set $\theta_{\s t}$ by fixing $\tan\beta = 5$ (as well as 
$\mu = 4\mstau$ as usual) and setting $A_{t}$ accordingly.
Note that this treatment of the parameters implicitly determines the mass of the lighter
sbottom, so further co-annihilation effects can take place that potentially
increase the yield. The Higgs mass was set to $m_h=126\GEV$ `by hand'. 
The result for the yield is, however, not sensitive to the actual value of the
Higgs mass. (The implications of the requirement to actually obtain this
Higgs mass from the radiative corrections in the stop sector are discussed
in section~\ref{sec:yieldMCscan}.)
 
As discussed before, strong interaction can already lead to a reduction of the yield 
by an order of magnitude. In the case of large stop-Higgs couplings the corresponding
processes become dominant and lead to a further significant reduction.
We obtain stau yields of $5\times10^{-14}$ and less than $10^{-16}$ for a mass splitting
of $m_{\s t_2}/m_{\s t_1}=1.5$ and $3$, respectively.
The lower panels of figure \ref{fig:stop-mix-coan} show the relative contributions
to the annihilation. Due to the exact mass-degeneracy of the stop with the stau, the
pair-annihilation processes of stops dominate over annihilation process involving the stau.
For a relatively small mass gap between the lighter and the heavier stop, 
$m_{\s t_2}/m_{\s t_1}=1.5$, i.e., for a moderate coupling 
$C[h,\s t_1,\s t_1] \lesssim m_{\s t_1}$, annihilation into gluino pairs and pairs of vector 
bosons are the dominant channels. For very large mass gaps, $m_{\s t_2}/m_{\s t_1}=3$, 
i.e., for large values of $C[h,\s t_1,\s t_1]$, the channel $\s t_1\s t_1\to hh$ becomes important. 
In this regime the leading contribution from $\s t_1\s t_1\to hh$  comes from the pair annihilation 
of stops via the $t$-channel diagram. This contribution involves two stop-stop-Higgs vertices. 
The cross section is therefore proportional to $C[h,\s t_1,\s t_1]^4$ while all contributions with 
an $s$-channel $h$ are only proportional to $C[h,\s t_1,\s t_1]^2$. For this reason the channels
$\s t_1 \s t_1\to VV$ and $\s t_1 \s t_1\to tt,bb$ become less important with larger mass gaps 
and larger left-right mixing of the stops.

The upper right panel of \Figref{fig:stop-mix-coan} shows the mixing parameter $|X_t|/M_{\text{s}}$, 
where $M_{\text{s}} = \sqrt{m_{\s t_1}m_{\s t_2}}$, corresponding to the lines drawn in the left panel. 
For $m_{\s t_2}/m_{\s t_1}=1.5$ and $3$ the mixing parameter is $|X_t|/M_{\text{s}} = 3$ and almost 
$14$, respectively. $|X_t|/M_{\text{s}}\simeq\sqrt{6}$ maximizes the positive radiative corrections to 
the Higgs mass \cite{Carena:1995wu,Heinemeyer:1999zf} and thus is preferred in the absence of 
overly large stop masses.

\subsubsection{Varying $m_A$ in the case of co-annihilation with mixed stops} \label{sec:stopmA}

\begin{figure}
\centering
\setlength{\unitlength}{1\textwidth}
\begin{picture}(0.85,0.38)
\put(-0.024,-0.0){ 
  \put(-0.02,0.015){\includegraphics[scale=1.1]{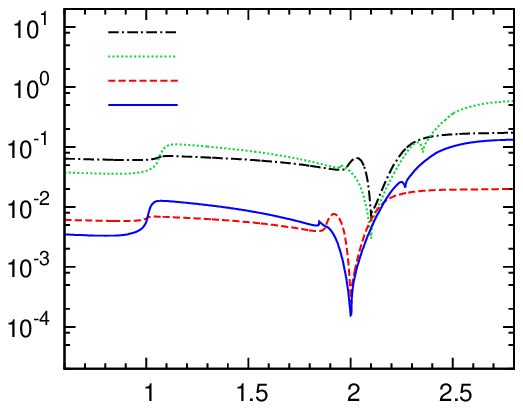} }
   \put(0.16,0.305){\tiny $\tan\beta\!=\!20$, 5\% off}
   \put(0.16,0.286){\tiny $\tan\beta\!=\!2$, 5\% off}
   \put(0.16,0.267){\tiny $\tan\beta\!=\!20$, deg.}
   \put(0.16,0.248){\tiny $\tan\beta\!=\!2$, deg.}
   \put(0.31,0.292){\footnotesize $\stau_1 \simeq \stauR$}
  \put(0.19,0.005){\footnotesize $m_H/\mstau$}
  \put(-0.015,0.087){\rotatebox{90}{\footnotesize $Y(m_i)/Y (m_i\simeq4\mstau)$}}
  }
\put(0.45,0.0){ 
  \put(-0.02,0.015){\includegraphics[scale=1.1]{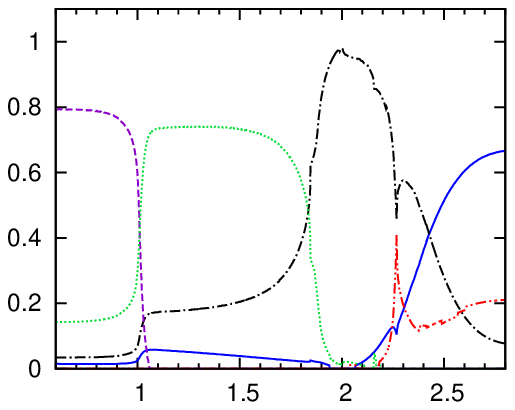} }
   \put(0.185,0.07){\tiny $\s t_1\!\s t_1\!\!\to\! gg$}
   \put(0.306,0.286){\tiny $\s t_1\!\s t_1\!\!\to\!\{t,b\}$}
   \put(0.08,0.258){\tiny $\s t_1\!\s t_1\!\!\to\!2\{H\!,\!A\}$}
   \put(0.136,0.205){\tiny $\s t_1\!\s t_1\!\!\to\!1\{H\!,\!A\}$}
   \put(0.274,0.14){\tiny others}
   \put(0.09,0.292){\footnotesize $\tan\beta\!=\!2$, deg.}
  \put(0.19,0.005){\footnotesize $m_H/\mstau$}
  \put(-0.02,0.11){\rotatebox{90}{\footnotesize rel.~contribution}}
  }
\end{picture}
\caption{Co-annihilation with a maximally left-right mixed stop ($\theta_{\s t} = \pi/4$ or $3\pi/4$) 
as a function of $m_H/\mstau$. Left panel: Relative yield for an exact degeneracy (blue, solid
and red dashed curves) as well as for $m_{\s t_1} = 1.05\mstau$ (denoted by `5\% off',
green, dotted and black, dot-dashed curves). We choose two sets of parameters, one with
$\tan\beta = 2$ and $A_t = 4\TEV$ (blue, solid and green, dashed curves) and one with 
$\tan\beta = 20$ and $A_t =- 4\TEV$ (red, dashed and black, dot-dashed curves). For both 
cases we choose $A_{\tau}=A_b=0$ and $\mu=m_i=4\TEV$, where $m_i$ stands for all other 
soft parameters not involved here. The soft parameters $m_{\s Q_3}$ and $m_{\s u_3}$
are determined by tree-level relations from the desired $m_{\s t_1}$ and $\theta_{\s t}$.
Right panel:~Relative contributions of the annihilation channels for the case $m_{\s t_1}=\mstau$,
$\tan\beta = 2$ and $A_t = 4\TEV$. The red curve is the contribution of all channels that are
not explicitly displayed.
}
\label{fig:stop-Hres-coan}
\end{figure}

If we relax our assumption $m_A \simeq 4 \mstau$ we can study the effects of heavy 
Higgs resonances and of opening up channels with heavy Higgs final states.
Figure \ref{fig:stop-Hres-coan} shows the yield in a co-annihilation scenario where the stop 
is maximally mixed, $\theta_{\s t} = \pi/4$ or $3\pi/4$, and $\mstau=1\TEV$ as a function of 
$m_H/\mstau$. We show the relative yield for an exact degeneracy (blue, solid and red 
dashed curves) as well as for $m_{\s t_1} = 1.05 \mstau$ (green, dotted and black, 
dot-dashed curves). We choose two sets of parameters, one with $\tan\beta = 2$ and 
$A_t = 4\TEV$ (blue, solid and green, dashed curves) and one with $\tan\beta = 20$ and 
$A_t =- 4\TEV$ (red, dashed and black, dot-dashed curves). For both cases we set 
$A_{\tau}=A_b=0$ and $\mu=4\TEV$. The soft parameters $m_{\s Q_3}$ and $m_{\s u_3}$
are determined by tree-level relations from the desired $m_{\s t_1}$ and $\theta_{\s t}$. 
Again, we use an iterative algorithm in order to control these parameters after spectrum 
generation.

In the right panel of figure~\ref{fig:stop-Hres-coan} we show the relative contributions of the
annihilation channels for the case $m_{\s t_1}=\mstau$, $\tan\beta = 2$ and $A_t = 4\TEV$. 
The red curve is the contribution of all channels that are not explicitly displayed. Its 
pronounced peak at around $2250\GEV$ is caused by the channel $\s t_1 \s b_1 \to t b$, 
which contributes around $38\%$. The mass of the sbottom is around $1150\GEV$.

Similar to the case of EWino co-annihilation, we obtain a strong reduction of the yield in the 
presence of a resonant pole. In contrast, we also see a decrease of the yield below the 
threshold for heavy Higgs final states. This effect is only pronounced for small $\tan\beta$ 
since the annihilation into final state heavy Higgses contributes significantly only for very 
large stop-Higgs couplings.

\subsubsection{Large stau mixing}\label{sec:largestaumixing}

\begin{figure}[tb]
\centering
\setlength{\unitlength}{1\textwidth}
\begin{picture}(0.85,0.78)
 \put(0.45,0){  
   \put(-0.03,0.025){\includegraphics[scale=1.1]{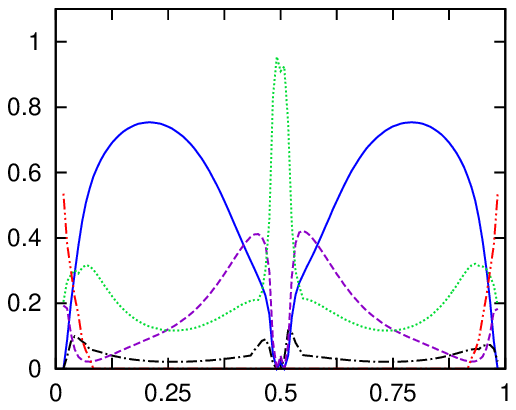}}
   \put(0.25,0.305){\footnotesize $m_{\s \tau_2}/\mstau\!=2$}
    \put(0.08,0.3){\tiny see left panel}
    \put(0.08,0.286){\tiny for legend}
  \put(0.2,0.005){\footnotesize $\thest/\pi$}
  \put(-0.02,0.11){\rotatebox{90}{\footnotesize rel.~contribution}}
  }
 \put(-0.024,0){ 
  \put(-0.03,0.025){\includegraphics[scale=1.1]{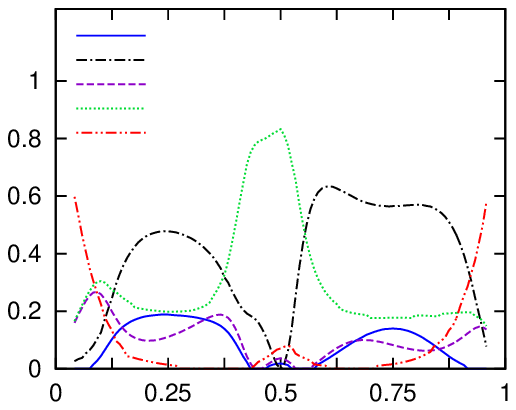}} 
   \put(0.13,0.312){\tiny $\stau\stau\!\to\! hh$}
   \put(0.13,0.293){\tiny $\stau\stau\!\to\! tt,bb$}
   \put(0.13,0.274){\tiny $\stau\stau\!\to\!WW$}
   \put(0.13,0.255){\tiny $\stau\stau\!\to\,$others}
   \put(0.13,0.239){\tiny others}
   \put(0.234,0.305){\footnotesize $m_{\s \tau_2}/\mstau\!=1.2$}
  \put(0.2,0.005){\footnotesize $\thest/\pi$}
  \put(-0.02,0.11){\rotatebox{90}{\footnotesize rel.~contribution}}
  }
 \put(0.46,0.39){ 
  \put(-0.03,0.025){\includegraphics[scale=1.1]{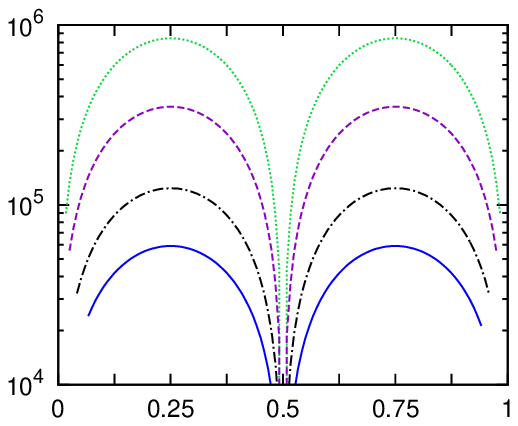}}
  \put(0.2,0.005){\footnotesize $\thest/\pi$}
  \put(-0.025,0.133){\rotatebox{90}{\footnotesize $|X_{\tau}|\,[\GEV\,]$}}
    \put(0.25,0.1){\tiny see left panel}
    \put(0.25,0.086){\tiny for legend}
  }
  \put(-0.024,0.39){ 
  \put(-0.03,0.025){\includegraphics[scale=1.1]{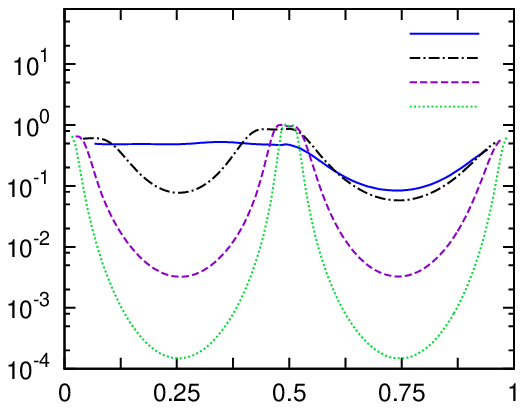} }
   \put(0.18,0.315){\tiny $m_{\s \tau_2}/\mstau=1.1$}
   \put(0.2818,0.296){\tiny $1.2$}
   \put(0.2818,0.277){\tiny $1.5$}
   \put(0.2818,0.258){\tiny $2.0$}
  \put(0.2,0.005){\footnotesize $\thest/\pi$}
  \put(-0.035,0.055){\rotatebox{90}{\footnotesize $Y/Y (\thest=0.5\,\pi,\,m_i\simeq4\mstau)$}}
  }
\end{picture}
\caption{%
Upper panels: Stau yield $Y$ (left panel) and the absolute value of $X_{\tau} = A_{\tau}-\mu\tan\beta$
(right panel) as a function of the stau mixing angle $\thest$
for $\mstau=1000\GEV$ and different choices of $m_{\stau_2}$ as specified in
the key of the upper left panel. We normalized the curves by the yield for a purely
right-handed lighter stau and $m_{\stau_2}/\mstau\simeq4$. 
The asymmetry around $\thest=\pi/2$ arises from the contribution of an
$s$-channel heavy Higgs (see text for details).
The lower panels show the relative contributions to the annihilation for
$m_{\s \tau_2}/\mstau=1.2$ (left panel) and $m_{\s \tau_2}/\mstau =2$ (right panel).
}
\label{fig:stauhiggs}
\end{figure}

We will now discuss large mixing in the stau sector itself. Accordingly we will switch off 
any avoidable effect of co-annihilation. In the decoupling limit, and for enhanced  
Higgs-stau couplings, these couplings read approximately
\begin{align}
C[h,\stau_{1},\stau_{1}] &\simeq
 \frac{g\, m_{\tau}}{2M_W} \left(A_\tau-\mu\tan\beta\right)\sin2\thest \equiv \hat C_{h,\stau_1}\,,\\ 
C[H,\stau_{1},\stau_{1}] &\simeq
 \frac{g\, m_{\tau}}{2M_W} \left(A_\tau\tan\beta+\mu\right)\sin2\thest\equiv \hat C_{H,\stau_1} \,.
\end{align}
We first vary the stau mixing angle while keeping $m_A \simeq 4 \mstau$.
Analogous to the case of the stop, we perform the scan for different choices of 
$m_{\stau_2}/\mstau$. We choose $\tan\beta = 20$, $\mu=4\mstau$ and achieve 
the required $X_\tau$ by choosing $A_{\tau}$ accordingly.
Figure \ref{fig:stauhiggs} shows that the yield can be reduced by several orders of magnitude
for  large mass splittings and significant left-right mixing, i.e., large couplings 
$C[h,\stau_{1},\stau_{1}]$. 
This result was first discussed in \cite{Ratz:2008qh,Pradler:2008qc}. In \cite{Pradler:2008qc} 
the authors equally scanned over $\thest$ but chose a fixed value for $X_\tau$. The results
obtained there are compatible with ours.
The upper right panel of \Figref{fig:stauhiggs} displays the size of $|X_\tau|$ which is required 
in order to provide the fixed ratio $m_{\stau_2}/\mstau$ when varying $\thest$. This reveals 
that low stau yields can only be obtained for very large values of $|X_\tau|$.

The upper left panel of figure \ref{fig:stauhiggs} shows that for a moderate 
mass ratio $m_{\stau_2}/\mstau$, the curves are not symmetric around 
$\thest=\pi/2$. This effect arises from the interference term of a heavy Higgs
in the $s$-channel of the annihilation processes $\stau_1\stau_1\to tt,bb,hh$.
While the coupling $C[h,\stau_{1},\stau_{1}] $ is completely
symmetric around $\thest=\pi/2$, $C[H,\stau_{1},\stau_{1}]$ is not. The asymmetry
could, however, be reduced or removed by another choice of $A_\tau$, $\mu$ and $\tan\beta$
to achieve the required $X_\tau$, or by a stronger decoupling of $m_A$.

In the lower panels of figure \ref{fig:stauhiggs} the dominant contributions to the annihilation
are shown. For $m_{\s \tau_2}/\mstau=1.1$, the channels $\stau_1\stau_1\to tt,bb$ are
the most important channels for large mixings. Theses channels involve one
stau-stau-Higgs coupling, their cross sections are thus proportional to $C[h,\stau_{1},\stau_{1}]^2 $
\cite{Endo:2010ya}. For $m_{\s \tau_2}/\mstau=2$, the channel $\stau_1\stau_1\to hh$
clearly dominates. Its leading contribution to the cross section in this regime is proportional 
to $C[h,\stau_{1},\stau_{1}]^4$ \cite{Ratz:2008qh,Pradler:2008qc}.

\subsubsection{Varying $m_A$ in the case of large stau mixing}\label{sec:survresonance}

\begin{figure}[h!]
\centering
\setlength{\unitlength}{1\textwidth}
\begin{picture}(0.85,0.86)
 \put(0.45,0){  
   \put(-0.03,0.025){\includegraphics[scale=1.1]{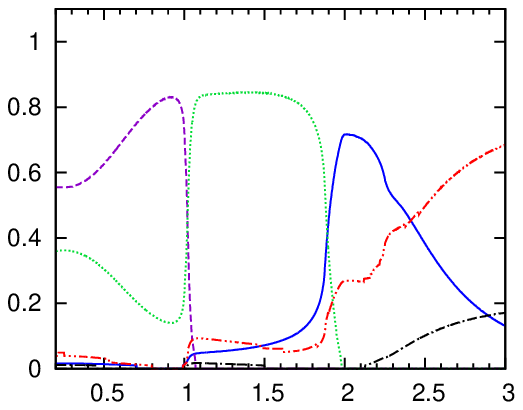}}
    \put(0.08,0.3){\tiny see left panel}
    \put(0.08,0.286){\tiny for legend}
    \put(0.235,0.305){\footnotesize $A_{\tau}/\mstau\!=-8$}
   \put(0.27,0.28){\footnotesize $\tan\beta=50$}
  \put(0.19,0.005){\footnotesize $m_H/\mstau$}
  \put(-0.02,0.11){\rotatebox{90}{\footnotesize rel.~contribution}}
  }
 \put(-0.024,0){ 
  \put(-0.03,0.025){\includegraphics[scale=1.1]{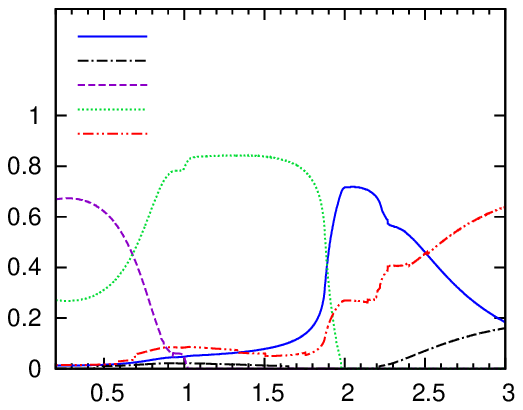}} 
   \put(0.134,0.313){\tiny $\stau_1\stau_1\!\to\! bb$}
   \put(0.134,0.294){\tiny $\stau_1\stau_1\!\to\! tt$}
   \put(0.134,0.275){\tiny $\stau_1\stau_1\!\to\!2\{H\!,\!A\}$}
   \put(0.134,0.257){\tiny $\stau_1\stau_1\!\to\!1\{H\!,\!A\}$}
   \put(0.134,0.238){\tiny others}
   \put(0.255,0.305){\footnotesize $A_{\tau}/\mstau\!=8$}
   \put(0.27,0.28){\footnotesize $\tan\beta=50$}
  \put(0.19,0.005){\footnotesize $m_H/\mstau$}
  \put(-0.02,0.11){\rotatebox{90}{\footnotesize rel.~contribution}}
  }
 \put(0.46,0.39){ 
  \put(-0.03,0.025){\includegraphics[scale=1.1]{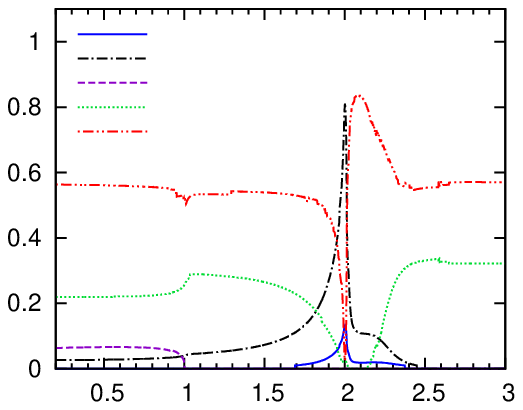}}
   \put(0.255,0.305){\footnotesize $A_{\tau}/\mstau\!=8$}
   \put(0.28,0.28){\footnotesize $\tan\beta=5$}
   \put(0.134,0.314){\tiny $\stau_1\stau_1\!\to\! bb$}
   \put(0.134,0.295){\tiny $\stau_1\stau_1\!\to\! tt$}
   \put(0.134,0.276){\tiny $\stau_1\stau_1\!\to\!2\{H\!,\!A\}$}
   \put(0.134,0.255){\tiny $\stau_1\stau_1\!\to\,$others}
   \put(0.134,0.239){\tiny coann.}
  \put(0.19,0.005){\footnotesize $m_H/\mstau$}
  \put(-0.02,0.11){\rotatebox{90}{\footnotesize rel.~contribution}}
  }
  \put(-0.024,0.39){ 
  \put(-0.03,0.025){\includegraphics[scale=1.1]{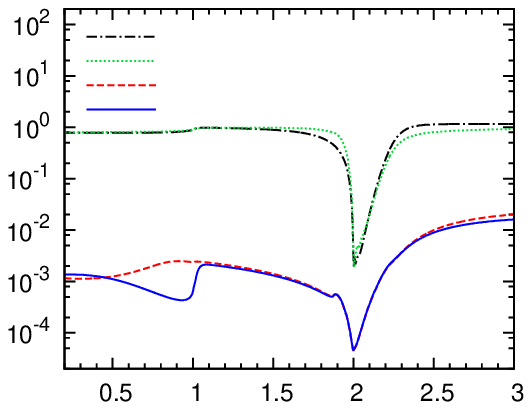} }
   \put(0.13,0.313){\tiny $A_{\tau}/\mstau\!\!=\!8,\tan\beta\!=\!5$}
   \put(0.13,0.294){\tiny $A_{\tau}/\mstau\!\!=\!-8,\tan\beta\!=\!5$}
   \put(0.13,0.275){\tiny $A_{\tau}/\mstau\!\!=\!8,\tan\beta\!=\!50$}
   \put(0.13,0.256){\tiny $A_{\tau}/\mstau\!\!=\!-8,\tan\beta\!=\!50$}
  \put(0.19,0.005){\footnotesize $m_H/\mstau$}
  \put(-0.035,0.055){\rotatebox{90}{\footnotesize $Y/Y (\thest=0.5\,\pi,\,m_i\simeq4\mstau)$}}
  }
\end{picture}
\caption{%
Upper left panel:~Stau yield $Y$ as a function of $m_H/\mstau$ for $\mstau=1000\GEV$,
$\thest=\pi/4$ and different choices of $A_{\tau}$ and $\tan\beta$. We normalized the curves 
by the yield for a purely right-handed lighter stau and $m_i/\mstau\simeq4$. Upper right and 
lower panels:~Relative contributions of the annihilation channels as a function of $m_H/\mstau$
for three of the curves displayed in the upper left panel.
}
\label{fig:mAscan}
\end{figure}

If we relax our assumption $m_A \simeq 4 \mstau$ the contributions with heavy 
Higgs intermediate or final states can dominate the annihilation cross section. 
On the one hand, the heavy Higgs can appear in the $s$-channel leading to a resonant 
pole in the propagator when $m_H\simeq2\mstau$ \cite{Pradler:2008qc}. On the other hand, heavy Higgses
can appear in the final state around or below threshold, i.e., when $m_h+m_H\lesssim2\mstau$ 
or $m_H\lesssim \mstau$.
The upper left panel of figure \ref{fig:mAscan} shows the yield for a maximally 
mixed stau, $\thest =\pi/4$, and $\mstau=1\TEV$ as a function of $m_H/\mstau$.
For small values of $\tan\beta$ the yield does not significantly deviate from the
one for a right-handed stau, except for the resonance where the yield is reduced
by up to more than two orders of magnitude. The upper right panel in figure \ref{fig:mAscan}
shows the relative contributions to the annihilation. For most of the displayed
range of $m_H/\mstau$ co-annihilation channels contribute the most (red dot-dot-dashed curve). 
This is caused by the relatively small $X_\tau$ that requires a small mass splitting of 
 $\mstau$, $m_{\stau_2}$ and $m_{\s\nu_\tau}$ in the presence of maximal mixing.
In the resonance, the channel $\stau_1\stau_1\to tt $ dominates (black dot-dashed curve). 
Note that the peak in the contribution of co-annihilation channels slightly above the
resonance $m_H/\mstau=2$ stems from the resonant annihilation of $\stau_2$ and $\s\nu_\tau$,
which are slightly heavier than the lighter stau.

For $\tan\beta=50$ we obtain a reduction of the yield by about four orders of magnitude.
This result is independent of the chosen sign of $A_\tau$ and therefore independent
of the sign of the coupling $C[H,\stau_{1},\stau_{1}]$. Since the couplings of the heavy Higgs 
to the bottom quark are proportional to $\tan\beta$ in the decoupling limit, the dominant
channel for $\tan\beta\gtrsim8$ is $\stau_1\stau_1\to bb$.
Since the coupling $C[H,\stau_{1},\stau_{1}]$ is also proportional to $\tan\beta$, 
for very small stau yields in the resonance region we typically obtain
$b\bar{b}$ final states. 

Another interesting observation can be made in the region $m_H/\mstau\lesssim1$.
Below the threshold for two heavy Higgses in the final state the stau yield is significantly
reduced in the case of a negative $A_\tau$ (blue solid curve) while this feature is not present for
positive $A_\tau$ (red dashed curve). (The other parameters are identical.) 
This asymmetry is due to an interference of the $t$-channel diagram $\stau_1\stau_1\to HH$ 
with the $s$-channel diagrams $\stau_1\stau_1\to h,H\to HH$. The diagram  
$\stau_1\stau_1\to H\to HH$ is sensitive to the sign of  $C[H,\stau_{1},\stau_{1}]$ and 
introduces a constructive (destructive) interference for $C[H,\stau_{1},\stau_{1}]$ 
negative (positive). When decreasing the mass of the heavy Higgs, this diagram is 
reduced by the increasing denominator of the heavy Higgs propagator.

\subsection{Differences in the scaling behavior} \label{sec:scalingdev}

\begin{figure}[h!]
\centering
\setlength{\unitlength}{1\textwidth}
\begin{picture}(0.43,0.38)
\put(-0.024,-0.0){ 
  \put(-0.02,0.02){
  \includegraphics[scale=1.1]{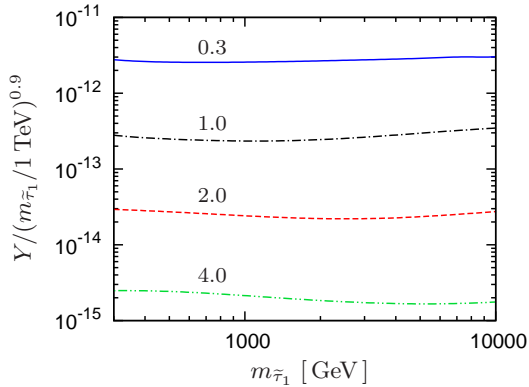} 
}
   \put(0.15,0.297){\scriptsize $0.3$}
   \put(0.15,0.229){\scriptsize $1.0$}
   \put(0.15,0.164){\scriptsize $2.0$}
   \put(0.15,0.092){\scriptsize $4.0$}
  \put(0.197,0.005){\footnotesize $\mstau\;[\GEV\,]$}
  \put(-0.018,0.11){\rotatebox{90}{\footnotesize $Y/(\mstau/1\TEV)^{0.9}$}}
  }

\end{picture}
\caption{%
Scaling behavior of the stau yield for stau pair annihilation in the presence of
maximal left-right mixing, $\thest=\pi/4$. We plot $Y/\mstau^{0.9}$ as a function
of $\mstau$ for four different (fixed) values of the rescaled coupling parameter
$\hat C_{h,\stau_1}/\mstau$.
}
\label{fig:stauhWWscaling}
\end{figure}

In all the processes discussed in sections \ref{sec:coannnomix} and \ref{sec:largemix} we have 
set the stau mass to $\mstau=1\TEV$. Point \ref{item:scalingbeh} in the list of observations in 
section~\ref{sec:surveygeneral} implies that for fixed ratios of SUSY masses and for an 
annihilation cross section that is independent of the masses of SM particles in the limit 
$a_\text{SM} \to 0$, the results can be extrapolated to any value of $\mstau$ by a simple
rescaling of the yield that is approximately linear in the stau mass. Indeed, we explicitly 
checked the scaling behavior of all limiting cases considered in section \ref{sec:coannnomix}
and found 
\beq
\label{eq:ypropall}
Y\propto\mstau^{\delta}\,,
\eeq
where $\delta\simeq0.9$ as in eqs.~\eqref{eq:YstauR} and \eqref{eq:YstauL}.
However, the argument does not apply for non-vanishing left-right mixing in the sfermion sector, 
since the limit $a_\text{SM}\to 0$ would imply $\sin2\theta_{\s f}\to 0$, 
cf.~\Eqref{eq:deltaM_NLOSP_LOSP} in the limit $m_{\tau}/\mstau\to 0$. The mixing term
introduces an explicit scale dependence, since it is proportional to the fermion mass.
However, \Eqref{eq:ypropall} holds approximately if we keep the ratios of parameters
fixed that govern the annihilation processes sensitive to the left-right mixing.
As an example, for large $m_H$, i.e, no resonance effects, the leading term for stau pair 
annihilation in the presence of large stau-Higgs couplings is annihilation into light Higgs final 
states via $t$-channel stau exchange, which behaves like \cite{Ratz:2008qh}
\beq
\tilde\sigma\propto \left(\hat C_{h,\stau_1}/\mstau\right)^4 ,
\label{eq:ratz}
\eeq
where we have additionally adopted the limit $m_h\ll\mstau$. In figure~\ref{fig:stauhWWscaling}
we plotted $Y/\mstau^{0.9}$ as a function of $\mstau$ between $300\GEV$ and $10\TEV$ 
for the case of maximal left-right mixing, $\thest=\pi/4$, and $\hat C_{h,\stau_1}/\mstau$ fixed to 
four different values. Although the scaling is slightly changed for $\hat C_{h,\stau_1}/\mstau\ge1$, 
\Eqref{eq:ypropall} remains a reasonable approximation. The slight deviations arise from the 
interplay of different contributions, most importantly stau pair annihilation into $h, W$ and $Z$.
For $\hat C_{h,\stau_1}/\mstau=0.3$, diagrams involving stau-Higgs couplings that introduce 
sensitivity to the left-right mixing are negligible. A similar behavior can be found for the other 
cases considered in section~\ref{sec:largemix} when holding $\hat C_{h,\s f_1}/m_{\s f_1}$ fixed.

Keeping $\hat C_{h,\stau_1}/\mstau$ and $\thest$ fixed implies that $m_{\stau_2}/\mstau$ and 
$m_{\s \nu_{\tau}}/\mstau$ vary with $\mstau$. This results in an increasing importance of
co-annihilation effects with the heavier stau and the tau sneutrino. This effect is only 
significant for $\hat C_{h,\stau_1}/\mstau\leq 2$, however. For $\hat C_{h,\stau_1}/\mstau=0.3$,
the co-annihilation effects in the extremely compressed stau sector lead to a net increase of the 
yield compared to the case of a single right-handed stau. Note that keeping the ratio 
$m_{\stau_2}/\mstau$ and $\thest$ constant would require $\hat C_{h,\stau_1}$ to increase 
proportionally to $\mstau^2$, which would result in large deviations from \Eqref{eq:ypropall}, 
in particular a net decrease of the yield with increasing $\mstau$. When raising $\mstau$ in 
this setup the required large values for $\hat C_{h,\stau_1}$ would quickly drive the model into 
phenomenologically unfeasible regions (see sections~\ref{sec:thbounds} and 
\ref{sec:constraintsappl}).

\subsection{Summary and classification of regions} \label{sec:classificationregions}

One important outcome of the performed survey is the fact that
in all regions the scaling behavior of the yield with $\mstau$ (and for fixed $a_i$ otherwise)
is approximately linear. Hence, the desire to achieve low stau yields points 
to low stau masses and in the same way to low masses for co-annihilating sparticles
close in mass to the stau. On the other hand, lower mass limits for the different sparticles 
can be derived from LHC searches. These searches potentially translate into lower limits on 
the stau yield for a given region. In order to be able to discuss the impact of these and further 
experimental and theoretical bounds on the yield in section \ref{sec:yieldMCscan}, we 
will now summarize and classify the most important phenomenologically
different regions we found in this section. 
 
Without co-annihilation effects and significant left-right mixing in the stau sector the
stau yield is roughly $Y\simeq10^{-12}$ for $\mstau=1\TEV$. The yield does not change 
order-of-magnitude-wise when introducing (nearly) mass-degenerate selectrons and 
smuons---slepton co-annihilation effects lead to a slight increase in the yield.
In this region in parameter space the sleptons dominantly annihilate into vector bosons and 
leptons. We will refer to this region as the \emph{bulk region}. The possibility of obtaining low 
stau yields in this region is simply restricted by the lower bound on the stau and slepton mass.

If we allow for EWino co-annihilation the effect on the yield ranges from a slight increase
(for bino co-annihilation) to a decrease by almost an order of magnitude (for wino co-annihilation)
with respect to the case of no co-annihilation. If we additionally approach the region of a
resonant $s$-channel propagator, $m_A\simeq 2\mne$, we found a net decrease of the yield 
by up to more than two orders of magnitude for maximally mixed EWinos. The main limiting 
factor for achieving small yields in the \emph{EWino co-annihilation region} is given by the 
lower limits on EWino masses in the long-lived stau scenario.

Similarly, lower bounds on the yield in the \emph{gluino co-annihilation region} and 
\emph{squark co-annihilation region} potentially arise from the respective bounds
on the gluino and squark masses in the long-lived stau scenario. In both scenarios
a reduction of the yield by around one order of magnitude could be achieved.

When considering scenarios where sfermions with large left-right mixing are involved in 
the annihilation processes, lower limits on the stau yield do not arise solely from the lower
limits on sparticle masses. In contrast, the involved Higgs-sfermion couplings depend on 
a priori free parameters of the theory which can only be restricted by theoretical bounds 
(from vacuum stability and unitarity) or indirect experimental bounds (e.g., precision and
flavor observables, MSSM Higgs searches). 
We distinguish two characteristic regions for which stau-Higgs couplings are important.
The \emph{Higgs final state region} and the \emph{Higgs resonant region} are characterized by
a dominant annihilation of staus into two final state Higgses (any combination of $h,H,A^0,H^\pm$)
and annihilation into bottom or top quarks, respectively. Both effects are also present in the case of 
co-annihilation of stops or sbottoms with large left-right mixing which we summarize as the 
\emph{3rd generation squark co-annihilation region}.

In the following section we will describe the various constraints on the model parameter space 
whose implications for the seven regions defined here will be applied in section~\ref{sec:yieldMCscan}.

\section{Implications of the first LHC runs}\label{sec:pMSSMsummary}

The LHC has brought important insights into the physics of elementary particles constraining
possible extensions of the SM\@. In the first runs of proton-proton collisions at center-of-mass
energies of $7$ and $8\TEV$, searches for the SM Higgs boson as well as for SUSY and other
theories beyond the SM have been in the focus.
Searches for additional Higgs bosons have imposed severe bounds on the
MSSM Higgs sector, especially for the region of low $m_A$. Furthermore, the determination 
of the couplings of the discovered Higgs boson to the SM particles may lead to indications for 
new physics that could serve as discriminators between different models beyond the SM.

We will here consider the latest results from the LHC experiments and further constraints
and will work out their implications for scenarios with a long-lived stau. 
A key ingredient of the analysis is the interpretation of the searches for 
heavy stable charged particles (HSCP) performed at the 
$7$ and $8\TEV$ LHC in the considered pMSSM parameter space. We will
include not only the searches for charged sleptons but also the searches for
$R$-hadrons, which can appear for small mass gaps between the gluino or squarks and the stau.
Those regions are of particular interest for us in order to cover scenarios with gluino and squark 
co-annihilation.
To interpret the collider bounds in the pMSSM parameter space, we have to compute the 
complete SUSY cross sections for each generated point in the parameter space. 
The enormous computing time for the calculation of the full SUSY cross sections
at next-to-leading-order (NLO) precision makes it necessary to find other 
methods allowing for a fast estimation of the cross sections suitable for a large number 
of points. We achieved this by developing a fast cross section estimator based on grids 
and interpolation routines. This is particularly important for EWino production which 
in principle depends on many parameters.

In the present section, our main goal  is to reveal the interplay between the constraints 
on the Higgs sector, the limits from HSCP searches, and other theoretical or
experimental constraints from flavor and precision observables.
To this end we will perform a Monte Carlo scan over the pMSSM parameter space.

\subsection{Monte Carlo scan in the 17-dimensional parameter space}
\label{sec:pMSSM}

The pMSSM is based on the following assumptions on the general MSSM: 
\emph{(i)}~$R$-parity is conserved, \emph{(ii)}~all complex phases in the soft 
breaking potential vanish, so that no new sources of $CP$ violation are introduced 
beyond contributions from the CKM matrix, \emph{(iii)}~sfermion mass matrices are 
diagonal in flavor space and the trilinear couplings are proportional to Yukawa 
couplings, so that no new sources of flavor violation are introduced,
\emph{(iv)}~universality and vanishing trilinear couplings for the first and second
generation sfermions are assumed. After imposing the electroweak symmetry breaking 
conditions, this leads to 19 free parameters.\footnote{%
A similar pMSSM parameter space with a gravitino LSP and generic NLSPs was
discussed in \cite{CahillRowley:2012cb}, where the authors focussed on recent 
search results at a $7/8\TEV$ LHC\@. However, the collider limits of the $8\TEV$ LHC
for HSCP were not taken into account there. 
}
To simplify the estimation of cross sections for collider bounds we further reduce the 
number of parameters by imposing
\beq
\label{eq:12sqdeg}
m_{\s Q_{1,2}}= m_{\s u_{1,2}}= m_{\s d_{1,2}}\,,
\eeq
which does not affect the qualitative discussion in the work. This way we are left with
a 17-parameter pMSSM, with all parameters defined at the TeV scale. 

We impose the following hard restrictions on the generated points. First, the 
lighter stau is taken to be the NLSP (and thus the lightest sparticle of the MSSM),
\beq
\label{eq:stauLOSPcon}
\stau_1 = \text{NLSP}\,.
\eeq
Second, at least one of the neutral $CP$-even Higgses lies within the LHC Higgs discovery 
window
\beq
\label{eq:higgswindow}
m_h\;\,\text{or/and} \;\,m_H \in [123;128]\GEV,
\eeq
which we will discuss in section  \ref{sec:HiggsWindow} in more detail.
The analysis performed here is independent of the nature of the LSP\@.
We merely assume that a very weakly interacting non-MSSM sparticle is the LSP
and that the lifetime of the stau is larger
than $\mathcal{O}(10^{-7}\SEC)$, i.e., it is long-lived
and thus leaves the LHC detectors before decaying.  Note that this does
not restrict our 17-dimensional parameter space since the LSP mass is an
independent parameter which can easily ensure this constraint and does
not have any further consequences for our analysis at this point. For the 
example of a gravitino LSP, $\tau_\stau \gtrsim 10^{-7}\SEC$ implies 
$m_{\widetilde G} \gtrsim 0.4\MEV$ for $\mstau = 1\TEV$.

Under these assumptions we perform a numerical random scan over pMSSM
parameter space and generate points according to the following procedure. 
\begin{enumerate}
\item \label{item:suspect}
After a random selection of the parameters at the low scale, we generate the physical 
masses as well as mixing angles using the spectrum generator \textsc{SuSpect}~2.41 
\cite{Djouadi:2002ze}. 
The input parameters and scan ranges are described in section \ref{sec:inputparams}.
Minimal requirements on the scan points are imposed on this stage---we only proceed with 
points obeying \Eqref{eq:stauLOSPcon} and the accepted output intervals
of $\mstau$, $m_{\s t_1}$, $m_{\s b_1}$ (see section \ref{sec:spec_gen}).
\item \label{item:FeynHiggs}
The Higgs sector spectrum is recalculated using \textsc{FeynHiggs}~2.9.2~\cite{Heinemeyer:1998yj}
and only points that fulfill \eqref{eq:higgswindow} are kept for the further steps. Furthermore, we
computed the coupling strength for the Higgs decay modes with \textsc{FeynHiggs}.
\item 
Decay widths and branching ratios are obtained from \textsc{SUSY-Hit}~1.2 \cite{Djouadi:2006bz}.
We used a modified version of \textsc{SDecay} that enables additional decay modes 
\cite{Kraml:2007sx}. All potentially important 3- and 4-body decay widths that are not
computed by this program are calculated with \textsc{Whizard} 2.1.1 \cite{Kilian:2007gr}.

\item 
For the computation of flavor 
observables and cosmological quantities we run \textsc{micrOMEGAs}~2.4.5~\cite{Belanger:2008sj}. 

\item 
For the computation of exclusion bounds from collider searches in the Higgs sector, 
performed at LEP, the Tevatron and the LHC, we run \textsc{HiggsBounds 4.0.0}  \cite{Bechtle:2011sb}.

\item \label{item:HSCP}
In order to derive the HSCP bounds and discuss the perspective for a future discovery
at the LHC, we determined all relevant cross sections for a center-of-mass energy of 7, 8 
and $14\TEV$. We computed the direct stau production via $s$-channel Higgses $h,H$
with \textsc{Whizard} 2.1.1 \cite{Kilian:2007gr}.
The cross sections for all other contributions are estimated via the fast interpolation
method described in \ref{sec:LHCbounds}. For the interpolation we use grids 
computed by \textsc{Prospino}~2.1 \cite{1997NuPhB.492...51B,1999PhRvL..83.3780B,Plehn:2004rp,Beenakker:1997ut}
as well as grids from the program package \textsc{NLLfast} 
\cite{Beenakker:2009ha,Beenakker:2010nq,Kulesza:2008jb,Kulesza:2009kq}.
\end{enumerate}

In the following, we describe \ref{item:suspect}--\;\!\ref{item:HSCP} in detail.

\subsubsection{Input parameters and scan ranges} \label{sec:inputparams}

As the 17 independent input parameters at the TeV scale we choose
\begin{equation}\label{eq:inputparams}
A_t,\,A_b,\,A_\tau;\,\mu,\,\tan\beta,\,m_A;M_1,\,M_2,\,M_3;\,
\theta_\stau,\,\mstau;\,\theta_\stopo,\,\mstop,\,\msbot;\,
m_{\s L_{1,2}},\,m_{\s e_{1,2}},\,m_{\s Q_{1,2}}.
\end{equation}
With this choice, we trade the soft parameters 
$m_{\s L_{3}}$, $m_{\s e_{3}}$, $m_{\s Q_{3}}$, $m_{\s u_{3}}$ and $m_{\s d_{3}}$ 
for the respective spectrum parameters, i.e., the masses and mixing angles
of the third-generation sfermions, and treat them as independent input parameters. 
This has two reasons. First, this way we achieve a better control over the third-generation 
sfermion masses in the presence of large mixings and thus by choosing appropriate scan 
ranges we avoid scanning over regions where the stau is not the NLSP or which are already 
forbidden by conservative model-independent collider bounds.
Second, spectra with large mixings are equally strongly represented as those with
small mixings. This has an important impact on our considerations of stau yield
which is potentially sensitive to the stau mixing angle. 

\begin{table}[bt]
\centering
\begin{tabular}{c|c|c}
Parameter & Interval input & Accepted output interval (if different) \\
 \hline
  $A_{t}$ & $[\,-10^4;10^4\,]$ &  \\
  $A_{b}$ & $[-8000; 8000]$ & \\
  $A_{\tau}$ &$[-8000; 8000]$ & \\ 
 \hline
   $\mu$ & $[-8000; 8000]$ & \\
 $m_A$ & $[100;4000]$ & \\
  $\tan{\beta}$ & $[1; 60]$ & \\
 \hline
  $\theta_{\tilde{\tau}}$ & $[10^{-4};\pi/2]^{\star}$ & $[0;\pi]$ \\
  $\mstau$ & $[200;2000]$ & [216;2200]\\
 \hline
  $\theta_{\tilde{t}}$ & $[10^{-4};\pi/2]^{\star}$ & $[0;\pi]$ \\
  $m_{\tilde{t}_1}$ & $[\max(\mstau,700);5000]$ & $[\max(\mstau,740);6000]$ \\
  $m_{\tilde{b}_1}$ & $[\max(\mstau,700);5000]$ & $[\max(\mstau,740);6000]$ \\ 
 \hline
  $m_{\s L_{1,2}}$ & $[\mstau;4000]$ & \\
  $m_{\s e_{1,2}}$ & $[\mstau;4000]$ & \\
$m_{\s Q_{1,2}}$& $[ \max(\mstau,1200);8000]$& \\
 \hline
  $M_{1}$ &  $[\mstau;4000]$  & \\
  $M_{2}$ &$[\mstau;4000]$  & \\
 $M_{3}$ & $[ \max(\mstau,1000);5000]$ &  \\
\end{tabular}
\caption{Parameter ranges for the 17-dimensional pMSSM scan. 
The second column shows the intervals of the randomly generated input parameters. In the third
generation sfermion sector we choose masses and mixing angles as input parameters and
determine the corresponding soft masses from these input parameters at tree-level. The third 
column displays the accepted intervals for these masses and mixing angles after computing the 
full spectrum including higher order corrections. All dimensionful parameters are given in GeV\@.
 \newline
\hspace*{0.5ex} {$^{\star}$The interval $[ 0;\pi/2]$ is mapped onto $[ 0;\pi/2]$ or $[\pi;\pi/2]$ according to the sign
of $X_{\tau} = A_{\tau}-\mu\tan\beta$, see section \ref{sec:spec_gen} for details.
In order to avoid numerical instabilities we choose $10^{-4}$ as a lower limit on scan range
of the mixing angles.}
}
\label{tab:scanlim}
\end{table}

If not stated otherwise, for all input parameters we choose linearly flat priors in the scan. 
The scan ranges are summarized in table \ref{tab:scanlim}. The
ranges are motivated by the requirement of a $\stau$ NLSP as well as conservative collider 
bounds on individual particles (see section~\ref{sec:spec_gen}). 
First generation squarks and sleptons are kept degenerate. 
In addition to this `blind' scan we performed dedicated scans accumulating more points
in certain sub-ranges which are of particular interest according to the results of section
\ref{sec:systscan}. Those dedicated regions are summarized in appendix \ref{sec:dedscan}.
If not stated otherwise we refer to the complete set of scan points including the 
dedicated scans. In total we generated $5\times10^5$ points.
Note that we do not attach any physical meaning to the density of generated points 
in parameter space.

\subsubsection{Spectrum generation} \label{sec:spec_gen}

After the random generation of the input parameters given in \eqref{eq:inputparams} 
we determine the soft masses of the third-generation sleptons and squarks,
$m^2_{\s{L}_3}$, $m^2_{\s{e}_3}$, $m^2_{\s{Q}_3}$, $m^2_{\s{u}_3}$ and
$m^2_{\s{d}_3}$, from the respective free parameters in \eqref{eq:inputparams}, 
using the tree-level relations \eqref{eq:mLLinv1} and \eqref{eq:mRRinv1}
(see appendix~\ref{app:con}) and analogous expressions for stops and sbottoms.
Points with negative mass squares are rejected at this point. From these input
parameters the SUSY spectrum is computed with \textsc{SuSpect}~2.41.
Points which do not fulfill \eqref{eq:stauLOSPcon} and \eqref{eq:higgswindow}
are rejected as well as points that do not lie within the accepted output
intervals for $\mstau$, $m_{\s t_1}$ and $m_{\s b_1}$ listed in table \ref{tab:scanlim}. 
The lower limits of these intervals are motivated by conservative collider bounds on 
individual sparticle masses in the long-lived stau scenario we derived earlier 
(see \cite{Heisig:2012ep,Heisig:2012zq}). 
However, we will see that they are well below the limits we will finally infer from the 
interpretation of the HSCP searches at 7 and $8\TEV$. Hence, these lower limits only 
serve to gain efficiency in generating valid points and have no impact on the physical 
results.

\begin{figure}[h!]
\centering
\setlength{\unitlength}{1\textwidth}
\begin{picture}(0.84,0.72)
  \put(-0.024,0.36){ 
 \put(-0.03,0.025){\includegraphics[scale=1.05]
{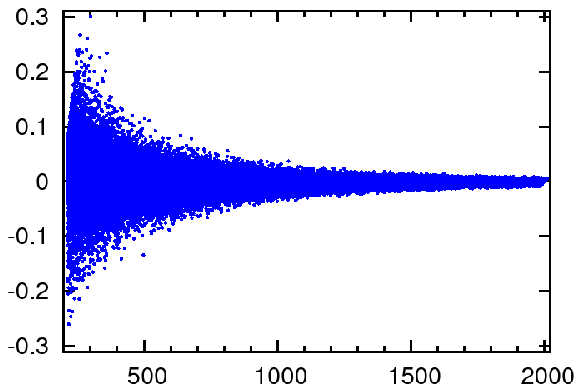}
}
  \put(0.17,0.0){\footnotesize $m_{\s \tau_1}^{\text{out}}\;[\GEV\,]$}
  \put(-0.04,0.125){\rotatebox{90}{\footnotesize $m_{\s \tau_1}^{\text{out}}/m_{\s \tau_1}^{\text{in}}-1$}}
  }
 \put(0.47,0.36){ 
  \put(-0.03,0.025){ 
\includegraphics[scale=1.05]{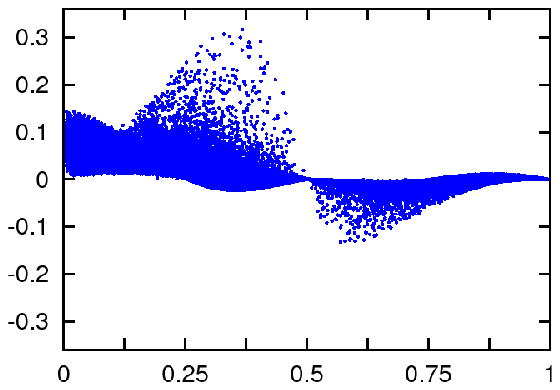}
} 
  \put(0.2,0.0){\footnotesize $\thest^{\text{out}}$}
  \put(-0.04,0.13){\rotatebox{90}{\footnotesize $\thest^{\text{out}}/\thest^{\text{in}}-1$}}
  }
 \put(-0.024,0){ 
  \put(-0.03,0.025){
  \includegraphics[scale=1.05]{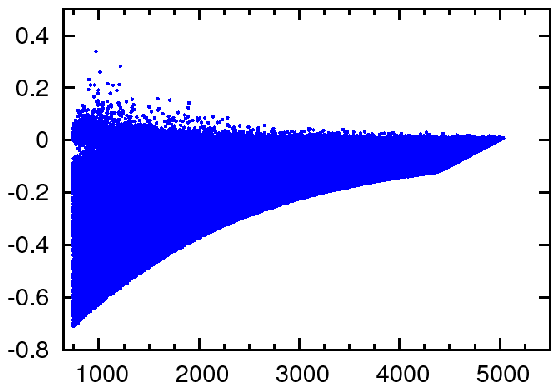} 
}
  \put(0.17,0.0){\footnotesize $m_{\s b_1}^{\text{out}}\;[\GEV\,]$}
  \put(-0.04,0.125){\rotatebox{90}{\footnotesize $m_{\s b_1}^{\text{out}}/m_{\s b_1}^{\text{in}}-1$}}
  }
 \put(0.47,0){  
  \put(-0.01,0.031){
  \includegraphics[scale=1.05]{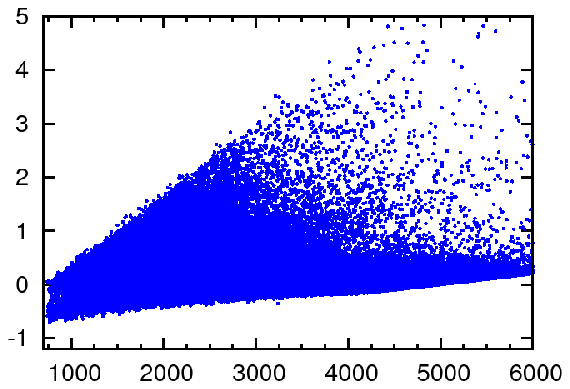}
}
  \put(0.17,0.0){\footnotesize $m_{\s t_1}^{\text{out}}\;[\GEV\,]$}
  \put(-0.04,0.125){\rotatebox{90}{\footnotesize $m_{\s t_1}^{\text{out}}/m_{\s t_1}^{\text{in}}-1$}}
  }

\end{picture}
\caption{%
Relative deviation of the output and input parameters depending on the value of the
output parameter for $\mstau$ (upper left panel), $\thest$ (upper right panel),
$m_{\s b_1}$ (lower left panel) and $m_{\s t_1}$ (lower right panel) for all generated points.
The input and output values refer to the values that have been chosen randomly and those 
that have been obtained from the spectrum generator including loop corrections, respectively.
}
\label{fig:paramcheck}
\end{figure}

\textsc{SuSpect} computes up to 2-loop corrections for the sparticle masses.
For illustration, figure~\ref{fig:paramcheck} shows the relative correction to the input parameters 
as a function of the output parameters computed by \textsc{SuSpect}.
For the stau mass and mixing angle, loop corrections that are taken into account in the 
computation via \textsc{SuSpect} are relatively small. The bulk of points acquire corrections
well below 10\%. However, deviations up to 30\% are present in the stau sector.
For the stop and sbottom mass higher order corrections are much more significant.
Especially in the case of the stop the output value for $m_{\s t_1}$ turns out to 
overshoot the intended value by several 100\%. However, as we only use the output
values for all further discussions, within the limitations of \textsc{SuSpect} we achieve
self-consistent spectra which we will use in the following discussion.

We re-compute the Higgs sector of the spectrum as well as the Higgs 
decay table with \textsc{FeynHiggs 2.9.2}. The value for the Higgs mass
$m_h$ computed by \FH\ is smaller than the value computed by \textsc{SuSpect}
for most of the parameter points. Since larger $m_h$ tend to be more
challenging to achieve, we consider the lower limit on the Higgs mass
to be more important. Thus, to be conservative we use the \textsc{FeynHiggs}
value. The resulting spectrum is used for the further analysis. All points that fulfill 
\eqref{eq:higgswindow} are recorded and count as generated points.

\subsubsection{Meeting the LHC Higgs window} \label{sec:HiggsWindow}

Within this work we interpret the discovered Higgs boson with a mass 
of $125.5 \pm 0.2 \,\text{(stat.)}{}^{+0.5}_{-0.6} \,\text{(syst.)}\GEV$ at ATLAS 
\cite{ATLAS-CONF-2013-014} and 
$125.7 \pm 0.3 \,\text{(stat.)} \pm 0.3 \,\text{(syst.)}\GEV$ at CMS 
\cite{CMS-PAS-HIG-13-005} 
as either the light or heavy neutral $CP$-even Higgs of the MSSM (or even as both
contributing to the signal). Accordingly, taking into account the theoretical 
uncertainty in the prediction of the Higgs mass (see, e.g., \cite{Degrassi:2002fi}), 
we demand \eqref{eq:higgswindow}.

The ranges of the input parameters have an effect on the distribution of the resulting Higgs masses.
Through loop corrections, the sparticle masses, especially
the stops, are intimately related to $m_h$.

In a neutralino LSP scenario, a pMSSM scan with flat priors and input parameter ranges 
just above the current collider bounds, the distribution for $m_{h}$ typically peaks at values 
below the interval \eqref{eq:higgswindow} and falls off over the interval towards large values 
(see, e.g., \cite{CahillRowley:2012rv}). 
This implies that for the case of $m_{h}\,\in\,[123;128]\,\GeV$ this window would mostly 
be populated towards its lower end reflecting the preference of the MSSM for a lighter $m_h$.
In this work we aim to avoid an asymmetric distribution of $m_h$ around
the experimental value since the allowed window is to 
account for the theoretical uncertainty in the computed Higgs mass. 
Instead, we choose to aim for a flat distribution in $m_h$ in our scan.
Hence, we allow for relatively large $A_t$ in this scan.
Remarkably, with the scan ranges given in table \ref{fig:paramcheck}
we achieve an almost flat distribution in $m_h$ over the interval
\eqref{eq:higgswindow}. This is partly due to the fact
that in the long-lived stau scenario stronger model-independent bounds on the
sparticle masses exist which shifted our scan ranges 
towards higher masses (see section \ref{sec:spec_gen}).
The blue line in the upper panel of figure \ref{fig:binned}
shows the distribution of the Higgs mass $m_h$ for the blind scan (the distribution 
for the complete set of points is virtually identical).

A second effect on the Higgs sector is induced by the allowed range for $m_A$. 
For the range chosen here\footnote{%
The range for $m_A$ has been chosen such as to allow for all phenomenologically 
interesting effects described in sections \ref{sec:mAEwinos}, \ref{sec:stopmA} and 
\ref{sec:survresonance}.
}
most parameter points end up in the decoupling limit avoiding to cover the region
where $m_H$ could make up the discovered Higgs.
In other words the ratio between the number of points with 
$m_h$ versus $m_H$ in the interval \eqref{eq:higgswindow} depends strongly on the 
chosen scan range for $m_A$. In order to have control over this
arbitrary bias we require $m_A<140\GEV$ for half of the generated points, i.e., 
half of the points in our scan lie explicitly not in the decoupling limit.
This way, around $65\%$ ($35\%$) of the generated points feature $m_{h}$ ($m_H$) to lie
in the interval \eqref{eq:higgswindow}. For around $0.7\%$ of the
points both Higgs bosons lie in this interval.

\subsubsection*{Selection effects induced by \eqref{eq:higgswindow}}

To obtain $m_h$ in the window \eqref{eq:higgswindow} demands the presence
of large radiative corrections on $m_h$ requiring an interplay of several parameters
that govern these radiative corrections, namely the masses and the mixing in the stop 
sector and furthermore---in descending order of importance---in the sbottom and stau 
sector. While the stop contributions to the Higgs mass are large (as demanded) and 
positive, the sbottom and stau contributions typically diminish the Higgs mass and can
be significant for negative $\mu M_3$ and large $\tan\beta$,
making it harder to satisfy \eqref{eq:higgswindow} \cite{Carena:2011aa}.

These features induce a selection effect resulting in a non-flat distribution in some of 
the input parameters that we initially scanned over with flat priors. Although we do not 
assign any physical meaning to the \emph{absolute} point density in the parameter 
space in our later results, it is, however, interesting to see in which way the flat priors 
are `bent' by the additional requirement \eqref{eq:higgswindow}. This shall be 
subject of a brief discussion in this subsection. For this discussion we consider 
the `blind' scan only.

\begin{figure}
\centering
\setlength{\unitlength}{1\textwidth}
\begin{picture}(0.84,0.35)
  \put(-0.024,0.0){ 
  \put(-0.03,0.025){
  \includegraphics[scale=1.05]{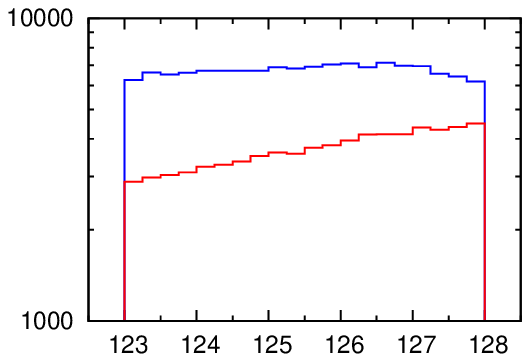}
}
  \put(0.17,0.01){\footnotesize $m_{h/H}\;[\GEV\,]$}
  \put(-0.04,0.125){\rotatebox{90}{\footnotesize \#\,points/bin}}
  }
 \put(0.47,0.0){ 
  \put(-0.03,0.025){\includegraphics[scale=1.05]{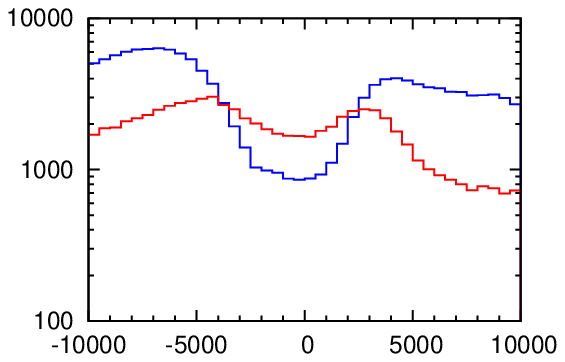}} 
  \put(0.17,0.01){\footnotesize $A_t\;[\GEV\,]$}
  \put(-0.04,0.13){\rotatebox{90}{\footnotesize \#\,points/bin}}
  }
\end{picture}
\caption{%
Binned distributions of all generated points (no additional constraints)
for the blind scan. Left panel:~The blue (red) histograms 
show the distribution in $m_h$  ($m_H$) for the subset of point with 
$m_h$  ($m_H$) in the window \eqref{eq:higgswindow}. Right panel:~Distribution
in $A_t$ (same color coding).
}
\label{fig:binned}
\end{figure}

The largest effect can be observed for $A_t$, which shows a clear preference for large 
absolute values, $|A_t|>3\TEV$, according to the large mixing required in order to 
obtain high $m_h$, see blue line in the right panel of figure \ref{fig:binned}. This effect 
is much less pronounced for those points where $m_H$ lies in the window 
\eqref{eq:higgswindow} (blue curve). Further, the distributions in $m_{\s b_1}$ and 
$m_{\s t_1}$ are bent towards disfavoring the upper and lower part of the allowed scan 
range, respectively. Interestingly, if we restrict $A_t$ to a smaller range (e.g., 
$|A_t|<3\TEV$ or less), the $m_{\s t_1}$ distribution changes to favor the lower part of 
the scan range. This is due to the large (relative) mixing required and shows that this 
mixing is in fact more important than the overall stop mass scale. The maximal radiative
correction is present for $X_t/\sqrt{m_{\s t _1}m_{\s t _2}}\simeq\sqrt{6}$. 
Other parameters that are affected by the requirement of the Higgs mass and
by the accepted output intervals for $\mstau$, $m_{\s t_1}$ and $m_{\s b_1}$ listed in
table \ref{tab:scanlim} are $\tan\beta$, disfavoring values below $\sim10$, $\mu$, 
peaking around $\pm2\TEV$ and the stop mixing angle, slightly disfavoring maximal 
left-right mixing, i.e., $\theta_{\s t}\simeq \pi/4$ or $3\pi/4$.
All other scan parameters stay flat up to a trivial drop towards small masses
as a direct consequence of \Eqref{eq:stauLOSPcon}.

\subsection{Interpretation of the HSCP searches in the pMSSM}\label{sec:LHCbounds}

Long-lived staus show up as heavy stable charged particles (HSCP) in the detectors at the 
LHC, i.e., they are recognized as muons but with two features that potentially allow for a 
discrimination against real muons: an anomalous time-of-flight (ToF) and an anomalous 
ionization loss ($\D E/\D x$). Both are accessible at the LHC experiments. 
So far, HSCP searches have been performed at ATLAS \cite{ATLAS:2012vd} 
(based on $4.7\ifb$ at $7\TEV$) and CMS \cite{CMS1305.0491} (based on $5.0\ifb$ at $7\TEV$ 
and $18.8\ifb$ at $8\TEV$) and no significant excess over background has been reported.
The null searches have been interpreted in a few long-lived stau scenarios: for a GMSB
scenario (ATLAS and CMS) as well for direct production of
mass-degenerate sleptons (ATLAS) and the direct production of staus only (CMS).
The latter analysis provides an almost model-independent lower bound on the stau mass 
of $339\GEV$.\footnote{%
The lower limit for the scan range of the stau mass in table \ref{tab:scanlim} has been 
motivated from the $7\TEV$ data \cite{Chatrchyan1205.0272} while taking the most 
conservative choice for the stau mixing angle \cite{Heisig:2012ep}.
}
We will here interpret the recent search of CMS 
in the framework of the 17-dimensional pMSSM\@. To do so,
we determine the cross sections for all relevant SUSY production
processes for each scan point, as described in section \ref{sec:fastest}.
The estimation of the cross section upper limit extracted from the search 
\cite{CMS1305.0491} will be described in section \ref{sec:appHSCPlimits}.

\subsubsection{Fast estimation of SUSY cross sections} \label{sec:fastest}

For each pMSSM point we determine the cross sections for various production channels
at the $7$, $8$ and $14\TEV$ LHC in order to estimate the viability of each point after 
the HSCP null-searches and to discuss the prospects for the LHC long-term run.
The computation of all potentially relevant SUSY cross sections at NLO precision is
time-consuming\footnote{%
As an example, the computation of the NLO cross sections with \textsc{Prospino} for 
the complete set of SUSY processes available in this program takes 
$\mathcal{O}(10\,\text{h})$ of computing time on a single 2.5\,GHz CPU.
}
and especially not convenient for the use in Monte Carlo scans containing a large 
number of points.  In order to achieve a sufficiently fast determination of the cross 
sections for each generated pMSSM point we develop a fast cross section estimation 
tool based on grids and interpolation routines. However, some production processes 
in principle involve many parameters requiring high-dimensional grids, which would 
mean to shift the problem of large computing time to the generation of the grids. 
Therefore, we exploit the potential for approximations wherever suitable. By 
factorizing the dependence on certain combinations of parameters we describe all 
channels approximately with a set of up to maximally three-dimensional grids. 
In the following we will list the respective parameterizations and  approximations 
chosen in the different sectors.

\subsubsection*{Slepton sector}

In the slepton sector we build up one- and two-dimensional grids in the corresponding
sparticle masses for the processes
$\s e_{\text{R}} \s e_{\text{R}}$, $\s e_{\text{L}} \s e_{\text{L}}$, $ \s \nu_e \s \nu_e$,
$\stau_1 \stau_1$, $\stau_2 \stau_2$, $\snutau \snutau$ and 
$\s e_{\text{L}} \s \nu_e$, $\stau_1 \stau_2$, $\stau_1 \snutau$, $\stau_2 \snutau$, respectively. 
For this purpose we compute the cross section for Drell-Yan (DY) production 
(via an $s$-channel $\gamma/Z$ or $W^{\pm}$)
with \textsc{Prospino}~\cite{1999PhRvL..83.3780B} at NLO\@. SUSY QCD contributions have 
been kept small by setting the mass of all colored sparticles to $5\TEV$. 
For the third-generation sleptons the left-right mixing introduces an additional variable
the cross section depends upon. Here, we make use of the fact that the dependence
on the stau mixing angle $\thest$ factorizes once the center-of-mass energy $\sqrt{\hat s}$
of the production process is well above $M_Z$ \cite{Heisig:2011dr}. This limit is 
easily reached for the rather heavy stau masses we are considering.
To be concrete, the cross section for the process $pp \to ij$, $\,i,j = \stau_1,\stau_2,\s\nu_\tau$, 
can be written in the form
\beq
\label{eq:stauDYfact}
\sigma (m_i,m_j,\thest) \simeq \mathcal{A}(m_i,m_j)\times \mathcal{B}(\thest)\,.
\eeq
We choose 
\begin{align}
&\!\!\mathcal{A}(m_i,m_j)=\sigma(m_i,m_j,\pi/4)\,, \\
&\;\,\mathcal{B}(\thest)=\frac{\sigma(m_{\text{ref}},m_{\text{ref}},\thest)}{\sigma(m_{\text{ref}},m_{\text{ref}},\pi/4)}\,.
\end{align}
where $m_{\text{ref}}=500\GEV$.

For the evaluation of the cross section in the scan we interpolate logarithmically 
over the cross section $\mathcal{A}$ and linearly in the correction factor $\mathcal{B}$.

\subsubsection*{EWino sector}

In the EWino sector we parametrize the cross sections by the underlying 
SUSY input parameters instead of the physical masses and mixings, namely
$M_1$, $M_2$, $\mu$, $\tan\beta$ and the common first- and second-generation 
squark soft mass $m_{\s Q_{1,2}}$. In order to describe the cross section as a 
function of these five parameters with maximally three-dimensional grids we factorize 
the dependence on these five parameters as follows. First, we decouple the bino from 
the spectrum and consider $M_1$ separately from $M_2$ and $\mu$. This is motivated 
by the hierarchy in the respective couplings: for degenerate $M_1$ and $M_2$ (or 
$\mu$) the bino contribution is relatively small. Second, we factorize the dependence on 
the squark masses. This dependence is introduced by $t$-channel squark diagrams 
which can lead to a significant net reduction of the cross section even though taking 
squark pair and associated squark production into account. This arises from a negative
interference between the DY production of EWinos and the $t$-channel contribution
and is relevant for intermediate mass gaps between $m_{\s Q_{1,2}}$ and $M_2$ 
where the squarks are still light enough to contribute in the $t$-channel but already too 
heavy to (over-) compensate the reduction by squark production.

We found that the complete cross section from neutralino and chargino production 
(including the associated squark-EWino production) can be well approximated by 
three functions, each depending on three parameters:
\beq\begin{split}
\label{eq:EWinoappr}
\sigma_{\text{EWino}}\simeq&\;\,
\sigma(pp\to\s\chi^0_1\s\chi^0_1)\!\left[M_1,m_{\s Q_{1,2}},\tan\beta\right]\\
&+\sigma(pp\to\s\chi_i\s\chi_j) \!\left[M_2,\mu,\tan\beta\right]
\times
R\!\left[\frac{\mu}{M_2},\frac{m_{\s Q_{1,2}}}{M_2},M_2\right] 
\end{split}\eeq
with
\beq
R\!\left[\frac{\mu}{M_2},\frac{m_{\s Q_{1,2}}}{M_2},M_2\right] \equiv
\frac{
\sigma(pp\to\s\chi_i\s\chi_j,\s\chi_i\sq)\!\left[\frac{\mu}{M_2},\frac{m_{\s Q_{1,2}}}{M_2},M_2\right]  
}{
\sigma(pp\to\s\chi_i\s\chi_j)\!\left[M_2,\mu\right] 
}\,,
\eeq
where $\s\chi_i = \s\chi^0_1,\s\chi^0_2,\s\chi^0_3,\s\chi^{\pm}_1,\s\chi^{\pm}_2$, and where 
$\s\chi_i\sq$ denotes the associated EWino-squark production. 
Where they are not displayed as an argument in the brackets, 
we set mass parameters to $5\TEV$ and $\tan\beta = 15$.
We computed the three grids, corresponding to the three functions in \Eqref{eq:EWinoappr},
with \textsc{Prospino} \cite{1999PhRvL..83.3780B} at NLO precision. To save 
computing time we only ran the NLO computation for a subset of points and extracted the
$K$-factor from the resulting coarser grid, under the assumption that the $K$-factor
varies more slowly with varying parameters than the cross section itself.

The functions have the weakest dependence on the last argument in each of the 
brackets in \Eqref{eq:EWinoappr}. Accordingly, we computed significantly fewer 
points in the corresponding directions in the grid space.
Contributions from associated gluino-EWino production were neglected.
For the generation of the spectrum from the SUSY parameters we used \textsc{SuSpect}~2.41, 
as we did for the generation of the pMSSM points in the Monte Carlo scan.
We interpolated logarithmically over the cross sections and linearly in the correction factor
$R$ as well as in the $K$-factors. With this description we found an agreement within a $15\%$ 
error with the full NLO computation with \textsc{Prospino} for a variety of very different spectra.

\subsubsection*{Squark and gluino sector}

For the production of third-generation squarks, contributions from $t$-channel gluino diagrams 
are small due to the small parton densities of the required heavy-flavor quarks. Furthermore, 
electroweak production is relatively unimportant. Hence, the relevant production channels are 
$\s t_1\s t_1$, $\s t_2\s t_2$, $\s b_1\s b_1$, $\s b_2\s b_2$ via an $s$-channel gluon diagram, a
$t$-channel squark diagram or the gluon-squark four-vertex. The production cross sections for 
these processes only depend on the mass of the respective squark alone. For the 7 and $8\TEV$ 
LHC cross sections we take the corresponding one-dimensional grids from \textsc{NLLfast}
\cite{Beenakker:2010nq} which include NLO and next-to-leading-log (NLL) corrections. For the
$14\TEV$ case we compute the grid with \textsc{Prospino} \cite{Beenakker:1997ut} at NLO\@.

For the first- and second-generation squark and gluino production, $\s g \s g$, $\s q \s g$, $\s q \s q$
and $\s q\s q^*$, we interpolate two-dimensional grids in the variables $\mgo$ and  
$m_{\s q} \equiv {(m_{\s u_{\text{L}}} m_{\s u_{\text{R}}} m_{\s d_{\text{L}}} m_{\s d_{\text{R}}})^{1/4}}$
which are taken from \textsc{NLLfast} \cite{Beenakker:2009ha} for the case of 7 and $8\TEV$
LHC cross sections and which we compute with \textsc{Prospino} \cite{1997NuPhB.492...51B} 
at NLO precision for the $14\TEV$ LHC cross sections. We interpolate logarithmically over the 
cross sections. The error from the interpolation is typically less than $1\%$.

The total cross section obtained from summing over all the processes described above 
was compared to the full cross section from $\textsc{Prospino}$ for a variety of different
spectra and found to agree within an error of typically $10\%$. For a few points we found errors 
up to $15\%$ where we underestimate the cross section computed by $\textsc{Prospino}$.

\subsubsection*{Stau production via intermediate Higgs}

In addition to the above channels we include the direct production of staus via an $s$-channel
Higgs intermediate state \cite{Lindert:2011td}. The channel $pp\to h \to \stau_1\stau_1$
can be important in the presence of large left-right mixing of the stau. Additionally, we take 
into account the heavy Higgs intermediate state $pp\to H \to \stau_1\stau_1$. As mentioned
earlier, for the general case (no decoupling limit) these processes depend on a variety of 
parameters. Accordingly, we compute the production cross section for these channels for each 
of the generated pMSSM points using the complete spectrum.
We perform the computation at the leading order via \text{Whizard} 2.1.1
\cite{Kilian:2007gr} where the effective gluon fusion vertex for the MSSM \cite{Spira:1995rr} 
has been implemented. We consider gluon-fusion and bottom-fusion. For the production 
via bottom-fusion we reweight the cross section according to the resummed bottom-Higgs 
coupling (for the leading contributions in $\tan\beta$), as described in appendix~\ref{sec:resum}. 
For this computation we employed the value for the correction to the bottom mass, $\Delta_b$, 
from \textsc{micrOMEGAs}.

\subsubsection{Estimation of cross section upper limits} \label{sec:appHSCPlimits}

As shown in \cite{Heisig:2012zq}, the signal efficiency\footnote{%
Signal efficiency denotes the product of detector efficiency and selection acceptance for 
signal events.
}
for the signatures of long-lived stau scenarios at the LHC is much less sensitive to the 
spectrum than, e.g., in the case for scenarios with neutral stable sparticles escaping the 
detector, where compressed or widely spread spectra are typically much harder to find.
In this reference it has been shown that for the production via colored sparticles the 
signal efficiency of long-lived staus only drops below roughly $20\%$ for widely spread 
spectra for which this production mechanism is no longer the dominant channel but is 
exceeded by the direct production of staus which provides higher signal efficiencies.
This way, the signal efficiency for the total SUSY production does not drop below
about $20\%$ in the mass ranges of interest for the LHC analysis, provided that
there is no long-lived sparticle other than the stau and thus all decay chains
terminate in the stau before traversing the sensitive parts of the detector.
The significant decrease of the signal efficiency for the production via 
colored sparticles for widely spread spectra is due to the potentially large boost of the 
stau developed in the decay of a very heavy colored sparticle. Staus with a velocity 
close to the speed of light, $\beta\simeq1$, are extremely difficult to discriminate against 
background muons since the discrimination heavily relies on a deviation from $\beta=1$.

Following this argument, electroweak production mechanisms, e.g., chargino production, 
offer even less potential to cause a drop in the overall signal efficiency. This is because, 
due to the smaller electroweak cross sections, the mass gap between the produced 
sparticle and the stau is smaller if the electroweak production process in question is 
demanded to give a significant contribution compared to the direct stau production.
This fact facilitates the estimation of the signal efficiencies (and for the resulting
cross section upper limits) requiring the extrapolation of the results given in 
\cite{CMS1305.0491} to a general pMSSM point. In the following we will describe 
this procedure in more detail.

If the decay of heavier sparticles into the stau is not prompt, the analysis becomes
more complicated. We will examine the case of long-lived colored sparticles
which we found to be the most relevant in this study. In particular, gluinos can become
long-lived even for relatively large mass gaps $\mgo-\mstau\gtrsim100\GEV$. The 
treatment of long-lived colored sparticles is described below.

\subsubsection*{Application for prompt decays into the stau}

We consider a point to be excluded at 95\% C.L. if the signal strength, 
$\sigma^{\text{limit}}/\sigma^{\text{th}}$, obeys
\beq
\frac{\sigma^{\text{limit}}}{\sigma^{\text{th}}} < 1 \,,
\eeq
where $\sigma^{\text{limit}}$ is the observed $95\,\%$ C.L.
upper cross section limit from the experiment and $\sigma^{\text{th}}$ is the 
theoretical prediction for the total cross section. 
$\sigma^{\text{limit}}$
is a model-dependent quantity. In the simplest case,
for a given spectrum, the upper cross section limit is 
determined by 
\beq
\sigma^{\text{limit}} = \frac{S}{\varepsilon_{\text{S}}\int\!\mathcal{L}}\,,
\eeq
where $S$ is the required number of expected signal events for the considered 
spectrum which allows for a 95\% C.L. exclusion in the presence of the observed 
number of (background) events. $\varepsilon_{\text{S}}$ is the signal efficiency 
for this spectrum and $\int\!\mathcal{L}$ is the integrated luminosity. $S$ and 
$\varepsilon_{\text{S}}$ both are affected by the applied cuts---the latter directly 
and the former via its background rejection capability. In HSCP searches the 
highest sensitivities are typically reached for cuts that supply $S= 3$ for a 95\% 
C.L.\ exclusion \cite{Heisig:2012zq}.

In the CMS analysis \cite{CMS1305.0491} the observed upper cross section limits
are given for the two benchmark models (GMSB model and direct DY production) 
for the $7$ and $8\TEV$ run as a function of the stau mass $\sigma^{\text{limit}}(\mstau)$. 
Here, we take the combined Tracker$+$ToF data. In order to estimate the signal strength 
for a point in our pMSSM parameter space we assign the upper cross section limits 
channel-wise: For the direct DY production of the lighter staus we apply the direct DY 
production cross section limits. For all other slepton production mechanisms, the EWino 
production and the production of third-generation squarks we applied the cross section 
limits from the GMSB model as a function of the stau mass. This is done under the 
assumption that the signal efficiencies and corresponding background rejection for these 
channels are similar to the GMSB model, which is based on the arguments given 
above.\footnote{%
For the GMSB model considered in \cite{CMS1305.0491} ($\mstau=308\GEV$) the 
EWino production contributes $53\%$ while the direct DY production of the lighter stau 
and all other sleptons make up $13\%$ and $33\%$ of the total SUSY cross section, 
respectively. The contribution from first- and second-generation squarks is negligible.
} 
For an arbitrary stau mass we interpolated linearly between the analysis points given in 
\cite{CMS1305.0491}. For stau masses above $500\GEV$ we will only be in the vicinity 
of the exclusion limit if we have a rather degenerate spectrum and thus an important 
strong production of sparticles. For these production modes the signal efficiency
can decrease due to difficulties in the triggering of very slow staus \cite{Heisig:2012zq}.
In order to account for these spectra we extrapolated the upper cross section limits by 
conservatively assuming $\sigma^{\text{limit}}= 3.0\fb$ for the $7\TEV$ run and 
$\sigma^{\text{limit}}= 1.0\fb$ for the $8\TEV$ run. These values are in accordance with
the signal efficiencies that have been reported in \cite{Heisig:2012zq} in the limit of 
mass-degenerate spectra where one stau has been required to have a velocity above 
$\beta =0.6$ in order to ensure an efficient triggering of such events. For the production 
of staus via first- and second-generation squarks and gluinos as well as for the direct 
production via an $s$-channel Higgs we take as a conservative estimate a constant 
$\sigma^{\text{limit}}= 3.0\fb$ ($1.0\fb$) for the $7\TEV$ ($8\TEV$) run.\footnote{%
For the direct production of staus via an $s$-channel neutral, $CP$-even Higgs ($h/H$), 
stau production near threshold is enhanced and so the fraction of very slow staus is large. 
For this channel the decreasing trigger efficiencies for smaller velocities (below $\sim0.6$) 
are expected to be the restricting factor of the signal efficiency. A detailed study of the 
signal efficiency in this channel is left for future work.
}
The signal strength is then obtained by 
\beq
\label{eq:signalstr}
\frac{\sigma^{\text{limit}}}{\sigma^{\text{th}}} = 
\left(\sum_{i} \sum_{k} \frac{\sigma^{\text{th}}_{ik}}{\sigma^{\text{limit}}_{ik}}\right)^{-1}\,,
\eeq
where $\sigma^{\text{th}}_{ik}$ is the computed cross section for the channel $i$ at the 
LHC energy $k$ and $\sigma^{\text{limit}}_{ik}$ is the corresponding estimated 
observed cross section upper limit for the respective channel.

\subsubsection*{Application to delayed decays}

For the application of collider limits to the present scenario, it is crucial to 
know if there are long-lived sparticles other than the stau which play 
a role in the production and decay at the collider. We therefore compute
the width of all sparticles. We used a modified version of \textsc{SDecay} 
\cite{Muhlleitner:2003vg,Kraml:2007sx} which includes all relevant 
3-body decays of sleptons into the lighter stau. We compute further
3- and 4-body decays of squarks and gluinos into the stau, relevant if 
$\msq<\mne$ and $\mgo<\msq,\mne$, with \textsc{Whizard}~2.1.1~\cite{Kilian:2007gr}. 

\begin{figure}[h!]
\centering
\setlength{\unitlength}{1\textwidth}
\begin{picture}(0.5,0.37)
\put(-0.024,0){
  \put(0.0,0.025){
  \includegraphics[scale=1.15]{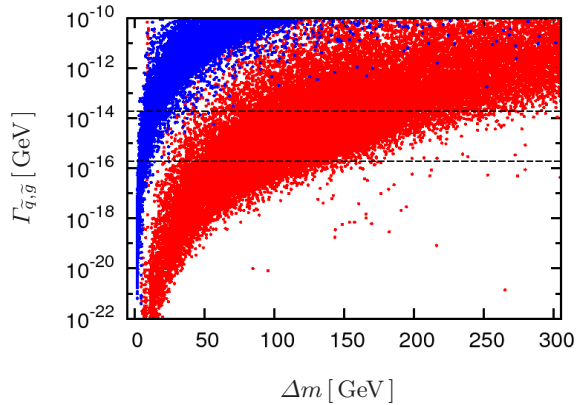}
} 
  \put(0.22,0.0){\footnotesize $\Delta m\,[\GEV\,]$}
  \put(-0.02,0.155){\rotatebox{90}{\footnotesize $\Gamma_{\sq,\go}\,[\GEV\,]$}}
  }
\end{picture}
\caption{Scatter plot displaying the dependence of the width of the squarks (blue points)
and gluino (red points) on the absolute mass difference to the stau, 
$\Delta m = \msq-\mstau$ and $\mgo-\mstau$, respectively. We consider all
squarks here, including stops and sbottoms. We only plot a point if
the corresponding width of the squark or gluino is the smallest width. 
The horizontal lines correspond to $\Gamma_{\go,\sq}=2\times10^{-14}\GEV$ 
and $\Gamma_{\go,\sq}=2\times10^{-16}\GEV$\@.
}
\label{fig:coloreddecay}
\end{figure}

Figure \ref{fig:coloreddecay} shows the mass gap between the squarks and the stau 
(blue points) as well as the  gluino and the stau (red points) versus the resulting decay 
width of the respective  sparticles. We only plot the points for which this width is the 
smallest among all widths of sparticles heavier than the stau. 
(This also ensures that parameter space points where both a squark and the gluino are 
long-lived appear only once.) For the gluino, even for mass gaps of up to $300\GEV$ we 
encountered points with small gluino widths which imply non-prompt decays into the stau. 
Note that these situation can only appear in the case that the masses of the squarks
\emph{and} EWinos are well above the gluino mass such that the $4$-body decays are 
suppressed by two off-shell propagators. For other situations the gluino width is typically 
much larger. We do not take into account loop-induced decay modes of gluino and squarks 
into staus leaving this for future investigations.

In order for the tracker analysis ($\D E/\D x$) to be efficient, the longitudinal and transversal 
impact parameter of the track candidates, $d_{z}$ and $d_{xy}$, are required to be smaller 
than $0.5\,\cm$ \cite{CMS1305.0491}. Bearing in mind that non-prompt decays typically
play a role in the case of rather small relative mass gaps between the heavier 
mother sparticle and the stau we do not expect a very pronounced kink in the track. 
We therefore consider a mother sparticle $X$ to be sufficiently short-lived to allow for 
the daughter stau to pass the tracker requirement, if
\beq
\label{eq:gammareq}
\Gamma_X>2\times10^{-14}\GEV\,.
\eeq
This corresponds to a decay length of $c\tau_X<1\,\cm$.\footnote{%
The decay length for a relativistic particle $X$ is $c\beta\gamma\tau_X$. However,
$\beta\gamma\simeq 1$ for $\beta\simeq0.7$. For heavy colored sparticles
produced close to threshold, $\beta\lesssim0.7$ is a typical velocity.}

For neutralinos and sneutrinos it requires very small mass gaps in order 
to violate \Eqref{eq:gammareq}. Consequently, these cases appear 
very rarely in our scan---$0.15\%$ of the points contain metastable neutralinos
while $0.6\%$ of the points contain metastable sneutrinos. The determination
of the appropriate collider limits for these cases requires a detailed analysis of all 
branching fraction and the consideration of various missing energy searches. Since 
these points are not of particular interest for this work we will leave the investigation 
of these cases for future work and will simply reject the corresponding points from 
the scan. For metastable charged sleptons other than the stau as well as
metastable charginos, we expect the analysis to be virtually identical, regardless
whether they decay into the stau or not, assuming that a possible kink in the track will 
not  significant change the sensitivity to the signature.

The case of metastable squarks and gluinos appears more frequently in our scan. 
We found that $5.8\%$ and $6.7\%$ of the points contain metastable squarks and 
gluinos, respectively. On the one hand, this relatively large fraction arises from the 
suppression of the required 3- and 4-body decays, on the other hand, it results from 
the dedicated scans, specifically accumulating points in the corresponding 
mass-degenerate  regions (see table \ref{tab:scandetailed}).
In the following we describe the treatment of metastable squarks and gluinos
in the determination of  the cross section upper limits.

If a metastable squark or gluino decays delayed, $\Gamma_{\go,\sq}<2\times10^{-14}\GEV$,
the stau is assumed not to be recognized in the tracker. Consequently, we only apply the 
ToF analysis taking into account the data from the muon chambers only. We refer
to this data as the `muon-only' analysis in the following. 
The cross section upper limits for the muon-only analysis have been reported for stops and 
gluinos only, where the direct  production of these sparticles is taken to be the only production 
mechanism. If we apply the muon-only analysis on long-lived staus we have to assume that
the kinematics of the staus are similar to the strongly interacting mother sparticles
that dominate the production. This is indeed the case for the small mass gaps that 
are required to cause the delayed decay of the stau. Furthermore, the detector response
of the drift-tubes in the muon chambers to an $R$-hadron carrying one unit of electric 
charge is virtually the same as for long-lived staus. Hence, we estimate the cross section 
upper limits for staus in the muon-only analysis by the limits derived for  stops. Note that 
the muon-only analysis has only be performed for the $8\TEV$ LHC run.

If the metastable colored sparticle has an even smaller decay width,
$\Gamma_{\go,\sq}<2\times10^{-16}\GEV$, corresponding to $c\tau_{\go,\sq}\gtrsim1\,$m, 
the muon-only analysis might not be applicable anymore. We therefore assume in this case
that the strongest sensitivity arises from the $R$-hadron itself that is recognized in the 
tracker.\footnote{%
The resulting sensitivity from the muon-only analysis and the $R$-hadron search is, in fact,
very similar. Hence, the analysis is not overly sensitive to the exact choice
of the width that separates the applicability of the muon-only and $R$-hadron
analysis.
}
Consequently, we apply the cross section upper limits from the corresponding $R$-hadron 
search where we conservatively choose the charge suppression model for the gluinos and 
squarks. To all production processes whose decay chains terminate in late decaying staus 
seen in the muon-only analysis or in $R$-hadron searches containing a gluino or squark the 
respective cross section limits are applied. By doing so, we implicitly assume that production 
modes of sparticles are only relevant if the mass gap between the produced sparticle to the 
respective sparticle seen in the detector is small or that the corresponding signal efficiencies 
do not depend strongly on the mass of the produced sparticles. The final signal strength is then 
determined by \Eqref{eq:signalstr}. (For those production processes that lead to a prompt decay 
into the stau we employ the Tracker$+$ToF analysis as described above.)

The  interpretation of the HSCP searches leads to very restrictive bounds on the sparticle 
masses.  For example, we did not find any allowed point in our scan with 
$m_{\s t_1},m_{\s b_1}\lesssim850\GEV$, $\msq\lesssim1400\GEV$ 
and $\mgo\lesssim1200\GEV$. Regarding the EWino sector, no point with 
$|\mu|,M_2\lesssim800\GEV$ survived the bounds.

\subsection{Further experimental constraints}\label{sec:bounds}

In this section we will discuss the implications of the most important 
experimental and theoretical constraints on the considered 17-parameter pMSSM
beyond direct SUSY searches considered in section \ref{sec:LHCbounds}.

\subsubsection{Constraints from Higgs searches at colliders}\label{sec:Higgssecbounds}

In addition to the condition \eqref{eq:higgswindow} we require that the scan points 
pass a variety of collider bounds from the Higgs searches at LEP, the Tevatron 
and the LHC imposed at the 95\% C.L\@. For the application of these bounds
we use the program package \textsc{HiggsBounds}~4.0.0~\cite{Bechtle:2013xfa},
which tests the compatibility of the predictions for the Higgs sector in a given model
against Higgs rates and masses measured in the mentioned experiments. We 
employed the full set of experimental results supplied by \HB. For the predictions for 
the spectrum of the MSSM Higgs sector \textsc{HiggsBounds} is linked to \FH~2.9.2.

The constraints have a large effect on our parameter space. Most importantly,
the bounds depend on $m_A$. Generically, we find that the parameter space is
constrained much more strongly for smaller values of $m_A$. Accordingly,
in the subset of points where the heavier CP-even Higgs takes the role of the 
SM-like Higgs, i.e., where $123\GEV<m_H<128\GEV$, nearly all points (99.88\%)
were rejected by the application of \textsc{HiggsBounds}. Most of these points 
(around 98\%) were rejected\footnote{%
Here we list the processes that contribute to the exclusion of a point most 
significantly as given in the output of \textsc{HiggsBounds}. Other processes 
may, however,  be similarly important.
}
by the CMS search for MSSM Higgs decays into tau pairs $(h,H,A^0)\to\tau\tau$ 
\cite{CMS-PAS-HIG-12-050}. The majority of the remaining points were excluded
by the search for Higgsstrahlung processes at LEP, where the Higgs is assumed 
to decay into $b\bar b$,  $(h,H,A^0)Z\to (b\bar b)Z$ \cite{Schael:2006cr}. 
Other processes are less important.

In the subset of points where the lighter CP-even Higgs plays the role of the SM-like Higgs, 
i.e., where $123\GEV<m_h<128\GEV$, around $27\%$ of the points were excluded. Again, 
for most of these (around 91\%) the CMS search for $(h,H,A^0)\to\tau\tau$ provides the 
highest significance. Further analyses of high importance are the search for 
$(h,H,A^0)\to ZZ \to \ell\ell\ell\ell$ at CMS \cite{CMS-PAS-HIG-13-002} and searches for a 
charged Higgs at CMS \cite{CMS-PAS-HIG-12-045}.
\begin{figure}
\centering
\setlength{\unitlength}{1\textwidth}
\begin{picture}(0.5,0.36)
\put(-0.024,0){
  \put(0.0,0.025){
  \includegraphics[scale=1.15]{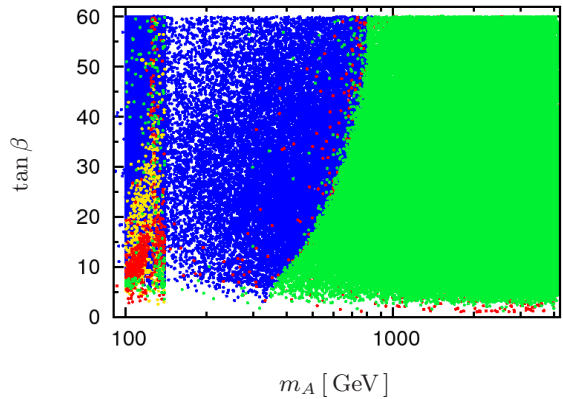}
} 
  \put(0.22,0.0){\footnotesize $m_A\,[\GEV\,]$}
  \put(-0.02,0.18){\rotatebox{90}{\footnotesize $\tan\beta$}}
  }
\end{picture}
\caption{%
Parameter points in the $m_A$-$\tan\beta$ plane. The blue points are rejected by 
the CMS search for $h,H,A^0\to\tau\tau$ processes \cite{CMS-PAS-HIG-12-050}, 
the yellow points are rejected by Higgsstrahlung processes $(h,H,A^0)Z\to (b\bar b)Z$ 
at LEP \cite{Schael:2006cr} and the red points are rejected by other searches. The 
green points have passed all exclusion limits as provided by \HB. (The bounds from 
HSCP searches have not been applied here.)
}
\label{fig:HBeffect}
\end{figure}

Figure \ref{fig:HBeffect} shows the allowed (green) and rejected points (blue,
yellow and red) in the $m_A$-$\tan\beta$ plane. Considering the rate of 
allowed versus rejected points in the different regions, the decoupling limit 
appears to be strongly favored by the current data.

\subsubsection{Constraints from flavor and precision observables}

Supersymmetric corrections to the mass of the $W$ boson impose another 
constraint on the parameter space.  Here, we use the experimental value 
$M_W = (80.385 \pm 0.015) \GEV$~\cite{Group:2012gb}.  Following
\cite{Bechtle:2012jw,Heinemeyer:2006px}, we increase the uncertainty by
a theory error of $15\MEV$, combine the uncertainties linearly and multiply
them by a factor of two in order to estimate the allowed range at the $95\%$ C.L. 
Thus, we apply the limit
\begin{equation} \label{eq:MW}
	M_W \in [80.325 ; 80.445] \GEV
\end{equation}
to the value calculated by \FH~2.9.2.

The flavor observables BR$(B \to X_s\gamma)$ and BR$(B_s^0\to\mu^+\mu^-)$ 
can be directly obtained from \MO. We use the world average
BR$(B \to X_s\gamma) = (3.43 \pm 0.21 \pm 0.07) \times 10^{-4}$
\cite{HFAGbsgAug12}. Treating the uncertainties as above we find the allowed 
range at the $95\%$ C.L.:
\begin{equation} \label{eq:BRbsg}
\text{BR}(B\to X_s\gamma) \in [2.87 ; 3.99] \times 10^{-4} \,.
\end{equation}
The rare $B_s^0$ decay has been observed with a branching ratio in
the $95\%$ C.L.\ range \cite{:2012ct,Chatrchyan:2013bka}
\begin{equation}\label{eq:BRbsmumu}
\text{BR}(B_s^0\to\mu^+\mu^-) \in [1.1 ; 6.4] \times 10^{-9} \,.
\end{equation}

Figure \ref{fig:pf_constr} illustrates the impact of these limits on the considered
pMSSM parameter space. The limit on $M_W$ rejects the largest
number of points. The lower panel shows that our choice \eqref{eq:MW}
ensures that the deviation of the $\rho$-parameter  from its SM value, $\Delta\rho$, 
does not exceed $0.0018$. The limit from $B \rightarrow X_s \gamma$ is particularly 
restrictive for the subset of points with $m_H$ in the LHC Higgs window as given by 
\eqref{eq:higgswindow}. Both flavor constraints imposed here favor large $m_A$.

\begin{figure}[bhp]
\centering
\setlength{\unitlength}{1\textwidth}
\begin{picture}(0.85,0.8)
  \put(-0.024,0.39){ 
  \put(-0.04,0.025){
  \includegraphics[scale=1.14]{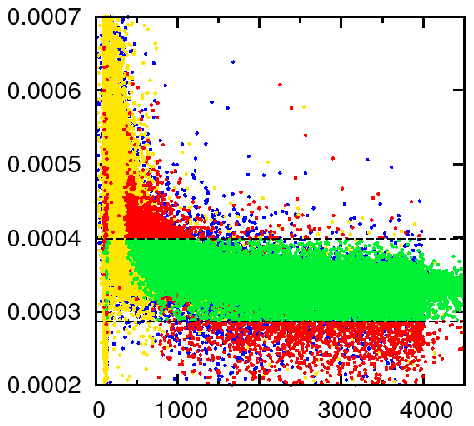}
 }
  \put(0.2,0.0){\footnotesize $m_A\;[\GEV\,]$}
  \put(-0.01,0.13){\rotatebox{90}{\footnotesize $\text{BR}(B \rightarrow X_s\,\gamma)$}}
  }
 \put(0.45,0.39){ 
  \put(-0.03,0.025){
  \includegraphics[scale=1.163]{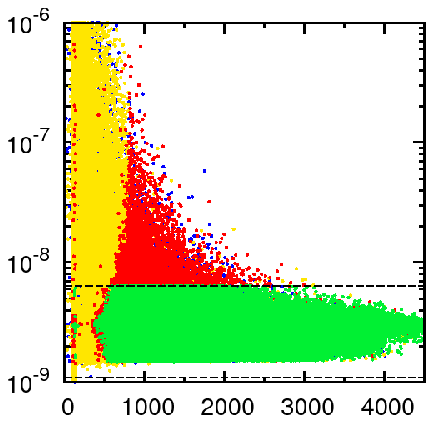}
}
  \put(0.2,0.0){\footnotesize $m_A\;[\GEV\,]$}
  \put(-0.00,0.121){\rotatebox{90}{\footnotesize $\text{BR}(B_s^0 \rightarrow \mu^+\,\mu^-)$}}
  }
 \put(0.205,0){ 
  \put(-0.016,0.025){
  \includegraphics[scale=1.163]{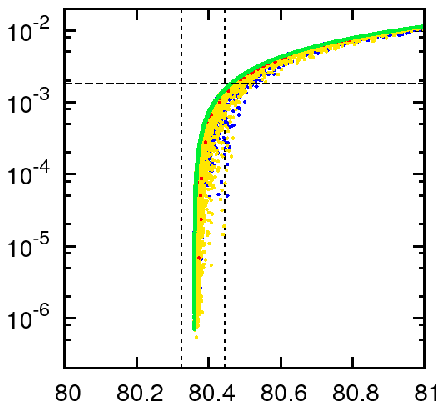}
} 
  \put(0.21,0.0){\footnotesize $M_W\;[\GEV\,]$}
  \put(0.02,0.18){\rotatebox{90}{\footnotesize $\Delta\rho$}}
  }
\end{picture}
\caption{%
Scatter plots displaying the effect of bounds from flavor and precision observables 
on the consider pMSSM parameter space. Upper panels:~Correlation between 
$\text{BR}(B \rightarrow X_s \gamma)$, $\text{BR}(B_s^0 \rightarrow \mu^+\,\mu^-)$ 
and $m_A$. The dashed lines denote the intervals \Eqref{eq:BRbsg} and 
\Eqref{eq:BRbsmumu}. Lower panel:~Correlation between the precision observables 
$\Delta\rho$ and $M_W$. The vertical and horizontal dashed lines denote the interval 
\Eqref{eq:MW} and $\Delta\rho= 0.0018$, respectively. We used the following color 
code. Blue: Rejected by the HSCP search. Yellow:~Passed the HSCP bounds. 
Red:~Additionally passed \HB. Green:~Additionally passed the flavor bounds. 
}
\label{fig:pf_constr}
\end{figure}

\subsection{Bounds from charge or color breaking minima}\label{sec:thbounds}

\begin{figure}[bp]
\centering
\setlength{\unitlength}{1\textwidth}
\begin{picture}(0.85,0.78)
  \put(-0.027,0.39){ 
  \put(-0.03,0.025){
  \includegraphics[scale=1.15]{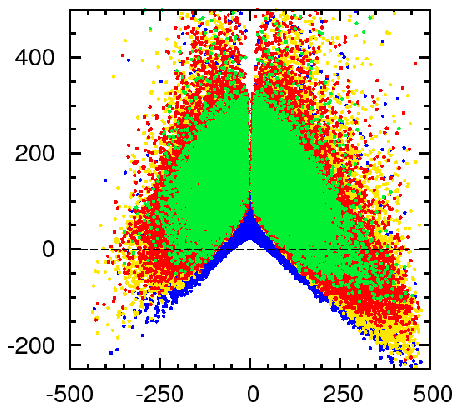} 
}
  \put(0.19,0.0){\footnotesize $\mu\tan\beta\;[\TEV\,]$}
  \put(0.01,0.112){\rotatebox{90}{\footnotesize RHS of \eqref{eq:CCBHisano}$\;[\TEV\,]$ }}
  }
   \put(0.44,0.39){ 
  \put(-0.03,0.025){
  \includegraphics[scale=1.15]{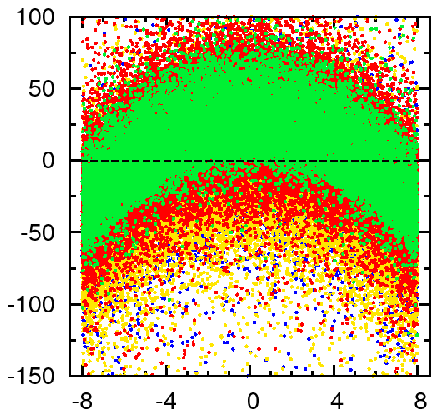}
}
  \put(0.19,0.0){\footnotesize $A_{\tau}\;[\TEV\,]$}
  \put(0.01,0.11){\rotatebox{90}{\footnotesize RHS of \eqref{eq:atauccb}$\;[\TEV^2\,]$ }}
  }
   \put(-0.027,0){ 
  \put(-0.03,0.025){
\includegraphics[scale=1.15]{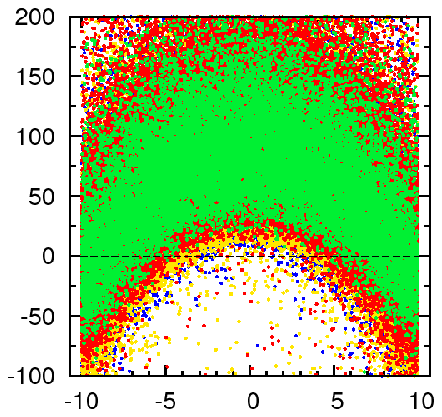}
} 
  \put(0.19,0.0){\footnotesize $A_b\;[\TEV\,]$}
  \put(0.01,0.11){\rotatebox{90}{\footnotesize RHS of \eqref{eq:abotccb}$\;[\TEV^2\,]$}}
  }
 \put(0.44,0){  
   \put(-0.03,0.025){
\includegraphics[scale=1.15]{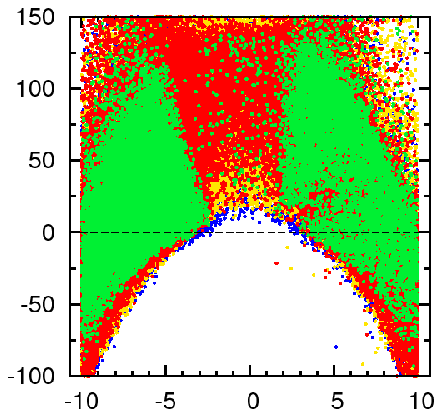}
}
  \put(0.19,0.0){\footnotesize $A_t\;[\TEV\,]$}
  \put(0.01,0.11){\rotatebox{90}{\footnotesize RHS of \eqref{eq:atopccb}$\;[\TEV^2\,]$}}
  }
\end{picture}
\caption{%
Scatter plots illustrating the CCB constraints on the considered parameter space. 
Points below the horizontal dashed lines are excluded by the corresponding CCB 
bound. We used the following color code. Blue: Rejected by the HSCP searches. 
Yellow:~Passed the HSCP bounds. Red:~Additionally passed \HB. Green:~Additionally 
passed the flavor and precision bounds.
}
\label{fig:th_constr}
\end{figure}

For large values of certain parameters, the MSSM scalar potential can 
acquire minima where U$(1)_\text{em}$ or SU$(3)_c$ is broken (charge or
color breaking, CCB).  For large $\tan\beta$, requiring the standard
electroweak vacuum to be stable or metastable with a lifetime larger
than the age of the universe implies an upper bound on the product
$\mu\tan\beta$ 
\cite{Rattazzi:1996fb,Hisano:2010re,Carena:2012mw,Kitahara:2013lfa}.
We use \cite{Kitahara:2013lfa},
\begin{align} \label{eq:CCBHisano}
0<& -|\mu \tan \beta_{\text{eff}}|  + 56.9 \sqrt{m_{\s L_3} m_{\s e_3}}
+ 57.1 \left(m_{\s L_3} +1.03\, m_{\s e_3} \right)   - 1.28 \times 10^4 \GEV
\nonumber\\
&+ \frac{1.67 \times 10^6 \GEV ^2 }{m_{\s L_3}+m_{\s e_3} }  
- 6.41 \times 10^7 \GEV ^3 \left ( \frac{1}{m_{\s L_3}^2  } + \frac{0.983}{m_{\s e_3}^2}  \right) ,
\end{align}
where
\begin{align}
\tan\beta_\text{eff} &\equiv \tan\beta \, \frac{1}{1 + \Delta_\tau} \,,
\\
\Delta_\tau &\simeq
-\frac{3g^2}{32\pi^2} \mu\tan\beta \, M_2 \, I(\msnutau,M_2,\mu)
+\frac{g^{\prime2}}{16\pi^2} \mu\tan\beta \, M_1 \,I(\mstau,\mstautwo,M_1)
\,,
\end{align}
and $I(a,b,c)$ is defined in \Eqref{eq:functI} in appendix~\ref{sec:resum}.
The quantity $\Delta_\tau$ describes the higher-order corrections to the
tau Yukawa coupling in the limit of large $\tan\beta$, analogous to the
corrections to the bottom Yukawa coupling discussed in
appendix~\ref{sec:resum}.  Depending on $m_A$ and $A_\tau$, the upper
bound \eqref{eq:CCBHisano} can become more stringent by approximately
$20\%$ \cite{Carena:2012mw}.

In order to take into account CCB constraints on the trilinear
couplings, we apply the simple conditions
\cite{Frere:1983ag,AlvarezGaume:1983gj,Claudson:1983et,Kounnas:1983td,Derendinger:1983bz}
\begin{align}
\label{eq:atauccb}
	0 &< -A_\tau^2+
	 3 \left( m^2_{\s{L}_3} + m^2_{\s{e}_3} + m^2_{H_d} + \mu^2 \right) ,\\
\label{eq:abotccb}
	0 &< -A_b^2+
	 3 \left( m^2_{\s{Q}_3} + m^2_{\s{d}_3} + m^2_{H_d} + \mu^2 \right) ,\\
\label{eq:atopccb}
	0 &< -A_t^2+
	 3 \left( m^2_{\s{Q}_3} + m^2_{\s{u}_3} + m^2_{H_u} + \mu^2 \right) .
\end{align}

We caution that the listed analytical constraints are not always reliable
\cite{Gunion:1987qv,Casas:1995pd,Ferreira:2000hg,Ferreira:2001tk,Camargo-Molina:2013sta}.
We impose them as a conservative first estimate, leaving a detailed numerical analysis 
employing the recently released program \textsc{Vevacious} \cite{Camargo-Molina:2013qva} 
for future work. Figure~\ref{fig:th_constr} shows the impact of the constraints 
\eqref{eq:CCBHisano} and (\ref{eq:atauccb}--\ref{eq:atopccb}) on the considered pMSSM 
parameter space. The bounds on the trilinear couplings are quite restrictive. Furthermore, 
we see that the chosen range for $A_\tau$ almost saturates the allowed region.

\section{Stau yields in the Monte Carlo scan} \label{sec:yieldMCscan}

The results of section \ref{sec:systscan} allowed us to identify all regions that potentially
lead to exceptionally small stau yields. In this section we will investigate the limiting factors
for low stau yields that could arise from various constraints. This is especially
important for regions that contain large Higgs-sfermion couplings  which are governed
by a priori free parameters of the theory. In the presence of large left-right mixings 
of the sfermions one can only constrain the possible values of the yield by
imposing constraints on the parameters that govern the Higgs-sfermion couplings.
Working out the impact of these constraints is the subject of the present section. 
Furthermore, we will quantify how HSCP searches constrain the possible values of 
the yield. These searches are especially constraining in the case of co-annihilation 
with colored sparticles. Therefore, we will utilize the pMSSM Monte Carlo scan introduced 
in section \ref{sec:pMSSMsummary}.

\subsection{Application of constraints} \label{sec:constraintsappl}

\begin{figure}[hp]
\centering
\setlength{\unitlength}{1\textwidth}
\begin{picture}(0.85,1.14)
  \put(-0.024,0.78){ 
  \put(-0.03,0.025){
\includegraphics[scale=1.15]{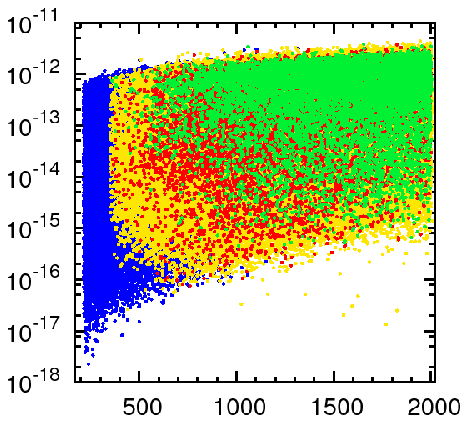}
}
  \put(0.195,0.0){\footnotesize $\mstau\;[\GEV\,]$}
  \put(0.0,0.2){\rotatebox{90}{\footnotesize $Y$}}
  \put(0.1615,0.0755){\scriptsize Effect of constraints}
  }
   \put(0.43,0.78){ 
  \put(-0.03,0.025){
\includegraphics[scale=1.15]{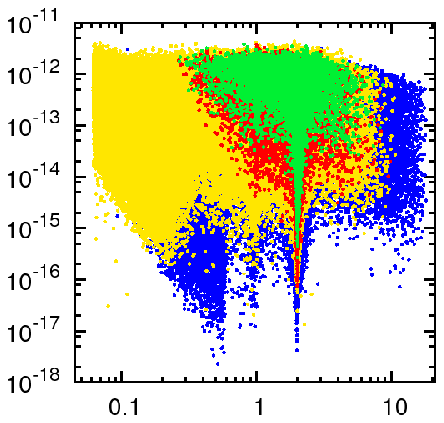}
}
  \put(0.202,0.0){\footnotesize $m_H/\mstau$}
  \put(0.0,0.2){\rotatebox{90}{\footnotesize $Y$}}
  \put(0.1615,0.0755){\scriptsize Effect of constraints}
  }
  \put(-0.024,0.39){ 
  \put(-0.03,0.025){
\includegraphics[scale=1.15]{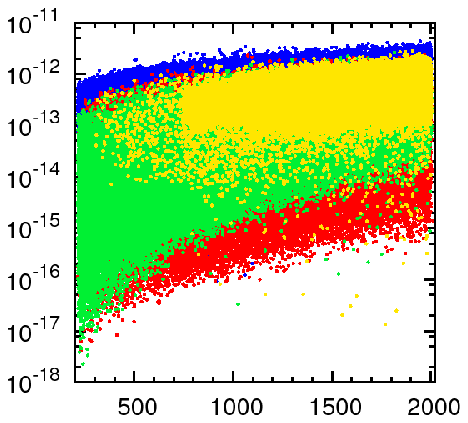} 
}
  \put(0.195,0.0){\footnotesize $\mstau\;[\GEV\,]$}
  \put(0.0,0.2){\rotatebox{90}{\footnotesize $Y$}}
  \put(0.13,0.0755){\scriptsize Dominant annihilation ch.}
  }
   \put(0.43,0.39){ 
  \put(-0.03,0.025){
\includegraphics[scale=1.15]{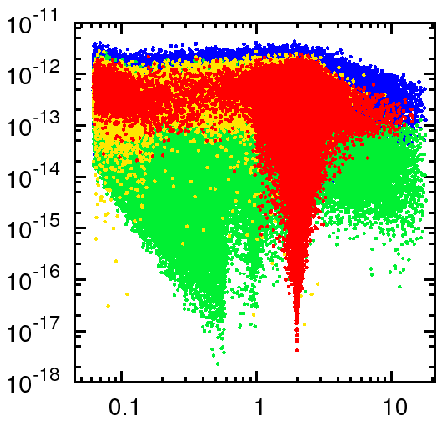}
}
  \put(0.202,0.0){\footnotesize $m_H/\mstau$}
  \put(0.0,0.2){\rotatebox{90}{\footnotesize $Y$}}
  \put(0.13,0.0755){\scriptsize Dominant annihilation ch.}
  }
 \put(-0.024,0){ 
  \put(-0.03,0.025){
\includegraphics[scale=1.15]{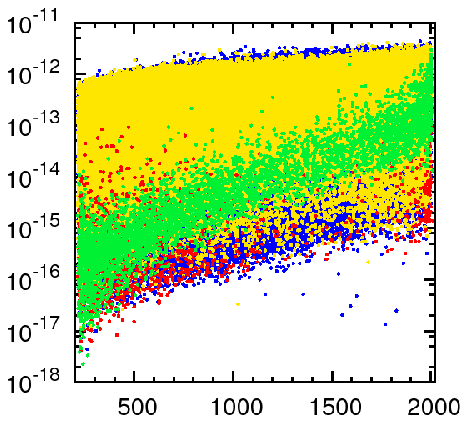}
} 
  \put(0.195,0.0){\footnotesize $\mstau\;[\GEV\,]$}
  \put(0.0,0.2){\rotatebox{90}{\footnotesize $Y$}}
  \put(0.135,0.0755){\scriptsize Dominant production ch.}
  }
 \put(0.43,0){  
   \put(-0.03,0.025){
\includegraphics[scale=1.15]{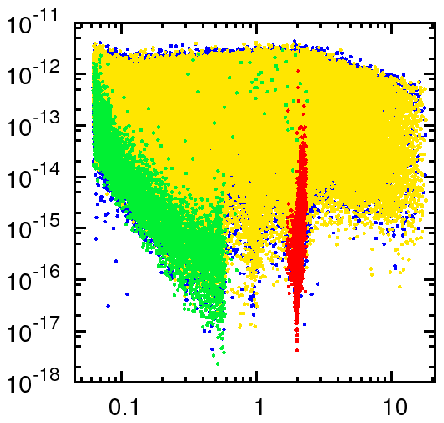}
}
  \put(0.202,0.0){\footnotesize $m_H/\mstau$}
  \put(0.0,0.2){\rotatebox{90}{\footnotesize $Y$}}
  \put(0.135,0.0755){\scriptsize Dominant production ch.}
}
\end{picture}
\caption{%
Distribution of scan points in the $\mstau$-$Y$ plane (left panels) and 
$m_H/\mstau$-$Y$ plane (right panels). Upper panels:~Effect of the constraints
on the parameter space. The blue, yellow and red points are rejected by 
the HSCP searches, \HB$+$FP constraints and CCB bounds, respectively. The 
green points pass all the constraints. Middle panels:~Dominant annihilation 
channels. The red, green and yellow points belong to the Higgs resonant region, 
Higgs final state region and co-annihilation regions, respectively. The blue 
points do not belong to one of these classes. Lower panels:~Production channels 
that contribute dominantly to the strength of the HSCP signal. For the green and 
red points direct stau production via a the light and heavy Higgs is dominant, 
respectively. The yellow points are dominated by other production processes in 
the stau sector. The blue points are dominated by other processes. Note that the 
point density is saturated in parts of the plane such that blue points are simply 
covered by the others, etc.}
\label{fig:allregconstr6}
\end{figure}

The upper panels of figure \ref{fig:allregconstr6} show the effect of the constraints
discussed in section~\ref{sec:pMSSMsummary}. The blue points are rejected by the HSCP
searches performed at the 7 and $8\TEV$ LHC (see section \ref{sec:LHCbounds} for details).
The most obvious result is that the HSCP searches reject all points with $\mstau\lesssim340\GEV$.
This is the most conservative bound on the stau mass in agreement with the bound reported in 
\cite{CMS1305.0491}.\footnote{%
The bound in \cite{CMS1305.0491} was obtained for an almost completely right-handed lighter 
stau. As a slightly smaller DY cross section for $\stau_1$ production is obtained for 
$\thest\neq\pi/2$ \cite{Heisig:2011dr}, we could expect allowed points lying $\mathcal{O}(10)\GEV$
below the limit of \cite{CMS1305.0491}.
However, such a value for $\thest$ requires either the heavier stau and tau sneutrino to be 
relatively light or the off-diagonal elements of the stau mass matrix, $X_\tau$, to be relatively 
large leading to an enhanced stau-Higgs coupling. In both cases additional contributions 
enhance the overall production rate. Hence, we do not find any allowed points below the limit of 
\cite{CMS1305.0491}.
}
For small stau yields the bound on the stau mass tends to become more restrictive---the border
between blue and yellow points shows a kink at around $Y=10^{-16}$. This feature can be 
understood as follows. In the region of small stau masses, small yields $Y\lesssim10^{-15}$ are 
typically achieved in the Higgs final state region (green points in the middle left panel in 
figure~\ref{fig:allregconstr6}) where the couplings to the Higgs are enhanced. For these points 
the production of staus via a light or heavy $CP$-even neutral Higgs at the LHC is typically the 
dominant contribution to the stau production (see green and red points, respectively, in the lower 
left corner of the lower left panel in figure~\ref{fig:allregconstr6}). This additional production 
mode raises the stau mass limit and forbids this region. Here we see a first correlation 
between the observable in the early universe and the measurements at the LHC\@. A similar 
effect occurs in the Higgs resonant region. This is best seen in the right panels of figure 
\ref{fig:allregconstr6}, where we plot the yield against $m_H/\mstau$. In the resonance peak,
$m_H/\mstau\simeq2$, very small stau yields are obtained. However, the very tip of this peak
is excluded by HSCP searches, to a large extent due to the resonant production
of staus via the heavy Higgs (see lower right panel of figure~\ref{fig:allregconstr6}). 
For co-annihilation scenarios the bounds on the sparticle masses restrict the possible 
stau yields according to the scaling of the yield with the stau mass. The yellow points in the 
middle left panel of figure \ref{fig:allregconstr6} show the domain of the co-annihilation regions 
in the $\mstau$-$Y$ plane.

The bounds from MSSM Higgs searches taken from \textsc{HiggsBounds} and the flavor and 
precision bounds (abbreviated by FP in the following) are particularly restrictive in the region 
of small $m_H/\mstau$. The yellow points in the upper panels of figure \ref{fig:allregconstr6}
are rejected by \HB\ and FP constraints. In fact the complete region of $m_H/\mstau\lesssim0.2$
is excluded by these bounds. For smaller yields $Y\lesssim10^{-13}$, even higher values of 
$m_H/\mstau$ are rejected by the \HB\ and FP constraints. This is partly due to the fact that the 
regions with smaller yields $Y\lesssim10^{-13}$ are dominated by the Higgs final state region 
and Higgs resonant region (see green and red points, respectively, 
in the middle panels of figure \ref{fig:allregconstr6}), which require large 
stau-Higgs couplings. These are more easily achieved with large values of $\tan\beta$
for which the constraints on $m_A$ from \HB\ become even stronger, see figure \ref{fig:HBeffect}.

Constraints from CCB reject points in all corners of the displayed planes. The 
constraints on $A_t$ and $A_b$  can affect points without co-annihilation effects with stops
or sbottoms and are therefore not necessarily related to the stau yield. However, 
a clear correlation is seen in the region of smallest stau yields. The CCB bounds
push up the minimal yield allowed by the HSCP bounds by about another order of magnitude,
see red points in the upper right panel of figure \ref{fig:allregconstr6}.

Note, finally, that the allowed points (green points in the upper panels of figure~\ref{fig:allregconstr6})
all lie within a relatively narrow band in $m_H/\mstau$. They span about four orders of magnitude
in the yield, $2\times10^{-16}\lesssim Y\lesssim 4\times10^{-12}$. 

In figures \ref{fig:classwise1} and \ref{fig:classwise2} we show the effect of
the constraints on the parameter space for the above defined regions separately.
The red points belong to the respective region, while the blue points belong
to the complete set of points. The pure colors denote the allowed points, while the 
pale points are excluded by one or more of the constraints. Points with yields smaller 
than $10^{-14}$ occur only for resonant annihilation via a heavy Higgs (see middle 
panels of \Figref{fig:classwise1}). Among these points, the smallest yields ($Y\lesssim 10^{-15}$) 
are achieved for dominant stau annihilation and no co-annihilation effects.
Away from the heavy Higgs resonance we find yields as small as $2\times 10^{-14}$ in the 
Higgs final state region with large stau-Higgs coupling $\hat{C}_{h,\stau_1}\sim1$
(cp.\ middle left and upper right panels of \Figref{fig:classwise1}), and slightly below 
$10^{-13}$ in the gluino and 3rd generation squark co-annihilation regions
(see upper left and middle right panels of \Figref{fig:classwise2}).

It is interesting to note that in the EWino co-annihilation region
(lower panels of \Figref{fig:classwise1}) and in the 3rd generation
squark co-annihilation region (middle and lower panels of \Figref{fig:classwise2})
stau yields down to roughly $5\times 10^{-15}$ are allowed. The smallest yields are 
again reached for resonant annihilation via a heavy Higgs. In these regions
no particular left-right mixing in the stau sector (and for EWino co-annihilation no 
particular left-right mixing in the sfermion sector at all) is required. Hence, these are 
the lowest values we found that could equally be realized in scenarios with a 
selectron or smuon NLSP.

The points with the largest yields almost always belong to the bulk region 
(see blue points in the middle panels of figure~\ref{fig:allregconstr6}). Note that there is a 
relatively sharp limit of existing points in the high yield end, in contrast to the lower end of 
the range featuring a few scattered points with very low yields. This is due to the fact that 
the potential to increase the yield is limited by the number of sparticles that could increase 
the yield by virtue of co-annihilation effects. In fact, the estimate given in \Eqref{eq:YstauR} 
lies approximately in the middle of the band of blue points (bulk region) in the middle left 
panel of figure \ref{fig:allregconstr6}. Thus, \Eqref{eq:YstauR} is not too far from the largest 
yields that can be achieved in the pMSSM.

The percentage of surviving points in the regions is
4.4\% in the bulk region, 0.18\% in the {Higgs final state region}, 
5.2\% in the {Higgs resonant region},
5.8\% in the EWino co-annihilation region,
1.1\% in the gluino co-annihilation region, 
3\% in the squark co-annihilation region and 
3.7\% in the 3rd generation squark co-annihilation region.

We plot the yield against the rescaled Higgs-sfermion coupling $\hat{C}_{\Phi,\s f}/m_{\s f}$
for the case of the stau as well as for the case of the stop and sbottom 
in the upper and middle right panels of figure \ref{fig:classwise1} as well as the lower
left and right panels of figure \ref{fig:classwise2}, respectively. In the latter case we
exemplarily plot $\hat{C}_{h,\s t_1}/m_{\s t_1}$ and $\hat{C}_{H,\s b_1}/m_{\s b_1}$; 
the couplings to the respective other Higgs behave roughly similarly.
Large values are typically excluded mainly by the
CCB bounds and precision observables as well as by flavor constraints.

Finally, we note that unitarity of the $S$-matrix sets further bounds on the involved
couplings, see, e.g., \cite{Griest:1989wd,Ratz:2008qh,Berger:2008ti}. The minimal yields
allowed by unitarity are roughly $Y\simeq 7\times 10^{-18}\,(\mstau/\TEV)$ 
\cite{Berger:2008ti} for stau-stau annihilation and 
$Y\simeq 4\times 10^{-17}\,(\mstau/\TEV)$ for third-generation squark co-annihilation, 
taking additional degrees of freedom into account. 
As the minimal yields allowed in the respective regions are more than an order of 
magnitude larger than these values we assume that the bounds from the requirement 
of unitarity are significantly weaker than the other bounds considered in this paper, 
especially those from CCB minima. However, a detailed analysis investigating the
particular annihilation processes relevant for our scenario and the effects of relaxing 
the approximations that were used in \cite{Berger:2008ti} appears worthwhile and may 
lead to more stringent bounds.

\begin{figure}[hp]
\centering
\setlength{\unitlength}{1\textwidth}
\begin{picture}(0.85,1.14)
  \put(-0.024,0.78){ 
  \put(-0.03,0.025){
\includegraphics[scale=1.15]{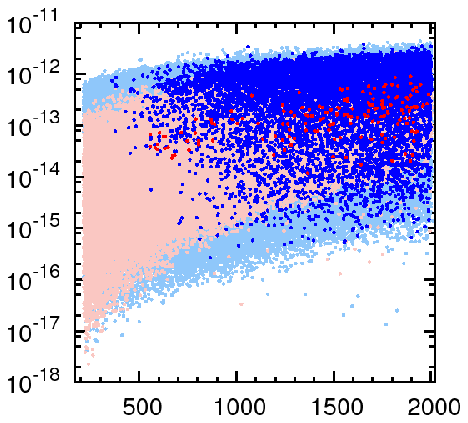} 
}
  \put(0.195,0.0){\footnotesize $\mstau\;[\GEV\,]$}
  \put(0.0,0.2){\rotatebox{90}{\footnotesize $Y$}}
  \put(0.112,0.08){\tiny\color{red} Higgs final state region}
  }
   \put(0.43,0.78){ 
  \put(-0.03,0.025){
  \includegraphics[scale=1.15]{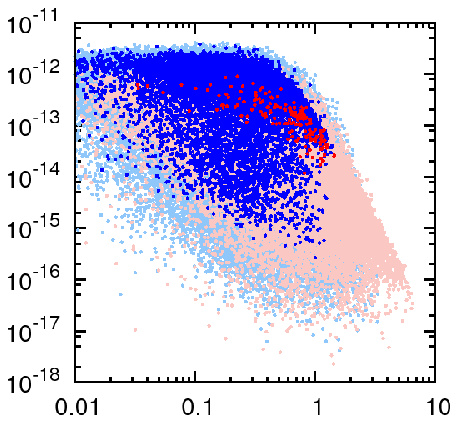}
 }
  \put(0.203,0.0){\footnotesize $\hat C_{h,\stau_1}/\mstau$}
  \put(0.0,0.2){\rotatebox{90}{\footnotesize $Y$}}
  \put(0.112,0.08){\tiny\color{red}  Higgs final state region}
  }
  \put(-0.024,0.39){ 
  \put(-0.03,0.025){
\includegraphics[scale=1.15]{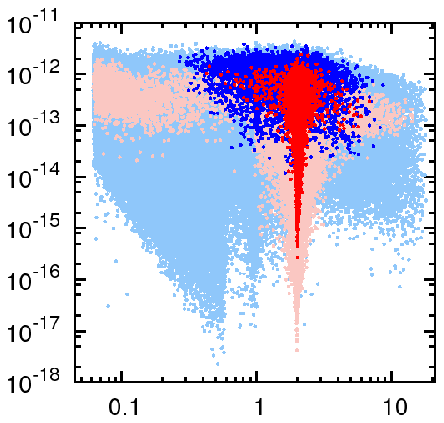}
}
  \put(0.202,0.0){\footnotesize $m_H/\mstau$}
  \put(0.0,0.2){\rotatebox{90}{\footnotesize $Y$}}
  \put(0.112,0.08){\tiny\color{red}  Higgs resonant region}
  }
   \put(0.43,0.39){ 
  \put(-0.03,0.025){
\includegraphics[scale=1.15]{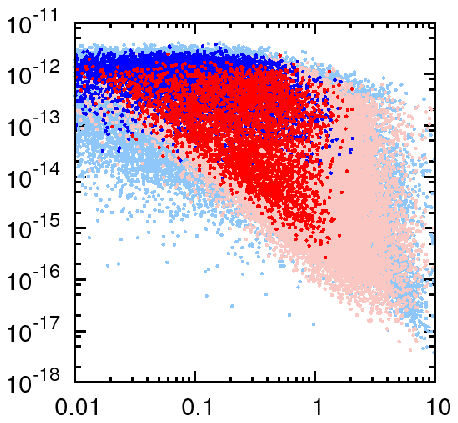}
}
  \put(0.201,0.0){\footnotesize $\hat C_{H,\stau_1}/\mstau$}
  \put(0.0,0.2){\rotatebox{90}{\footnotesize $Y$}}
  \put(0.112,0.08){\tiny\color{red}  Higgs resonant region}
  }
 \put(-0.024,0){ 
  \put(-0.03,0.025){
\includegraphics[scale=1.15]{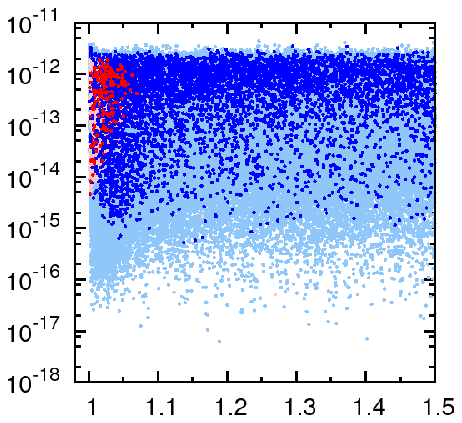}
} 
  \put(0.205,0.0){\footnotesize $\mne/\mstau$}
  \put(0.0,0.2){\rotatebox{90}{\footnotesize $Y$}}
  \put(0.112,0.08){\tiny\color{red}  EWino co-annihilation region}
  }
 \put(0.43,0){  
   \put(-0.03,0.025){
\includegraphics[scale=1.15]{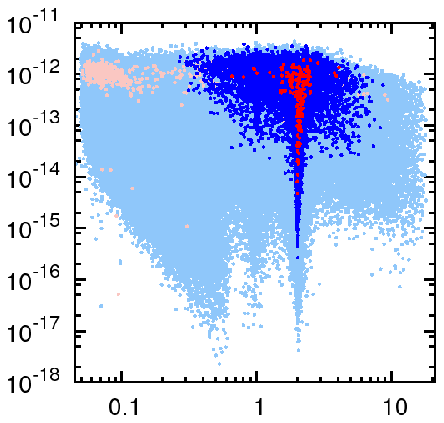}
}
  \put(0.202,0.0){\footnotesize $m_A/\mstau$}
  \put(0.0,0.2){\rotatebox{90}{\footnotesize $Y$}}
  \put(0.112,0.08){\tiny\color{red}  EWino co-annihilation region}
}
\end{picture}
\caption{Allowed points (pure colors) and rejected points (pale colors)
in the specified annihilation regions (red points) and in the full set (blue points).
Upper panels:~Higgs final state region.
Middle panels:~Higgs resonant region.
Lower panels:~EWino co-annihilation region. 
The regions are defined in section \ref{sec:classificationregions}.
}
\label{fig:classwise1}
\end{figure}

\begin{figure}[hp]
\centering
\setlength{\unitlength}{1\textwidth}
\begin{picture}(0.85,1.14)
  \put(-0.024,0.78){ 
  \put(-0.03,0.025){
\includegraphics[scale=1.15]{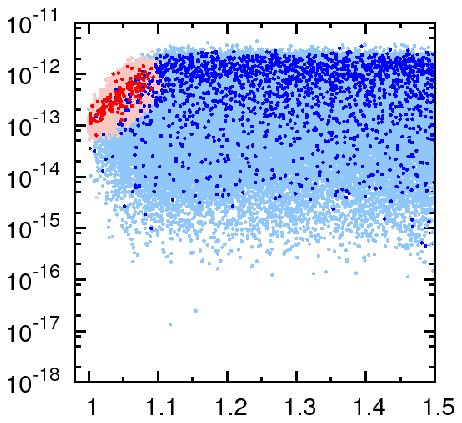} 
}
  \put(0.205,0.0){\footnotesize $\mgo/\mstau$}
  \put(0.0,0.2){\rotatebox{90}{\footnotesize $Y$}}
  \put(0.112,0.08){\tiny\color{red}  Gluino co-annihilation region}
  }
   \put(0.43,0.78){ 
  \put(-0.03,0.025){
\includegraphics[scale=1.15]{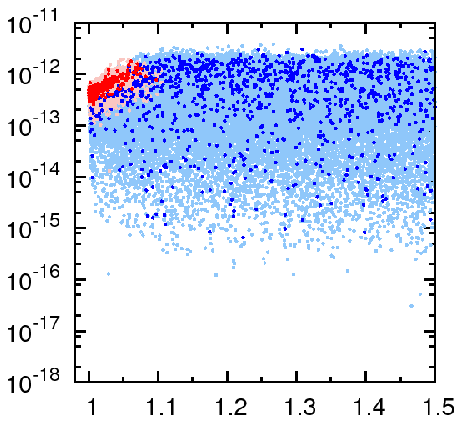}

}
  \put(0.205,0.0){\footnotesize $\msq/\mstau$}
  \put(0.0,0.2){\rotatebox{90}{\footnotesize $Y$}}
  \put(0.112,0.08){\tiny\color{red}  Squark co-annihilation region}
  }
  \put(-0.024,0.39){ 
  \put(-0.03,0.025){
\includegraphics[scale=1.15]{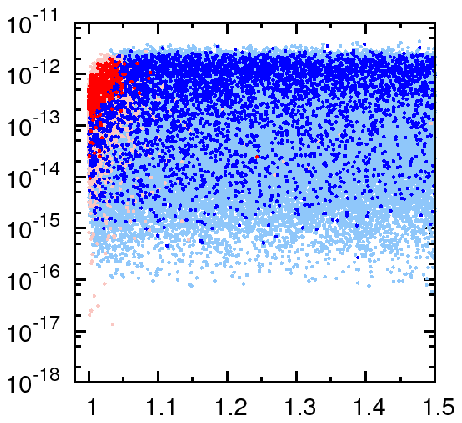} 
}
  \put(0.155,0.0){\footnotesize $\min(m_{\s b_1},m_{\s t_1})/\mstau$}
  \put(0.0,0.2){\rotatebox{90}{\footnotesize $Y$}}
  \put(0.112,0.08){\tiny\color{red}  3rd generation squark co-ann.}
  }
   \put(0.43,0.39){ 
  \put(-0.03,0.025){
\includegraphics[scale=1.15]{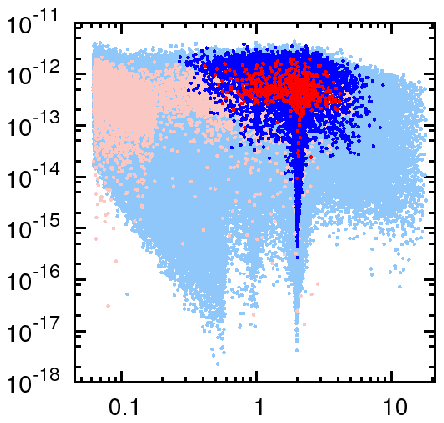}
}
  \put(0.202,0.0){\footnotesize $m_H/\mstau$}
  \put(0.0,0.2){\rotatebox{90}{\footnotesize $Y$}}
  \put(0.112,0.08){\tiny\color{red}  3rd generation squark co-ann.}
  }
 \put(-0.024,0){ 
  \put(-0.03,0.025){
\includegraphics[scale=1.15]{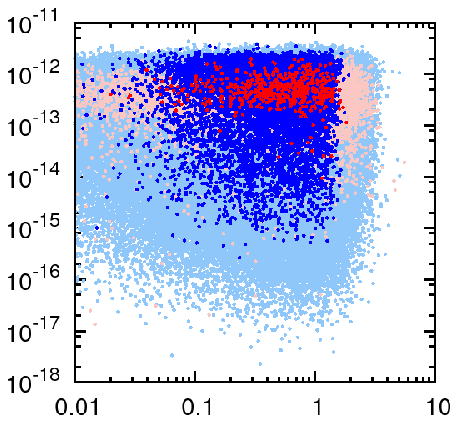}
} 
  \put(0.205,0.0){\footnotesize $\hat C_{h,\s t_1}/m_{\s t_1}$}
  \put(0.0,0.2){\rotatebox{90}{\footnotesize $Y$}}
  \put(0.112,0.08){\tiny\color{red}  3rd generation squark co-ann.}
  }
 \put(0.43,0){  
   \put(-0.03,0.025){
\includegraphics[scale=1.15]{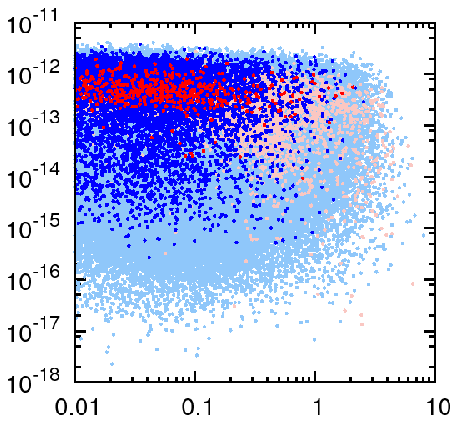}
}
  \put(0.203,0.0){\footnotesize $\hat C_{H,\s b_1}/m_{\s b_1}$}
  \put(0.0,0.2){\rotatebox{90}{\footnotesize $Y$}}
  \put(0.112,0.08){\tiny\color{red}  3rd generation squark co-ann.}
}
\end{picture}
\caption{Allowed points (pure colors) and rejected points (pale colors) in the specified 
annihilation regions (red points) and in the full set (blue points). Upper left panel:~Gluino 
co-annihilation region. Upper right panel:~Co-annihilation with the first- and 
second-generation squarks. Middle and lower panels:~Co-annihilation with sbottoms 
and stops. The regions are defined in section \ref{sec:classificationregions}.
}
\label{fig:classwise2}
\end{figure}

\subsection{Prospects to narrow down the stau yield at the LHC}
\label{sec:dhstrate}

In the case of a discovery at the upcoming LHC runs it would be desirable to 
determine the stau yield from the  LHC data and conclude on the viability of the 
underlying cosmological model. This is a difficult task as the yield depends upon 
various parameters with very different accessibility at the LHC\@. As a first step, we 
discuss in this subsection how one might be able to determine the parameter space 
region the scenario belongs to, which allows to narrow down the allowed range 
for the stau yield. The discussion remains qualitative and is not intended to be 
exhaustive.

The points in the scan that are close to the exclusion limit from the HSCP
searches at $7$ and $8\TEV$ typically provide a SUSY cross section at the 
$14\TEV$ LHC run of $\sigma_{14\TEV}^{\text{SUSY}}\simeq100\fb$.
This gives us a rough idea of the prospects for studying long-lived stau
scenarios at the LHC\@. For instance, with $300\ifb$ we obtain a total amount
of $3\times10^4$ SUSY events for these points.
In fact, due to the prominent signature of staus at the LHC, we could already
learn a lot about the spectrum from much fewer events.
First, already at the stage of discovering a long-lived stau scenario by 
the measurement of charged highly ionizing tracks in the detector,
we are provided with a good determination of the stau mass with a precision
around $15\%$ \cite{Chatrchyan1205.0272}. In the search for long-lived staus,
discovery is expected to take place on the basis of very few observed signal 
events \cite{Heisig:2012zq}, which translates into a total amount of $\mathcal{O}(10)$ 
produced stau pairs.\footnote{%
The discovery reach for an integrated luminosity 
of $300\ifb$ at the $14\TEV$ LHC is $\mstau\simeq700\GEV$ for the most 
conservative case of minimal direct DY production and up to $\mstau\simeq3\TEV$ for 
the case where the stau, the gluino and the squarks are close in mass \cite{Heisig:2012zq}.
The exclusion reach is similar.
}
Second, the cross section for direct stau production differs from that for the production 
of colored sparticles with a similar mass by around five orders of magnitude. Indeed, 
in our scan the SUSY production cross section for a given stau mass spans four to five 
orders of magnitude, where the lower edge corresponds to points with dominant direct 
DY production while the upper edge corresponds to scenarios with a very small mass 
splitting between the staus and the colored sparticles, in particular the first and second 
generation squarks. Thus, from the relatively precise determination of the stau mass 
and a rough idea of the production cross section one might, already at the stage of 
discovery, be able to decide whether the data is compatible with a gluino or squark 
co-annihilation scenario or not. Namely, if the stau is relatively light such that the 
number of observed events is compatible with direct DY production, a co-annihilation
scenario that could provide low yields could be excluded. On the other hand, if the 
stau is relatively heavy with respect to the measured rate of events such that dominant 
direct DY production is excluded, there are a variety of possibilities that could apply. 
This is in particular true for the intermediate range of production rates which could be 
compatible with stop or EWino co-annihilation or resonant stau annihilation via a 
heavy Higgs. In this case, more data is needed to distinguish between different 
scenarios. Let us briefly comment on two of them.

The first case concerns closely mass-degenerate staus and gluinos or squarks. 
As shown in section~\ref{sec:appHSCPlimits}, here the appearance of delayed 
decays is a quite common feature,  at least in the absence of other nearly 
mass-degenerate sparticles. Provided a very good understanding of the
detector, such a scenario could hence be identified by the appearance
of charge flipping tracks or other peculiarities that could occur due to the
presence of long-lived or late-decaying $R$-hadrons in the detector.

Another scenario, which is particularly interesting, is the Higgs resonant region.
Due to the appearance of the equally resonant production channel at the LHC 
this scenario provides a distinct signature \cite{Lindert:2011td}. We have seen
from the lower right panel in figure~\ref{fig:allregconstr6} that this production
channel can indeed be the dominant production channel of staus at the LHC
particularly in the region of low stau yields. As discussed in \cite{Lindert:2011td},
the velocity distribution of staus arising from the $s$-channel Higgs diagram
peaks at significantly lower velocities than, for instance, that for direct DY production.
Although challenging for the trigger settings (see, e.g., \cite{Heisig:2012zq}),
this signature can provide a way to distinguish the resonant $s$-channel Higgs
region from other regions. Furthermore, the invariant mass of these events would 
reveal a distinct peak at twice the stau mass once more data is accumulated. 
Note that this signal is quite clean with minimal dilution by background. 
Consequently, such a peak might be visible with a comparatively small number of 
events.

\section{Conclusions}\label{sec:Conclusion}

In this work we presented a thorough survey for possible values for the stau yield in the 
framework of the MSSM with a long-lived stau NLSP\@. Focussing on the
mass region that might still be accessible to a discovery at a long-term LHC run at $14\TEV$,
we pinned down the various possibilities for obtaining small stau yields in the pMSSM 
parameter space.
In particular we showed the different possibilities to lower the stau yield by co-annihilation 
effects, resonance effects, enhanced Higgs-sfermion couplings and combinations thereof. 
We were able to determine the following configurations with an increasing potential to 
achieve low stau yields. In the absence of any left-right mixing in the stau sector a light 
neutralino in the $t$-channel of the annihilation diagram can lead to a decrease in the yield 
with respect to the decoupled neutralino case, typically by a factor of about 2.

In contrast, a co-annihilating bino as well as co-annihilating first- and second-generation
sleptons increase the yield, again by factors of roughly 2.
Scenarios with squark and gluino co-annihilation can lead to a decrease of the yield
by a factor of $\mathcal{O}(10)$.
We found that a decrease of the yield by significantly more than one order of magnitude 
can only be achieved through annihilation processes which involve large Higgs-sfermion 
couplings, a resonant Higgs in the $s$-channel or both. 

In order to evaluate the phenomenological viability of the considered parameter space 
regions we performed a Monte Carlo scan over the 17-dimensional pMSSM with
the stau being the lightest among the MSSM sparticles. We interpreted the
Higgs boson recently discovered at the LHC as one of the $CP$-even neutral Higgses 
of the MSSM\@. By restricting $m_A$ to small values we forced around half of the scan 
points to explicitly lie outside the decoupling limit in order to cover interesting effects of 
large mixing in the Higgs sector. However, we found that almost all of these points are 
rejected by MSSM Higgs searches, most strongly by recent LHC searches.
We placed special emphasis on interpreting the current LHC limits for heavy stable 
charged particles. Data from the 7 and $8\TEV$ LHC runs were taken into account. 
Further, we explicitly included the possibility of long-lived colored sparticles 
appearing due to phase space suppression. We found that long-lived gluinos can 
appear for mass gaps up to $\Delta m \lesssim300\GEV$ if all 2- and 3-body decays 
are kinematically forbidden. Accordingly, we included the $R$-hadron searches 
performed by CMS in our analysis. The obtained results imply conservative mass limits 
on some of the model parameters. These limits most importantly constrain the yield 
in co-annihilation regions.
Furthermore, we showed the effects of the constraints from collider searches for MSSM 
Higgs signals, from flavor and precision observables as well as from CCB bounds on the 
allowed values of the stau yield in different regions. 

We found that all points with stau yields $Y\lesssim10^{-14}$ that feature a dominant 
annihilation into Higgs final states were excluded by these bounds.
Points with $Y<10^{-14}$ only survived in the vicinity of the resonant pole of the Higgs 
propagator at $ m_A\simeq2\mstau$. However, we encountered different scenarios
with this feature. For staus with a large left-right mixing their annihilation via an 
$s$-channel heavy Higgs provides the most effective way to achieve low stau yields, 
which can reach roughly $2\times10^{-16}$. For cases without mixing in the stau sector, 
we found two other possibilities to obtain small stau yields: co-annihilation with EWinos 
with a significant higgsino admixture as well as co-annihilation with stops or sbottoms 
with considerable left-right mixing---in both cases annihilation near the resonant pole 
of an $s$-channel Higgs is required.
We found allowed points down to $Y\simeq5\times10^{-15}$ and $Y\simeq10^{-14}$ 
in the former and latter case, respectively.

Thus, our results show that the current constraints on the parameter
space of the MSSM with a long-lived stau NLSP still allow for a stau
relic abundance small enough to satisfy the strict bounds from big bang
nucleosynthesis.  The smallness of the corresponding region in parameter
space suggests distinct features that will be probed in the upcoming LHC
run.

\subsection*{Acknowledgements}

We would like to thank Torsten Bringmann, Guillaume Chalons, Laura Covi, Werner Porod, 
Loic Quertenmont, J\"urgen Reuter, Oscar St\aa l, Frank Steffen, Dominik St\"ockinger, 
Georg Weiglein and Martin Winkler for very helpful discussions.
This work was supported by the German Research Foundation (DFG) via the
Junior Research Group `SUSY Phenomenology' and an `SFB Fellowship' within the Collaborative
Research Center 676 `Particles, Strings and the Early Universe'. BP also acknowledges 
the support of the State of S\~{a}o Paulo Research Foundation (FAPESP).

\section*{Appendix}
\begin{appendix}

\section{Cosmological quantities}\label{sec:cosmo}

In this appendix, we briefly introduce the cosmological quantities used in section 
\ref{sec:surveygeneral}. A more comprehensive description can be found in
\cite{Griest:1990kh,Gondolo:1990dk,Kolb:1990vq,Edsjo:1997bg}. We consider the 
total number density which is the sum of all number densities of supersymmetric 
particles. It is governed by the Boltzmann equation
\begin{equation}
\label{eq:Boltzmann}
  \frac{\D n}{\D t} =
  -3Hn - \sveff \left( n^2 - n_\text{eq}^2 \right) ,
\end{equation}
where $n_\text{eq}$ is the number density in thermal equilibrium and $H$ is the 
Hubble parameter. In terms of the yield $Y \equiv n/s$ and $x \equiv m/T$, this 
equation can be rewritten as~\cite{Gondolo:1990dk}
\begin{equation} 
 \frac{\D Y}{\D x} = \sqrt{\frac{8 \bar g}{45}} \pi\Mp \frac{m}{x^2} \,
 \sveff \left( Y_\text{eq}^2 - Y^2 \right) ,
\end{equation}
which leads to \Eqref{eq:boltzsolY} in the freeze-out approximation, i.e., after 
neglecting $Y_\text{eq}$. Here $s$ is the entropy density, $m$ is the mass of 
the lightest MSSM sparticle and $\bar{g}$ is a degrees of freedom parameter,
\begin{equation}
  \sqrt{\bar g} = \frac{g_{*S}}{\sqrt{g_{*}}}
  \left( 1+\frac{T}{3g_{*S}} \frac{\D g_{*S}}{\D T}
  \right) .
\end{equation}
The parameters $g_*$ and $g_{*S}$ count the effective numbers of relativistic 
degrees of freedom according to
\begin{align}
 g_*&=\sum_{\text{bosons}}g_i\left( \frac{T_i}{T}\right)^4 +
  \frac{7}{8}\sum_{\text{fermions}}g_i\left( \frac{T_i}{T}\right)^4,
\\
 g_{*S}&=\sum_{\text{bosons}}g_i\left( \frac{T_i}{T}\right)^3 +
  \frac{7}{8}\sum_{\text{fermions}}g_i\left( \frac{T_i}{T}\right)^3 .
\end{align}
The thermally averaged annihilation cross section is defined by
\beq
\label{eq:sigmaveffdef}
 \sveff = \sum_{ij} \langle
  \sigma_{ij}v_{ij} \rangle \frac{n_{i}^{\text{eq}}}{n^{\text{eq}}}
  \frac{n_{j}^{\text{eq}}}{n^{\text{eq}}}\,,
\eeq
where the sum runs over all supersymmetric initial state particles $i,j$. Further,
$n_{i,j}^\text{eq}$ and $n^\text{eq}$ are the individual and total equilibrium 
number densities, respectively. 
The thermal average $\langle\sigma_{ij}v_{ij}\rangle$ is given by
\beq
\label{eq:sigmaijvijdef}
  \langle \sigma_{ij}v_{ij} \rangle = \frac{\int \!\D^3
  \vec{p}_{i}\D^3\vec{p}_{j} \,
  f_{i}f_{j}\sigma_{ij}v_{ij}}
  {\int \!\D^3\vec{p}_{i}\D^3\vec{p}_{j}\, f_{i}f_{j}}\,,
\eeq
where $\vec{p}_i$ and $f_i$ are the three-momentum and the equilibrium 
phase-space density of particle $i$, respectively. Further, $v_{ij}$ is the 
M\o ller velocity, defined by
\beq
\label{eq:molvdef}
  v_{ij} = \frac{\sqrt{(p_{i} \cdot p_{j})^2-m_{i}^2 m_{j}^2}}{E_{i} E_{j}}\,.
\eeq

\subsubsection*{Yield $Y$ and density fraction $\Omega$}

The relation between the yield and the density fraction $\Omega$ of a relic 
particle is
\begin{equation}
\Omega=\frac{\rho_0}{\rho_\text{c}}=\frac{m Y s_0}{\rho_\text{c}}\,,
\end{equation}
where $\rho_0$ would be the current density of the relic if it had not decayed, 
$\rho_c$ is the critical density, and $s_0$ is the current entropy density of the 
universe. Inserting the numerical values \cite{Beringer:1900zz} yields
\begin{equation}\label{eq:masteryield_micro}
Y = 3.747 \times 10^{-9}\,\Omega h^2\,\frac{\GeV}{m}\,.
\end{equation}
This expression is used to compute the yield from the output of \MO.

\section{Mixing in the stau sector}
\label{app:con}

Considering real parameters, we denote the stau mass matrix by
\begin{equation} \label{eq:stau-matrix}
    \mathcal{M}_{\stau}^2 = 
    \begin{pmatrix}
       m_{\text{LL}}^2   & m_{\tau} X_{\tau} \\
       m_{\tau} X_{\tau} & m_{\text{RR}}^2      
    \end{pmatrix} = 
    \mathcal{R}_{\stau}^T
    \begin{pmatrix}
       \mstau^2 & 0 \\
       0 & \mstautwo^2   
    \end{pmatrix}
    \mathcal{R}_{\stau}\,,
\end{equation}
where
\begin{align}
\label{eq:matrix-entries}
    m_\text{LL}^2 &=  
    m^2_{\s{L}_3} + m_\tau^2 +
    \left( T^3_\tau - Q_\tau \sin^2\theta_{\text{W}} \right) M_Z^2 \cos{2\beta} \,,
\\
\label{eq:mRR}
    m_{\text{RR}}^2 &=
    m^2_{\s{e}_3} + m_\tau^2 +
    Q_\tau \sin^2\theta_{\text{W}} M_Z^2 \cos{2\beta} \,,
\\
    X_\tau &= A_\tau - \mu \tan\beta \,,
\end{align}
with $T^3$ and $Q$ referring to the weak isospin and the electric charge, 
respectively. The stau mixing matrix reads
\begin{equation}
    \label{eq:mixing-matrix}
    \mathcal{R}_{\stau} =
    \begin{pmatrix}
       \cos\thest & \sin\thest \\
       -\sin\thest & \cos\thest
    \end{pmatrix}\,.
\end{equation}
The lighter mass eigenstate $\stau_1$ is then given by
\begin{equation}
\stau_1 = \cos\theta_\stau\,\stau_\text{L}+\sin\theta_\stau\,\stau_\text{R} \,.
\end{equation}
Choosing $0 \leq \thest < \pi$ as in \cite{Lindert:2011td}, $\thest=0$ corresponds
to $\stau_1=\stau_\text{L}$, whereas $\thest=\pi/2$ corresponds to 
$\stau_1=\stau_\text{R}$.  In these special cases, $X_\tau=0$.  Maximal mixing 
is obtained at $\thest=\pi/4$ and or $\thest=3\pi/4$.

From the above equations, we can infer
\begin{align}
\label{eq:mstau-mRRmLL}
    \mstau^2 &=
    \frac{1}{2} \left[ m_\text{RR}^2 + m_\text{LL}^2 -
    \sqrt{(m_\text{RR}^2-m_\text{LL}^2)^2 + 4m_\tau^2X_\tau^2} \right] ,
\\
\label{eq:mstautwo-mRRmLL}
    \mstautwo^2 &=
    \frac{1}{2} \left[ m_\text{RR}^2 + m_\text{LL}^2 +
    \sqrt{(m_\text{RR}^2-m_\text{LL}^2)^2 + 4m_\tau^2X_\tau^2} \right] ,
\\
\label{eq:deltaM_NLOSP_LOSP}
    \sin 2\thest &=
    \frac{2 m_\tau X_\tau}{\mstau^2 - \mstautwo^2} \,,
\\
\label{eq:deltaM_stauRRLL}
    \tan 2\thest &= \frac{2 m_\tau X_\tau}{m_\text{LL}^2 - m_\text{RR}^2}\,,
\\
\label{eq:mLLinv1}
    m_\text{LL}^2 &= \mstau^2 - m_\tau X_\tau \tan\thest \,,
\\
\label{eq:mRRinv1}
    m_\text{RR}^2 &= \mstau^2 - m_\tau X_\tau \cot\thest \,.
\end{align}

By inserting \Eqref{eq:mLLinv1} into \Eqref{eq:matrix-entries} and
\Eqref{eq:mRRinv1} into \Eqref{eq:mRR}, respectively, we can express
$m_{\s{L}_3}$ and $m_{\s{e}_3}$ in terms of $\mstau$, $X_\tau$, $\thest$
and $\tan\beta$ at tree level.  Thus, all tree-level input parameters
for \textsc{SuSpect} are derived from the scan parameters in
table~\ref{tab:scanlim}.

The corresponding expressions for the third-generation squarks are
obtained by obvious replacements, except for
\begin{align}
    X_t &= A_t - \mu \cot\beta \,, \\
    X_b &= A_b - \mu \tan\beta \,.
\end{align}

\section{Sfermion-sfermion-Higgs couplings} \label{sec:sfersferhiggs}

In the MSSM, the couplings of the lighter mass eigenstates of the third
generation sfermions, $\stau_1,\s b_1$ and $\s t_1$, to the $CP$-even
neutral Higgses $h$ and $H$ are given by 
\bea
 C[h,\stau_{1},\stau_{1}] &
 =& \frac{g}{2M_W}\left\{m_{\tau}s_{2\theta_\stau}\left[\mu\frac{c_\alpha}{c_\beta}+A_\tau\frac{s_\alpha}{c_\beta}\right]  
\right.\nonumber \\
\label{eq:c_h_ll}
& &  \left.+ 2m_\tau^{2}\frac{s_\alpha}{c_\beta} + M_{W}^{2}s_{\alpha+\beta}\left[(t_{\text{w}}^{2}-1)c_{\theta_\stau}^{2}
-  2t_{\text{w}}^{2}s_{\theta_\stau}^{2}\right] \right\}, \\
C[H,\stau_{1},\stau_{1}] &
=& \frac{g}{2M_W}\left\{m_{\tau}s_{2\theta_\stau}\left[\mu\frac{s_\alpha}{c_\beta}-A_\tau\frac{c_\alpha}{c_\beta}\right] 
\right.\nonumber \\ 
& & \left. - 2m_\tau^{2}\frac{c_\alpha}{c_\beta} 
                          - M_{W}^{2}c_{\alpha+\beta}\left[(t_{\text{w}}^{2}-1)c_{\theta_\stau}^{2}-
                            2t_{\text{w}}^{2}s_{\theta_\stau}^{2}\right] \right\}, \\
C[h,\sboto_{1},\sboto_{1}] &
=& \frac{g}{2M_W}\left\{m_{b}s_{2\theta_\sboto}\left[\mu\frac{c_\alpha}{c_\beta}+A_b\frac{s_\alpha}{c_\beta}\right] 
\right.\nonumber \\ 
& & \left. + 2m_b^{2}\frac{s_\alpha}{c_\beta} 
                           - \frac{M_{W}^{2}}{3}s_{\alpha+\beta}\left[(t_{\text{w}}^{2}+3)c_{\theta_\sboto}^{2}+
                                2t_{\text{w}}^{2}s_{\theta_\sboto}^{2}\right] \right\},
\\
C[H,\sboto_{1},\sboto_{1}] &
=& \frac{g}{2M_W}\left\{ m_{b}s_{2\theta_\sboto}\left[\mu\frac{s_\alpha}{c_\beta}-A_b\frac{c_\alpha}{c_\beta}\right] 
\right.\nonumber \\ 
& & \left. - 2m_b^{2}\frac{s_\alpha}{c_\beta} 
                           + \frac{M_{W}^{2}}{3}c_{\alpha+\beta}\left[(t_{\text{w}}^{2}+3)c_{\theta_\sboto}^{2}+
                                2t_{\text{w}}^{2}s_{\theta_\sboto}^{2}\right] \right\}, \\
C[h,\stopo_{1},\stopo_{1}] &=
& \frac{-g}{2M_W}\left\{ m_{t}s_{2\theta_\stopo}\left[\mu\frac{s_\alpha}{s_\beta}+A_t\frac{c_\alpha}{s_\beta}\right] 
\right.\nonumber \\ 
& & \left. + 2m_t^{2}\frac{c_\alpha}{s_\beta} 
                           + \frac{M_{W}^{2}}{3}s_{\alpha+\beta}\left[(t_{\text{w}}^{2}-3)c_{\theta_\stopo}^{2}-
                                4t_{\text{w}}^{2}s_{\theta_\stopo}^{2}\right] \right\}, \\
C[H,\stopo_{1},\stopo_{1}] &
=& \frac{g}{2M_W}\left\{ m_{t}s_{2\theta_\stopo}\left[\mu\frac{c_\alpha}{s_\beta}-A_t\frac{s_\alpha}{s_\beta}\right] 
\right.\nonumber \\ 
& & \left. - 2m_t^{2}\frac{s_\alpha}{s_\beta} 
                           + \frac{M_{W}^{2}}{3}c_{\alpha+\beta}\left[(t_{\text{w}}^{2}-3)c_{\theta_\stopo}^{2}-
                                4t_{\text{w}}^{2}s_{\theta_\stopo}^{2}\right] \right\}, 
\eea
where we have abbreviated $c_\alpha \equiv \cos\alpha$, $s_\alpha \equiv \sin\alpha$ 
and $t_{\text{w}} \equiv \tan\theta_{W}$. In the decoupling limit $m_A \gg M_Z$
these expressions simplify according to $\alpha\to\beta-\frac{\pi}{2}$, cf., e.g.,
\cite{Martin:1997ns}. The first terms in the above equations are the leading 
contributions in the parameter space regions with enhanced sfermion-Higgs couplings.

\section{Resummation of the Higgs-bottom couplings}  \label{sec:resum}

The tree-level Higgs-bottom couplings in the MSSM read (see, e.g., \cite{Haber:1997dt})
\begin{align}
h_{hb\bar b}^\text{tree}&= -\frac{m_b}{v}\frac{\sin\alpha}{\cos\beta} =
 -\frac{m_b}{v}\big[\sin(\beta-\alpha)-\tan\beta\cos(\beta-\alpha)\big]
\\
h_{Hb\bar b}^\text{tree}&= \frac{m_b}{v}\frac{\cos\alpha}{\cos\beta} =
\frac{m_b}{v}\big[\cos(\beta-\alpha)+\tan\beta\sin(\beta-\alpha)\big]
\end{align}
Radiative corrections to these couplings can be significant
\cite{Hall:1993gn,Hempfling:1993kv,Carena:1994bv,Pierce:1996zz}. 
For positive $\mu$ and $A_t$, they typically lead to a suppression of the couplings. 
The leading $\tan\beta$-enhanced terms can be resummed to all orders in 
perturbation theory \cite{Carena:1999py,Heinemeyer:2004xw} leading to the 
approximate relative corrections \cite{Carena:1998gk}
\begin{align}
   \frac{h_{hb\bar b}}{h_{hb\bar b}^{\text{tree}}}
   &\simeq 1-\frac{\Delta(m_b)}{1+\Delta(m_b)}\left(1+\frac{1}{\tan\alpha\tan\beta}\right) ,
\\
   \frac{h_{Hb\bar b}}{h_{Hb\bar b}^{\text{tree}}}
   &\simeq 1-\frac{\Delta(m_b)}{1+\Delta(m_b)}\left(1-\frac{\tan\alpha}{\tan\beta}\right) .
\end{align}
The leading contributions to $\Delta(m_b)$ come from the gluino-sbottom loop and
from the charged higgsino-stop, wino-stop and wino-sbottom loops and are given by
\cite{Carena:1999py}
\beq
\begin{split}
\Delta(m_b)\simeq
  \frac{2\alpha_{\text{s}}}{3\pi} \mgo \mu \tan\beta\,
  I(m_{\s b_1}, m_{\s b_2}, \mgo) + \frac{h_t^2}{16\pi^2} \mu A_t \tan \beta\,
  I(m_{\s t_1}, m_{\s t_2},\mu)\\
 - \frac{g_2^2}{16\pi^2} \mu M_2 \tan \beta\left[
  \cos^2\theta_{\s t} \,I(m_{\s t_1}, M_2,\mu)
  +\sin^2\theta_{\s t} \,I(m_{\s t_2},M_2,\mu)\phantom{\frac12}\right.\\
  \left. +\frac12\cos^2\theta_{\s b} \,I(m_{\s b_1}, M_2,\mu)
  +\frac12\sin^2\theta_{\s b} \,I(m_{\s b_2},M_2,\mu)\right]\,,
\end{split}
\eeq
where
\beq
  I(a,b,c) = \frac{1}{(a^2-b^2)(b^2-c^2)(a^2-c^2)}
  \left[ a^2b^2 \log\frac{a^2}{b^2} + b^2c^2 \log\frac{b^2}{c^2} +
    c^2a^2 \log\frac{c^2}{a^2} \right]. 
\label{eq:functI}
\eeq
Note that in the decoupling limit, $\alpha\simeq\beta-\pi/2$, so the $hb\bar b$ 
coupling remains SM-like even in the presence of large values for $\Delta(m_b)$ 
since $\tan\alpha\tan\beta\simeq-1$. Thus, the correction vanishes. The 
$Hb\bar b$ coupling reads
\beq
h_{Hb\bar b}= \frac{m_b}{v}\tan\beta
\left[1-\frac{\Delta(m_b)}{1+\Delta(m_b)}\left(1+\cot^2\beta\right)\right]
\eeq
in the decoupling limit.

\section{Ranges for the dedicated scans} \label{sec:dedscan}

In table \ref{tab:scandetailed} we list all dedicated scan regions for the 17-dimensional 
pMSSM scan introduced in section \ref{sec:pMSSMsummary}. The dedicated scan 
regions are motivated by the results of section~\ref{sec:systscan}. In each region, the 
table displays those parameters that are constrained to a smaller range than given in
table \ref{tab:scanlim}. All parameters that are not listed for a particular range are 
scanned over according to table \ref{tab:scanlim}. Further, the lower and upper limits 
for the parameters listed in table \ref{tab:scanlim} are respected in any case.

\begin{table}[b!]
\centering
\renewcommand{\arraystretch}{1.3}
\begin{tabular}{c|c|c|c|c}
Dedicated scan region & Pct. & Parameter & Range & Prior  \\
 \hline
  blind scan & $43.3\%$ & -- & --  & --\\ 
 \hline
  $\stau_1$-$H$ resonance & $4.8\%$& $\left|m_A/\mstau-2\right| $& $[5\times10^{-4};0.5]$ & log \\  
                              & & $\sgn(m_A/\mstau-2)$& $\{-1,1\}$ & -- \\  
 \hline
  $\stau_1$-$H$ threshold & $6.7\%$& $\left|m_A/\mstau-1\right| $& $[10^{-3};2]$ & log \\  
                              & & $\sgn(m_A/\mstau-1)$& $\{-1,1\}$ & -- \\    
 \hline
 $M_1$-$\mu$ co-ann.\ resonance& $4.9\%$ & $M_1/\mstau-1 $& $[5\times10^{-4};0.5]$ & log \\  
                              & &  $\left|\mu/M_1-1\right| $& $[2\times10^{-4};0.2]$ & log \\  
                              & & $\sgn(\mu/M_1-1)$& $\{-1,1\}$ & -- \\    
                              & &  $\left|m_A/\sqrt{\mu M_1}-2\right| $& $[5\times10^{-4};0.5]$ & log \\  
                              & & $\sgn(m_A/\sqrt{\mu M_1}-2)$& $\{-1,1\}$ & -- \\    
 \hline
 $M_2$-$\mu$ co-ann.\ resonance& $4.9\%$ & $M_2/\mstau-1 $& $[5\times10^{-4};0.5]$ & log \\  
                              & &  $\left|\mu/M_2-1\right| $& $[2\times10^{-4};0.2]$ & log \\  
                              & & $\sgn(\mu/M_2-1)$& $\{-1,1\}$ & -- \\    
                              & &  $\left|m_A/\sqrt{\mu M_2}-2\right| $& $[5\times10^{-4};0.5]$ & log \\  
                              & & $\sgn(m_A/\sqrt{\mu M_2}-2)$& $\{-1,1\}$ & -- \\   
 \hline
  $\s t_1$ co-annihilation & $4.3\%$ & $m_{\s t_1}/\mstau-1 $& $[5\times10^{-4};0.5]$ & log \\  
 \hline
  $\s t_1$ co-ann.\ resonance & $2.2\%$ & $m_{\s t_1}/\mstau-1 $& $[5\times10^{-4};0.5]$ & log \\  
                               & &  $|m_A/m_{\s t_1}-2| $& $[5\times10^{-4};0.25]$ & log \\  
                              & & $\sgn(m_A/m_{\s t_1}-2)$& $\{-1,1\}$ & -- \\  
 \hline
  $\s b_1$ co-annihilation & $9.1\%$ & $m_{\s b_1}/\mstau-1 $& $[5\times10^{-4};0.5]$ & log \\  
 \hline
$\s b_1$ co-ann.\ resonance & $2.3\%$ & $m_{\s b_1}/\mstau-1 $& $[5\times10^{-4};0.5]$ & log \\  
                               & &  $|m_A/m_{\s b_1}-2| $& $[5\times10^{-4};0.25]$ & log \\  
                              & & $\sgn(m_A/m_{\s b_1}-2)$& $\{-1,1\}$ & -- \\  
  \hline
$\go$ co-annihilation & $10.7\%$ & $\mgo/\mstau-1 $& $[5\times10^{-4};0.5]$ & log\\
  \hline
$\sq$ co-annihilation & $6.8\%$ & $\msq/\mstau-1 $& $[5\times10^{-4};0.5]$ & log
\end{tabular}
\caption{Summary of all scan regions, the corresponding percentage of points, and the
parameters whose scan ranges deviate from the ones given in table \ref{tab:scanlim}.
All parameters not listed are scanned over  according to table \ref{tab:scanlim}.
We generated a total amount of $5\times10^5$ points in the 17-dimensional 
parameter space.
}
\label{tab:scandetailed}
\end{table}

\end{appendix}

\clearpage

\clearpage
\addcontentsline{toc}{section}{References}
\bibliographystyle{utphys}
\bibliography{staus}

\providecommand{\href}[2]{#2}\begingroup\raggedright\begin{thebibliography}{100}

\bibitem{Falomkin:1984eu}
I.~V. Falomkin, G.~B. Pontecorvo, M.~G. Sapozhnikov, M.~Y. Khlopov,
  F.~Balestra, and G.~Piragino, ``Low-energy $\bar p\; ^4${He} annihilation and
  problems of the modern cosmology, {GUT} and {SUSY} models'',
  \href{http://dx.doi.org/10.1007/BF02831163}{{\em Nuovo Cim.} {\bf A79} (1984)
   193--204}.
[Yad.\ Fiz.\ {\bf 39} (1984), 990].

\bibitem{Ellis:1984eq}
J.~R. Ellis, J.~E. Kim, and D.~V. Nanopoulos, ``{Cosmological Gravitino
  Regeneration and Decay}'',
\href{http://dx.doi.org/10.1016/0370-2693(84)90334-4}{{\em Phys. Lett.} {\bf
  B145} (1984)  181}.

\bibitem{Ellis:1984er}
J.~R. Ellis, D.~V. Nanopoulos, and S.~Sarkar, ``{The Cosmology of Decaying
  Gravitinos}'',
\href{http://dx.doi.org/10.1016/0550-3213(85)90306-2}{{\em Nucl. Phys.} {\bf
  B259} (1985)  175}.

\bibitem{Bolz:1998ek}
M.~Bolz, W.~Buchm{\"u}ller, and M.~Pl{\"u}macher, ``{Baryon asymmetry and dark
  matter}'', \href{http://dx.doi.org/10.1016/S0370-2693(98)01342-2}{{\em Phys.
  Lett.} {\bf B443} (1998)  209--213},
\href{http://arxiv.org/abs/hep-ph/9809381}{{\tt arXiv:hep-ph/9809381}}.

\bibitem{Fayet:1981sq}
P.~Fayet, ``Experimental consequences of supersymmetry'', in {\em Proceedings
  of the XVIth Rencontre de Moriond}, J.~{Tran Thanh Van}, ed., vol.~1,
  pp.~347--367.
\newblock Editions Frontieres,
1981.
\newblock

\bibitem{Pagels:1981ke}
H.~Pagels and J.~R. Primack, ``{Supersymmetry, Cosmology and New TeV
  Physics}'',
\href{http://dx.doi.org/10.1103/PhysRevLett.48.223}{{\em Phys. Rev. Lett.} {\bf
  48} (1982)  223}.

\bibitem{Moroi:1993mb}
T.~Moroi, H.~Murayama, and M.~Yamaguchi, ``Cosmological constraints on the
  light stable gravitino'',
\href{http://dx.doi.org/10.1016/0370-2693(93)91434-O}{{\em Phys. Lett.} {\bf
  B303} (1993)  289--294}.

\bibitem{Pospelov:2006sc}
M.~Pospelov, ``Particle physics catalysis of thermal {B}ig {B}ang
  {N}ucleosynthesis'',
  \href{http://dx.doi.org/10.1103/PhysRevLett.98.231301}{{\em Phys. Rev. Lett.}
  {\bf 98} (2007)  231301},
\href{http://arxiv.org/abs/hep-ph/0605215}{{\tt arXiv:hep-ph/0605215}}.

\bibitem{Jedamzik:2007qk}
K.~Jedamzik, ``{Bounds on long-lived charged massive particles from Big Bang
  nucleosynthesis}'',
  \href{http://dx.doi.org/10.1088/1475-7516/2008/03/008}{{\em JCAP} {\bf 0803}
  (2008)  008},
\href{http://arxiv.org/abs/0710.5153}{{\tt arXiv:0710.5153 [hep-ph]}}.

\bibitem{Kawasaki:2008qe}
M.~Kawasaki, K.~Kohri, T.~Moroi, and A.~Yotsuyanagi, ``{Big-Bang
  Nucleosynthesis and Gravitino}'',
  \href{http://dx.doi.org/10.1103/PhysRevD.78.065011}{{\em Phys. Rev.} {\bf
  D78} (2008)  065011},
\href{http://arxiv.org/abs/0804.3745}{{\tt arXiv:0804.3745 [hep-ph]}}.

\bibitem{Asaka:2000zh}
T.~Asaka, K.~Hamaguchi, and K.~Suzuki, ``{Cosmological gravitino problem in
  gauge mediated supersymmetry breaking models}'',
  \href{http://dx.doi.org/10.1016/S0370-2693(00)00959-X}{{\em Phys. Lett.} {\bf
  B490} (2000)  136--146},
\href{http://arxiv.org/abs/hep-ph/0005136}{{\tt arXiv:hep-ph/0005136}}.

\bibitem{Pradler:2008qc}
J.~Pradler and F.~D. Steffen, ``{Thermal relic abundances of long-lived
  staus}'', \href{http://dx.doi.org/10.1016/j.nuclphysb.2008.10.009}{{\em Nucl.
  Phys.} {\bf B809} (2009)  318--346},
\href{http://arxiv.org/abs/0808.2462}{{\tt arXiv:0808.2462 [hep-ph]}}.

\bibitem{Ratz:2008qh}
M.~Ratz, K.~Schmidt-Hoberg, and M.~W. Winkler, ``{A Note on the primordial
  abundance of stau NLSPs}'',
  \href{http://dx.doi.org/10.1088/1475-7516/2008/10/026}{{\em JCAP} {\bf 0810}
  (2008)  026},
\href{http://arxiv.org/abs/0808.0829}{{\tt arXiv:0808.0829 [hep-ph]}}.

\bibitem{CMS1305.0491}
CMS Collaboration, S.~Chatrchyan {\em et al.}, ``Searches for long-lived
  charged particles in pp collisions at $\sqrt{s}=7$ and 8 {TeV}'',
  \href{http://dx.doi.org/10.1007/JHEP07(2013)122}{{\em JHEP} {\bf 1307} (2013)
   122},
\href{http://arxiv.org/abs/1305.0491}{{\tt arXiv:1305.0491 [hep-ex]}}.

\bibitem{Buchmuller:2006tt}
W.~Buchm{\"u}ller, K.~Hamaguchi, M.~Ibe, and T.~T. Yanagida, ``{Eluding the BBN
  constraints on the stable gravitino}'',
  \href{http://dx.doi.org/10.1016/j.physletb.2006.07.070}{{\em Phys. Lett.}
  {\bf B643} (2006)  124--126},
\href{http://arxiv.org/abs/hep-ph/0605164}{{\tt arXiv:hep-ph/0605164}}.

\bibitem{Lee:1977ua}
B.~W. Lee and S.~Weinberg, ``{Cosmological Lower Bound on Heavy Neutrino
  Masses}'',
\href{http://dx.doi.org/10.1103/PhysRevLett.39.165}{{\em Phys. Rev. Lett.} {\bf
  39} (1977)  165--168}.

\bibitem{Binetruy:1983jf}
P.~Binetruy, G.~Girardi, and P.~Salati, ``{Constraints on a system of two
  neutral fermions from cosmology}'',
\href{http://dx.doi.org/10.1016/0550-3213(84)90161-5}{{\em Nucl. Phys.} {\bf
  B237} (1984)  285}.

\bibitem{Bernstein:1985th}
J.~Bernstein, L.~S. Brown, and G.~Feinberg, ``The cosmological heavy neutrino
  problem revisited'',
\href{http://dx.doi.org/10.1103/PhysRevD.32.3261}{{\em Phys. Rev.} {\bf D32}
  (1985)  3261}.

\bibitem{Srednicki:1988ce}
M.~Srednicki, R.~Watkins, and K.~A. Olive, ``{Calculations of Relic Densities
  in the Early Universe}'',
\href{http://dx.doi.org/10.1016/0550-3213(88)90099-5}{{\em Nucl. Phys.} {\bf
  B310} (1988)  693}.

\bibitem{Griest:1990kh}
K.~Griest and D.~Seckel, ``{Three exceptions in the calculation of relic
  abundances}'',
\href{http://dx.doi.org/10.1103/PhysRevD.43.3191}{{\em Phys. Rev.} {\bf D43}
  (1991)  3191--3203}.

\bibitem{Gondolo:1990dk}
P.~Gondolo and G.~Gelmini, ``{Cosmic abundances of stable particles: Improved
  analysis}'',
\href{http://dx.doi.org/10.1016/0550-3213(91)90438-4}{{\em Nucl. Phys.} {\bf
  B360} (1991)  145--179}.

\bibitem{Kolb:1990vq}
E.~W. Kolb and M.~S. Turner, ``{The Early universe}'',
{\em Front. Phys.} {\bf 69} (1990)  1--547.

\bibitem{Edsjo:1997bg}
J.~Edsj{\"o} and P.~Gondolo, ``{Neutralino relic density including
  coannihilations}'', \href{http://dx.doi.org/10.1103/PhysRevD.56.1879}{{\em
  Phys. Rev.} {\bf D56} (1997)  1879--1894},
\href{http://arxiv.org/abs/hep-ph/9704361}{{\tt arXiv:hep-ph/9704361}}.

\bibitem{Belanger:2008sj}
G.~Belanger, F.~Boudjema, A.~Pukhov, and A.~Semenov, ``{Dark matter direct
  detection rate in a generic model with micrOMEGAs 2.2}'',
  \href{http://dx.doi.org/10.1016/j.cpc.2008.11.019}{{\em Comput. Phys.
  Commun.} {\bf 180} (2009)  747--767},
\href{http://arxiv.org/abs/0803.2360}{{\tt arXiv:0803.2360 [hep-ph]}}.

\bibitem{Berger:2008ti}
C.~F. Berger, L.~Covi, S.~Kraml, and F.~Palorini, ``{The Number density of a
  charged relic}'', \href{http://dx.doi.org/10.1088/1475-7516/2008/10/005}{{\em
  JCAP} {\bf 0810} (2008)  005},
\href{http://arxiv.org/abs/0807.0211}{{\tt arXiv:0807.0211 [hep-ph]}}.

\bibitem{Djouadi:2002ze}
A.~Djouadi, J.-L. Kneur, and G.~Moultaka, ``{SuSpect: A Fortran code for the
  supersymmetric and Higgs particle spectrum in the MSSM}'',
  \href{http://dx.doi.org/10.1016/j.cpc.2006.11.009}{{\em Comput. Phys.
  Commun.} {\bf 176} (2007)  426--455},
\href{http://arxiv.org/abs/hep-ph/0211331}{{\tt arXiv:hep-ph/0211331}}.

\bibitem{Steffen:2006hw}
F.~D. Steffen, ``{Gravitino dark matter and cosmological constraints}'',
  \href{http://dx.doi.org/10.1088/1475-7516/2006/09/001}{{\em JCAP} {\bf 0609}
  (2006)  001},
\href{http://arxiv.org/abs/hep-ph/0605306}{{\tt arXiv:hep-ph/0605306}}.

\bibitem{Gherghetta:1998tq}
T.~Gherghetta, G.~Giudice, and A.~Riotto, ``{Nucleosynthesis bounds in gauge
  mediated supersymmetry breaking theories}'',
  \href{http://dx.doi.org/10.1016/S0370-2693(98)01527-5}{{\em Phys. Lett.} {\bf
  B446} (1999)  28--36},
\href{http://arxiv.org/abs/hep-ph/9808401}{{\tt arXiv:hep-ph/9808401}}.

\bibitem{Belyaev:2012qa}
A.~Belyaev, N.~D. Christensen, and A.~Pukhov, ``{CalcHEP 3.4 for collider
  physics within and beyond the Standard Model}'',
  \href{http://dx.doi.org/10.1016/j.cpc.2013.01.014}{{\em Comput. Phys.
  Commun.} {\bf 184} (2013)  1729--1769},
\href{http://arxiv.org/abs/1207.6082}{{\tt arXiv:1207.6082 [hep-ph]}}.

\bibitem{Drees:1992am}
M.~Drees and M.~M. Nojiri, ``{The Neutralino relic density in minimal $N=1$
  supergravity}'', \href{http://dx.doi.org/10.1103/PhysRevD.47.376}{{\em Phys.
  Rev.} {\bf D47} (1993)  376--408},
\href{http://arxiv.org/abs/hep-ph/9207234}{{\tt arXiv:hep-ph/9207234}}.

\bibitem{Arnowitt:1993mg}
R.~L. Arnowitt and P.~Nath, ``{Cosmological constraints and SU(5) supergravity
  grand unification}'',
  \href{http://dx.doi.org/10.1016/0370-2693(93)90883-J}{{\em Phys. Lett.} {\bf
  B299} (1993)  58--63},
\href{http://arxiv.org/abs/hep-ph/9302317}{{\tt arXiv:hep-ph/9302317}}.

\bibitem{Baer:1997ai}
H.~Baer and M.~Brhlik, ``{Neutralino dark matter in minimal supergravity:
  Direct detection versus collider searches}'',
  \href{http://dx.doi.org/10.1103/PhysRevD.57.567}{{\em Phys. Rev.} {\bf D57}
  (1998)  567--577},
\href{http://arxiv.org/abs/hep-ph/9706509}{{\tt arXiv:hep-ph/9706509}}.

\bibitem{Baer:2000jj}
H.~Baer, M.~Brhlik, M.~A. D{\'i}az, J.~Ferrandis, P.~Mercadante, P.~Quintana,
  and X.~Tata, ``{Yukawa unified supersymmetric $SO(10)$ model: Cosmology, rare
  decays, and collider searches}'',
  \href{http://dx.doi.org/10.1103/PhysRevD.63.015007}{{\em Phys. Rev.} {\bf
  D63} (2000)  015007},
\href{http://arxiv.org/abs/hep-ph/0005027}{{\tt arXiv:hep-ph/0005027}}.

\bibitem{Ellis:2001msa}
J.~R. Ellis, T.~Falk, G.~Ganis, K.~A. Olive, and M.~Srednicki, ``{The CMSSM
  parameter space at large $\tan\beta$}'',
  \href{http://dx.doi.org/10.1016/S0370-2693(01)00541-X}{{\em Phys. Lett.} {\bf
  B510} (2001)  236--246},
\href{http://arxiv.org/abs/hep-ph/0102098}{{\tt arXiv:hep-ph/0102098}}.

\bibitem{Lahanas:2001yr}
A.~B. Lahanas and V.~C. Spanos, ``{Implications of the pseudoscalar Higgs boson
  in determining the neutralino dark matter}'',
  \href{http://dx.doi.org/10.1007/s100520100861}{{\em Eur. Phys. J.} {\bf C23}
  (2002)  185--190},
\href{http://arxiv.org/abs/hep-ph/0106345}{{\tt arXiv:hep-ph/0106345}}.

\bibitem{Carena:1995wu}
M.~S. Carena, M.~Quiros, and C.~E.~M. Wagner, ``{Effective potential methods
  and the Higgs mass spectrum in the MSSM}'',
  \href{http://dx.doi.org/10.1016/0550-3213(95)00665-6}{{\em Nucl. Phys.} {\bf
  B461} (1996)  407--436},
\href{http://arxiv.org/abs/hep-ph/9508343}{{\tt arXiv:hep-ph/9508343}}.

\bibitem{Heinemeyer:1999zf}
S.~Heinemeyer, W.~Hollik, and G.~Weiglein, ``{Constraints on $\tan \beta$ in
  the MSSM from the upper bound on the mass of the lightest Higgs boson}'',
  \href{http://dx.doi.org/10.1088/1126-6708/2000/06/009}{{\em JHEP} {\bf 0006}
  (2000)  009},
\href{http://arxiv.org/abs/hep-ph/9909540}{{\tt arXiv:hep-ph/9909540}}.

\bibitem{Endo:2010ya}
M.~Endo, K.~Hamaguchi, and K.~Nakaji, ``{Probing High Reheating Temperature
  Scenarios at the LHC with Long-Lived Staus}'',
  \href{http://dx.doi.org/10.1007/JHEP11(2010)004}{{\em JHEP} {\bf 11} (2010)
  004},
\href{http://arxiv.org/abs/1008.2307}{{\tt arXiv:1008.2307 [hep-ph]}}.

\bibitem{CahillRowley:2012cb}
M.~W. Cahill-Rowley, J.~L. Hewett, S.~Hoeche, A.~Ismail, and T.~G. Rizzo,
  ``{The new look pMSSM with neutralino and gravitino LSPs}'',
  \href{http://dx.doi.org/10.1140/epjc/s10052-012-2156-1}{{\em Eur. Phys. J.}
  {\bf C72} (2012)  2156},
\href{http://arxiv.org/abs/1206.4321}{{\tt arXiv:1206.4321 [hep-ph]}}.

\bibitem{Heinemeyer:1998yj}
S.~Heinemeyer, W.~Hollik, and G.~Weiglein, ``{FeynHiggs: A Program for the
  calculation of the masses of the neutral $CP$-even Higgs bosons in the
  MSSM}'', \href{http://dx.doi.org/10.1016/S0010-4655(99)00364-1}{{\em Comput.
  Phys. Commun.} {\bf 124} (2000)  76--89},
\href{http://arxiv.org/abs/hep-ph/9812320}{{\tt arXiv:hep-ph/9812320}}.

\bibitem{Djouadi:2006bz}
A.~Djouadi, M.~M. M{\"u}hlleitner, and M.~Spira, ``{Decays of supersymmetric
  particles --- The program SUSY-HIT}'', {\em Acta Phys. Polon.} {\bf B38}
  (2007)  635--644,
\href{http://arxiv.org/abs/hep-ph/0609292}{{\tt arXiv:hep-ph/0609292}}.

\bibitem{Kraml:2007sx}
S.~Kraml and D.~T. Nhung, ``{Three-body decays of sleptons in models with
  non-universal Higgs masses}'',
  \href{http://dx.doi.org/10.1088/1126-6708/2008/02/061}{{\em JHEP} {\bf 0802}
  (2008)  061},
\href{http://arxiv.org/abs/0712.1986}{{\tt arXiv:0712.1986 [hep-ph]}}.

\bibitem{Kilian:2007gr}
W.~Kilian, T.~Ohl, and J.~Reuter, ``{WHIZARD: Simulating Multi-Particle
  Processes at LHC and ILC}'',
  \href{http://dx.doi.org/10.1140/epjc/s10052-011-1742-y}{{\em Eur. Phys. J.}
  {\bf C71} (2011)  1742}, \href{http://arxiv.org/abs/0708.4233}{{\tt
  arXiv:0708.4233 [hep-ph]}}.
\url{http://projects.hepforge.org/whizard/}.

\bibitem{Bechtle:2011sb}
P.~Bechtle, O.~Brein, S.~Heinemeyer, G.~Weiglein, and K.~E. Williams,
  ``{HiggsBounds 2.0.0: Confronting Neutral and Charged Higgs Sector
  Predictions with Exclusion Bounds from LEP and the Tevatron}'',
  \href{http://dx.doi.org/10.1016/j.cpc.2011.07.015}{{\em Comput. Phys.
  Commun.} {\bf 182} (2011)  2605--2631},
\href{http://arxiv.org/abs/1102.1898}{{\tt arXiv:1102.1898 [hep-ph]}}.

\bibitem{1997NuPhB.492...51B}
W.~Beenakker, R.~H{\"o}pker, M.~Spira, and P.~M. Zerwas, ``{Squark and gluino
  production at hadron colliders}'',
  \href{http://dx.doi.org/10.1016/S0550-3213(97)80027-2}{{\em Nucl. Phys.} {\bf
  B492} (1997)  51--103}, \href{http://arxiv.org/abs/hep-ph/9610490}{{\tt
  arXiv:hep-ph/9610490}}.
  \url{http://www.thphys.uni-heidelberg.de/~plehn/index.php?show=prospino}.

\bibitem{1999PhRvL..83.3780B}
W.~{Beenakker}, M.~{Klasen}, M.~{Kr{\"a}mer}, T.~{Plehn}, M.~{Spira}, and P.~M.
  {Zerwas}, ``{Production of Charginos, Neutralinos, and Sleptons at Hadron
  Colliders}'', \href{http://dx.doi.org/10.1103/PhysRevLett.83.3780}{{\em Phys.
  Rev. Lett.} {\bf 83} (1999)  3780--3783},
  \href{http://arxiv.org/abs/hep-ph/9906298}{{\tt arXiv:hep-ph/9906298}}.
  Erratum \href{http://dx.doi.org/10.1103/PhysRevLett.100.029901}{{\em ibid.}
  {\bf 100} (2008) 029901}.

\bibitem{Plehn:2004rp}
T.~Plehn, ``{Measuring the MSSM Lagrangean}'', {\em Czech. J. Phys.} {\bf 55}
  (2005)  B213--B220,
\href{http://arxiv.org/abs/hep-ph/0410063}{{\tt arXiv:hep-ph/0410063}}.

\bibitem{Beenakker:1997ut}
W.~Beenakker, M.~Kr{\"a}mer, T.~Plehn, M.~Spira, and P.~M. Zerwas, ``{Stop
  production at hadron colliders}'',
  \href{http://dx.doi.org/10.1016/S0550-3213(98)00014-5}{{\em Nucl. Phys.} {\bf
  B515} (1998)  3--14},
\href{http://arxiv.org/abs/hep-ph/9710451}{{\tt arXiv:hep-ph/9710451}}.

\bibitem{Beenakker:2009ha}
W.~Beenakker, S.~Brensing, M.~Kr{\"a}mer, A.~Kulesza, E.~Laenen, and
  I.~Niessen, ``{Soft-gluon resummation for squark and gluino
  hadroproduction}'',
  \href{http://dx.doi.org/10.1088/1126-6708/2009/12/041}{{\em JHEP} {\bf 0912}
  (2009)  041},
\href{http://arxiv.org/abs/0909.4418}{{\tt arXiv:0909.4418 [hep-ph]}}.

\bibitem{Beenakker:2010nq}
W.~Beenakker, S.~Brensing, M.~Kr{\"a}mer, A.~Kulesza, E.~Laenen, and
  I.~Niessen, ``{Supersymmetric top and bottom squark production at hadron
  colliders}'', \href{http://dx.doi.org/10.1007/JHEP08(2010)098}{{\em JHEP}
  {\bf 1008} (2010)  098},
\href{http://arxiv.org/abs/1006.4771}{{\tt arXiv:1006.4771 [hep-ph]}}.

\bibitem{Kulesza:2008jb}
A.~Kulesza and L.~Motyka, ``{Threshold resummation for squark-antisquark and
  gluino-pair production at the LHC}'',
  \href{http://dx.doi.org/10.1103/PhysRevLett.102.111802}{{\em Phys. Rev.
  Lett.} {\bf 102} (2009)  111802},
\href{http://arxiv.org/abs/0807.2405}{{\tt arXiv:0807.2405 [hep-ph]}}.

\bibitem{Kulesza:2009kq}
A.~Kulesza and L.~Motyka, ``{Soft gluon resummation for the production of
  gluino-gluino and squark-antisquark pairs at the LHC}'',
  \href{http://dx.doi.org/10.1103/PhysRevD.80.095004}{{\em Phys. Rev.} {\bf
  D80} (2009)  095004},
\href{http://arxiv.org/abs/0905.4749}{{\tt arXiv:0905.4749 [hep-ph]}}.

\bibitem{Heisig:2012ep}
J.~Heisig, ``Long-lived staus at the {L}{H}{C}'', in {\em Proceedings of the
  XLVIIth Rencontres de Moriond}, E.~Aug{\'e} {\em et al.}, eds., pp.~487--490.
\newblock ARISF, 2012.
\newblock
\href{http://arxiv.org/abs/1207.3058}{{\tt arXiv:1207.3058 [hep-ph]}}.
\newblock

\bibitem{Heisig:2012zq}
J.~Heisig and J.~Kersten, ``{Long-lived staus from strong production in a
  simplified model approach}'',
  \href{http://dx.doi.org/10.1103/PhysRevD.86.055020}{{\em Phys. Rev.} {\bf
  D86} (2012)  055020},
\href{http://arxiv.org/abs/1203.1581}{{\tt arXiv:1203.1581 [hep-ph]}}.

\bibitem{ATLAS-CONF-2013-014}
{{ATLAS} Collaboration}, ``Combined measurements of the mass and signal
  strength of the {H}iggs-like boson with the {A}{T}{L}{A}{S} detector using up
  to 25 fb$^{-1}$ of proton-proton collision data'',  ATLAS-CONF-2013-014,
  CERN, Geneva, Mar, 2013.
\newblock \url{http://cds.cern.ch/record/1523727}.

\bibitem{CMS-PAS-HIG-13-005}
{{CMS} Collaboration}, ``Combination of standard model {H}iggs boson searches
  and measurements of the properties of the new boson with a mass near
  $125\,${G}e{V}'',  CMS-PAS-HIG-13-005, CERN, Geneva, 2013.
\newblock \url{http://cds.cern.ch/record/1542387}.

\bibitem{Degrassi:2002fi}
G.~Degrassi, S.~Heinemeyer, W.~Hollik, P.~Slavich, and G.~Weiglein, ``{Towards
  high precision predictions for the MSSM Higgs sector}'',
  \href{http://dx.doi.org/10.1140/epjc/s2003-01152-2}{{\em Eur. Phys. J.} {\bf
  C28} (2003)  133--143},
\href{http://arxiv.org/abs/hep-ph/0212020}{{\tt arXiv:hep-ph/0212020}}.

\bibitem{CahillRowley:2012rv}
M.~W. Cahill-Rowley, J.~L. Hewett, A.~Ismail, and T.~G. Rizzo, ``{The Higgs
  Sector and Fine-Tuning in the pMSSM}'',
  \href{http://dx.doi.org/10.1103/PhysRevD.86.075015}{{\em Phys. Rev.} {\bf
  D86} (2012)  075015},
\href{http://arxiv.org/abs/1206.5800}{{\tt arXiv:1206.5800 [hep-ph]}}.

\bibitem{Carena:2011aa}
M.~Carena, S.~Gori, N.~R. Shah, and C.~E.~M. Wagner, ``{A 125 GeV SM-like Higgs
  in the MSSM and the $\gamma \gamma$ rate}'',
  \href{http://dx.doi.org/10.1007/JHEP03(2012)014}{{\em JHEP} {\bf 1203} (2012)
   014},
\href{http://arxiv.org/abs/1112.3336}{{\tt arXiv:1112.3336 [hep-ph]}}.

\bibitem{ATLAS:2012vd}
ATLAS Collaboration, G.~Aad {\em et al.}, ``{Searches for heavy long-lived
  sleptons and $R$-hadrons with the ATLAS detector in $pp$ collisions at
  $\sqrt{s} = 7$ TeV}'',
  \href{http://dx.doi.org/10.1016/j.physletb.2013.02.015}{{\em Phys. Lett.}
  {\bf B720} (2013)  277--308},
\href{http://arxiv.org/abs/1211.1597}{{\tt arXiv:1211.1597 [hep-ex]}}.

\bibitem{Chatrchyan1205.0272}
CMS Collaboration, S.~Chatrchyan {\em et al.}, ``{Search for heavy long-lived
  charged particles in pp collisions at $\sqrt{s}=7$ TeV}'',
  \href{http://dx.doi.org/10.1016/j.physletb.2012.06.023}{{\em Phys. Lett.}
  {\bf B713} (2012)  408--433},
\href{http://arxiv.org/abs/1205.0272}{{\tt arXiv:1205.0272 [hep-ex]}}.

\bibitem{Heisig:2011dr}
J.~Heisig and J.~Kersten, ``{Production of long-lived staus in the Drell-Yan
  process}'', \href{http://dx.doi.org/10.1103/PhysRevD.84.115009}{{\em Phys.
  Rev.} {\bf D84} (2011)  115009},
\href{http://arxiv.org/abs/1106.0764}{{\tt arXiv:1106.0764 [hep-ph]}}.

\bibitem{Lindert:2011td}
J.~M. Lindert, F.~D. Steffen, and M.~K. Trenkel, ``{Direct stau production at
  hadron colliders in cosmologically motivated scenarios}'',
  \href{http://dx.doi.org/10.1007/JHEP08(2011)151}{{\em JHEP} {\bf 1108} (2011)
   151},
\href{http://arxiv.org/abs/1106.4005}{{\tt arXiv:1106.4005 [hep-ph]}}.

\bibitem{Spira:1995rr}
M.~Spira, A.~Djouadi, D.~Graudenz, and P.~M. Zerwas, ``{Higgs boson production
  at the LHC}'', \href{http://dx.doi.org/10.1016/0550-3213(95)00379-7}{{\em
  Nucl. Phys.} {\bf B453} (1995)  17--82},
\href{http://arxiv.org/abs/hep-ph/9504378}{{\tt arXiv:hep-ph/9504378}}.

\bibitem{Muhlleitner:2003vg}
M.~M{\"u}hlleitner, A.~Djouadi, and Y.~Mambrini, ``{SDECAY: A Fortran code for
  the decays of the supersymmetric particles in the MSSM}'',
  \href{http://dx.doi.org/10.1016/j.cpc.2005.01.012}{{\em Comput. Phys.
  Commun.} {\bf 168} (2005)  46--70},
\href{http://arxiv.org/abs/hep-ph/0311167}{{\tt arXiv:hep-ph/0311167}}.

\bibitem{Bechtle:2013xfa}
P.~Bechtle, S.~Heinemeyer, O.~St\r{a}l, T.~Stefaniak, and G.~Weiglein,
  ``{HiggsSignals: Confronting arbitrary Higgs sectors with measurements at the
  Tevatron and the LHC}'',
  \href{http://dx.doi.org/10.1140/epjc/s10052-013-2711-4}{{\em Eur. Phys. J.}
  {\bf C74} (2014)  2711},
\href{http://arxiv.org/abs/1305.1933}{{\tt arXiv:1305.1933 [hep-ph]}}.

\bibitem{CMS-PAS-HIG-12-050}
{CMS Collaboration}, ``{Search for MSSM Neutral Higgs Bosons Decaying to Tau
  Pairs in pp Collisions}'',  CMS-PAS-HIG-12-050, CERN, Geneva, Nov, 2012.
\newblock \url{http://cds.cern.ch/record/1493521}.

\bibitem{Schael:2006cr}
ALEPH Collaboration, DELPHI Collaboration, L3 Collaboration, OPAL
  Collaboration, LEP Working Group for Higgs Boson Searches, S.~Schael {\em et
  al.}, ``{Search for neutral MSSM Higgs bosons at LEP}'',
  \href{http://dx.doi.org/10.1140/epjc/s2006-02569-7}{{\em Eur. Phys. J.} {\bf
  C47} (2006)  547--587},
\href{http://arxiv.org/abs/hep-ex/0602042}{{\tt arXiv:hep-ex/0602042}}.

\bibitem{CMS-PAS-HIG-13-002}
{CMS Collaboration}, ``{Properties of the {H}iggs-like boson in the decay $H
  \to ZZ \to 4\ell$ in pp collisions at $\sqrt{s} =7$ and $8\,${T}e{V}}'',
  CMS-PAS-HIG-13-002, CERN, Geneva, Mar, 2013.
\newblock \url{http://cds.cern.ch/record/1523767}.

\bibitem{CMS-PAS-HIG-12-045}
{{CMS} Collaboration}, ``{Combination of standard model Higgs boson searches
  and measurements of the properties of the new boson with a mass near 125
  GeV}'',  CMS-PAS-HIG-12-045, CERN, Geneva, Nov, 2012.
\newblock \url{http://cds.cern.ch/record/1494149}.

\bibitem{Group:2012gb}
CDF Collaboration, D0 Collaboration, {Tevatron Electroweak Working Group},
  ``{2012 Update of the Combination of CDF and D0 Results for the Mass of the
  $W$ Boson}'',
\href{http://arxiv.org/abs/1204.0042}{{\tt arXiv:1204.0042 [hep-ex]}}.

\bibitem{Bechtle:2012jw}
P.~Bechtle, S.~Heinemeyer, O.~St\r{a}l, T.~Stefaniak, G.~Weiglein, and
  L.~Zeune, ``{MSSM Interpretations of the LHC Discovery: Light or Heavy
  Higgs?}'', \href{http://dx.doi.org/10.1140/epjc/s10052-013-2354-5}{{\em Eur.
  Phys. J.} {\bf C73} (2013)  2354},
\href{http://arxiv.org/abs/1211.1955}{{\tt arXiv:1211.1955 [hep-ph]}}.

\bibitem{Heinemeyer:2006px}
S.~Heinemeyer, W.~Hollik, D.~St{\"o}ckinger, A.~M. Weber, and G.~Weiglein,
  ``{Precise prediction for $M_W$ in the MSSM}'',
  \href{http://dx.doi.org/10.1088/1126-6708/2006/08/052}{{\em JHEP} {\bf 0608}
  (2006)  052},
\href{http://arxiv.org/abs/hep-ph/0604147}{{\tt arXiv:hep-ph/0604147}}.

\bibitem{HFAGbsgAug12}
Heavy Flavor Averaging Group, Y.~Amhis {\em et al.}, ``{World Average Branching
  Fraction for $B \to X_s \gamma$}'', , Aug, 2012.
\newblock
  \url{http://www.slac.stanford.edu/xorg/hfag/rare/2012/radll/btosg.pdf}.

\bibitem{:2012ct}
LHCb Collaboration, R.~Aaij {\em et al.}, ``{First Evidence for the Decay
  $B_s^0 \to \mu^+ \mu^-$}'',
  \href{http://dx.doi.org/10.1103/PhysRevLett.110.021801}{{\em Phys. Rev.
  Lett.} {\bf 110} (2013)  021801},
\href{http://arxiv.org/abs/1211.2674}{{\tt arXiv:1211.2674 [hep-ex]}}.

\bibitem{Chatrchyan:2013bka}
CMS Collaboration, S.~Chatrchyan {\em et al.}, ``{Measurement of the $B_s^0 \to
  \mu^+ \mu^-$ branching fraction and search for $B^0 \to \mu^+ \mu^-$ with the
  CMS Experiment}'',
  \href{http://dx.doi.org/10.1103/PhysRevLett.111.101804}{{\em Phys. Rev.
  Lett.} {\bf 111} (2013)  101804},
\href{http://arxiv.org/abs/1307.5025}{{\tt arXiv:1307.5025 [hep-ex]}}.

\bibitem{Rattazzi:1996fb}
R.~Rattazzi and U.~Sarid, ``{Large $\tan\beta$ in gauge-mediated SUSY-breaking
  models}'', \href{http://dx.doi.org/10.1016/S0550-3213(97)00363-5}{{\em Nucl.
  Phys.} {\bf B501} (1997)  297--331},
\href{http://arxiv.org/abs/hep-ph/9612464}{{\tt arXiv:hep-ph/9612464}}.

\bibitem{Hisano:2010re}
J.~Hisano and S.~Sugiyama, ``{Charge-breaking constraints on left-right mixing
  of stau's}'', \href{http://dx.doi.org/10.1016/j.physletb.2010.12.013}{{\em
  Phys. Lett.} {\bf B696} (2011)  92--96},
  \href{http://arxiv.org/abs/1011.0260}{{\tt arXiv:1011.0260 [hep-ph]}}.
Erratum \href{http://dx.doi.org/10.1016/j.physletb.2013.01.018}{{\em ibid.}
  {\bf B719} (2013) 472}.

\bibitem{Carena:2012mw}
M.~Carena, S.~Gori, I.~Low, N.~R. Shah, and C.~E.~M. Wagner, ``{Vacuum
  stability and Higgs diphoton decays in the MSSM}'',
  \href{http://dx.doi.org/10.1007/JHEP02(2013)114}{{\em JHEP} {\bf 1302} (2013)
   114},
\href{http://arxiv.org/abs/1211.6136}{{\tt arXiv:1211.6136 [hep-ph]}}.

\bibitem{Kitahara:2013lfa}
T.~Kitahara and T.~Yoshinaga, ``{Stau with large mass difference and
  enhancement of $h\to\gamma\gamma$ decay rate in the MSSM}'',
  \href{http://dx.doi.org/10.1007/JHEP05(2013)035}{{\em JHEP} {\bf 1305} (2013)
   035},
\href{http://arxiv.org/abs/1303.0461}{{\tt arXiv:1303.0461 [hep-ph]}}.

\bibitem{Frere:1983ag}
J.~Fr{\`e}re, D.~Jones, and S.~Raby, ``{Fermion masses and induction of the
  weak scale by supergravity}'',
\href{http://dx.doi.org/10.1016/0550-3213(83)90606-5}{{\em Nucl. Phys.} {\bf
  B222} (1983)  11}.

\bibitem{AlvarezGaume:1983gj}
L.~Alvarez-Gaum{\'e}, J.~Polchinski, and M.~B. Wise, ``{Minimal low-energy
  supergravity}'',
\href{http://dx.doi.org/10.1016/0550-3213(83)90591-6}{{\em Nucl. Phys.} {\bf
  B221} (1983)  495}.

\bibitem{Claudson:1983et}
M.~Claudson, L.~J. Hall, and I.~Hinchliffe, ``{Low-energy supergravity: False
  vacua and vacuous predictions}'',
\href{http://dx.doi.org/10.1016/0550-3213(83)90556-4}{{\em Nucl. Phys.} {\bf
  B228} (1983)  501}.

\bibitem{Kounnas:1983td}
C.~Kounnas, A.~B. Lahanas, D.~V. Nanopoulos, and M.~Quir{\'o}s, ``{Low-energy
  behaviour of realistic locally-supersymmetric grand unified theories}'',
\href{http://dx.doi.org/10.1016/0550-3213(84)90545-5}{{\em Nucl. Phys.} {\bf
  B236} (1984)  438}.

\bibitem{Derendinger:1983bz}
J.~Derendinger and C.~A. Savoy, ``{Quantum effects and
  $\text{SU}(2)\times\text{U}(1)$ breaking in supergravity gauge theories}'',
\href{http://dx.doi.org/10.1016/0550-3213(84)90162-7}{{\em Nucl. Phys.} {\bf
  B237} (1984)  307}.

\bibitem{Gunion:1987qv}
J.~Gunion, H.~Haber, and M.~Sher, ``{Charge/Color Breaking Minima and
  $A$-Parameter Bounds in Supersymmetric Models}'',
\href{http://dx.doi.org/10.1016/0550-3213(88)90168-X}{{\em Nucl. Phys.} {\bf
  B306} (1988)  1}.

\bibitem{Casas:1995pd}
J.~Casas, A.~Lleyda, and C.~Mu{\~n}oz, ``{Strong constraints on the parameter
  space of the MSSM from charge and color breaking minima}'',
  \href{http://dx.doi.org/10.1016/0550-3213(96)00194-0}{{\em Nucl. Phys.} {\bf
  B471} (1996)  3--58},
\href{http://arxiv.org/abs/hep-ph/9507294}{{\tt arXiv:hep-ph/9507294}}.

\bibitem{Ferreira:2000hg}
P.~Ferreira, ``{A full one-loop charge symmetry breaking effective
  potential}'', \href{http://dx.doi.org/10.1016/S0370-2693(01)00552-4}{{\em
  Phys. Lett.} {\bf B509} (2001)  120--130},
  \href{http://arxiv.org/abs/hep-ph/0008115}{{\tt arXiv:hep-ph/0008115}}.
Erratum \href{http://dx.doi.org/10.1016/S0370-2693(01)01029-2}{{\em ibid.} {\bf
  B518} (2001) 333}.

\bibitem{Ferreira:2001tk}
P.~Ferreira, ``{Minimization of a one-loop charge breaking effective
  potential}'', \href{http://dx.doi.org/10.1016/S0370-2693(01)00716-X}{{\em
  Phys. Lett.} {\bf B512} (2001)  379--391},
  \href{http://arxiv.org/abs/hep-ph/0102141}{{\tt arXiv:hep-ph/0102141}}.
Erratum \href{http://dx.doi.org/10.1016/S0370-2693(01)01028-0}{{\em ibid.} {\bf
  B518} (2001) 334}.

\bibitem{Camargo-Molina:2013sta}
J.~Camargo-Molina, B.~O'Leary, W.~Porod, and F.~Staub, ``{Stability of the
  CMSSM against sfermion VEVs}'',
  \href{http://dx.doi.org/10.1007/JHEP12(2013)103}{{\em JHEP} {\bf 1312} (2013)
   103},
\href{http://arxiv.org/abs/1309.7212}{{\tt arXiv:1309.7212 [hep-ph]}}.

\bibitem{Camargo-Molina:2013qva}
J.~E. Camargo-Molina, B.~O'Leary, W.~Porod, and F.~Staub, ``{Vevacious: A Tool
  For Finding The Global Minima Of One-Loop Effective Potentials With Many
  Scalars}'', \href{http://dx.doi.org/10.1140/epjc/s10052-013-2588-2}{{\em Eur.
  Phys. J.} {\bf C73} (2013)  2588},
\href{http://arxiv.org/abs/1307.1477}{{\tt arXiv:1307.1477 [hep-ph]}}.

\bibitem{Griest:1989wd}
K.~Griest and M.~Kamionkowski, ``{Unitarity Limits on the Mass and Radius of
  Dark Matter Particles}'',
\href{http://dx.doi.org/10.1103/PhysRevLett.64.615}{{\em Phys. Rev. Lett.} {\bf
  64} (1990)  615}.

\bibitem{Beringer:1900zz}
Particle Data Group, J.~Beringer {\em et al.}, ``{Review of Particle
  Physics}'',
\href{http://dx.doi.org/10.1103/PhysRevD.86.010001}{{\em Phys. Rev.} {\bf D86}
  (2012)  010001}.

\bibitem{Martin:1997ns}
S.~P. Martin, ``{A Supersymmetry primer}'',
\href{http://arxiv.org/abs/hep-ph/9709356}{{\tt arXiv:hep-ph/9709356}}.

\bibitem{Haber:1997dt}
H.~E. Haber, ``{Higgs boson masses and couplings in the minimal supersymmetric
  model}'',
\href{http://arxiv.org/abs/hep-ph/9707213}{{\tt arXiv:hep-ph/9707213}}.

\bibitem{Hall:1993gn}
L.~J. Hall, R.~Rattazzi, and U.~Sarid, ``{Top quark mass in supersymmetric
  SO(10) unification}'', \href{http://dx.doi.org/10.1103/PhysRevD.50.7048}{{\em
  Phys. Rev.} {\bf D50} (1994)  7048--7065},
\href{http://arxiv.org/abs/hep-ph/9306309}{{\tt arXiv:hep-ph/9306309}}.

\bibitem{Hempfling:1993kv}
R.~Hempfling, ``{Yukawa coupling unification with supersymmetric threshold
  corrections}'',
\href{http://dx.doi.org/10.1103/PhysRevD.49.6168}{{\em Phys. Rev.} {\bf D49}
  (1994)  6168--6172}.

\bibitem{Carena:1994bv}
M.~S. Carena, M.~Olechowski, S.~Pokorski, and C.~E.~M. Wagner, ``{Electroweak
  symmetry breaking and bottom-top Yukawa unification}'',
  \href{http://dx.doi.org/10.1016/0550-3213(94)90313-1}{{\em Nucl. Phys.} {\bf
  B426} (1994)  269--300},
\href{http://arxiv.org/abs/hep-ph/9402253}{{\tt arXiv:hep-ph/9402253}}.

\bibitem{Pierce:1996zz}
D.~M. Pierce, J.~A. Bagger, K.~T. Matchev, and R.-j. Zhang, ``{Precision
  corrections in the minimal supersymmetric standard model}'',
  \href{http://dx.doi.org/10.1016/S0550-3213(96)00683-9}{{\em Nucl. Phys.} {\bf
  B491} (1997)  3--67},
\href{http://arxiv.org/abs/hep-ph/9606211}{{\tt arXiv:hep-ph/9606211}}.

\bibitem{Carena:1999py}
M.~S. Carena, D.~Garcia, U.~Nierste, and C.~E.~M. Wagner, ``{Effective
  Lagrangian for the $\bar{t} b H^{+}$ interaction in the MSSM and charged
  Higgs phenomenology}'',
  \href{http://dx.doi.org/10.1016/S0550-3213(00)00146-2}{{\em Nucl. Phys.} {\bf
  B577} (2000)  88--120},
\href{http://arxiv.org/abs/hep-ph/9912516}{{\tt arXiv:hep-ph/9912516}}.

\bibitem{Heinemeyer:2004xw}
S.~Heinemeyer, W.~Hollik, H.~Rzehak, and G.~Weiglein, ``{High-precision
  predictions for the MSSM Higgs sector at $\mathcal{O}(\alpha_b \alpha_s)$}'',
  \href{http://dx.doi.org/10.1140/epjc/s2005-02112-6}{{\em Eur. Phys. J.} {\bf
  C39} (2005)  465--481},
\href{http://arxiv.org/abs/hep-ph/0411114}{{\tt arXiv:hep-ph/0411114}}.

\bibitem{Carena:1998gk}
M.~S. Carena, S.~Mrenna, and C.~Wagner, ``{MSSM Higgs boson phenomenology at
  the Tevatron collider}'',
  \href{http://dx.doi.org/10.1103/PhysRevD.60.075010}{{\em Phys. Rev.} {\bf
  D60} (1999)  075010},
\href{http://arxiv.org/abs/hep-ph/9808312}{{\tt arXiv:hep-ph/9808312}}.

\end{thebibliography}\endgroup

\end{document} 

